\documentclass[twocolumn]{aastex62}

\submitjournal{ApJ}
\usepackage{amsmath}
\usepackage{comment}
\usepackage{cleveref}

\definecolor{SecureBlue}{HTML}{E3F2FD}   
\definecolor{ProbYellow}{HTML}{FFF9C4}   
 \usepackage[table]{xcolor}

\crefformat{section}{\S#2#1#3} 
\crefformat{subsection}{\S#2#1#3}
\crefformat{subsubsection}{\S#2#1#3}

\shortauthors{Rinaldi et al.}

\begin{document}

\title{\bf Filling the Gap in Cluster Evolution: JWST’s Glimpse into a Young, Star-Forming Cluster at Cosmic Noon}

\newcommand{\gsim}{{\;\raise0.3ex\hbox{$>$\kern-0.75em\raise-1.1ex\hbox{$\sim$}}\;}}

\correspondingauthor{Pierluigi Rinaldi}
\email{prinaldi@stsci.edu}

\author[0000-0002-5104-8245]{Pierluigi Rinaldi}
\affiliation{Space Telescope Science Institute, 3700 San Martin Drive, Baltimore, Maryland 21218, USA}
\affiliation{Steward Observatory, University of Arizona, 933 North Cherry Avenue, Tucson, AZ 85721, USA}

\author[0000-0002-8909-8782]{Stacey Alberts}
\affiliation{AURA for the European Space Agency (ESA), Space Telescope Science Institute, 3700 San Martin Dr., Baltimore, MD 21218, USA}
\affiliation{Steward Observatory, University of Arizona, 933 North Cherry Avenue, Tucson, AZ 85721, USA}

\author[0000-0001-9262-9997]{Christopher N. A. Willmer}
\affiliation{Steward Observatory, University of Arizona, 933 North Cherry Avenue, Tucson, AZ 85721, USA}

\author[0000-0001-6301-3667]{Courtney Carreira}
\affiliation{Department of Astronomy and Astrophysics, University of California, Santa Cruz, 1156 High Street, Santa Cruz, CA 95064, USA}

\author[0000-0003-2919-7495]{Christina C. Williams}
\affiliation{NSF National Optical-Infrared Astronomy Research Laboratory, 950 North Cherry Avenue, Tucson, AZ 85719, USA}

\author[0000-0000-0000-0000]{Gaël Noirot}
\affiliation{Space Telescope Science Institute, 3700 San Martin Drive, Baltimore, Maryland 21218, USA}

\author[0009-0001-4012-3043]{Carys J. E. Gilbert}
\affiliation{Department of Astronomy, University of Cape Town, Private Bag X3, Rondebosch 7701, South Africa}

\author[0000-0002-8651-9879]{Andrew J. Bunker}
\affiliation{Department of Physics, University of Oxford, Denys Wilkinson Building, Keble Road, Oxford OX13RH, UK}

\author[0000-0003-0215-1104]{William M. Baker}
\affiliation{DARK, Niels Bohr Institute, University of Copenhagen, Jagtvej 128, DK-2200 Copenhagen, Denmark}

\author[0000-0003-3419-538X]{Luigi Barchiesi} 
\affiliation{Department of Astronomy, University of Cape Town, Private Bag X3, Rondebosch 7701, South Africa}

\author[0000-0001-7673-2257]{Zhiyuan Ji}
\affiliation{Steward Observatory, University of Arizona, 933 North Cherry Avenue, Tucson, AZ 85721, USA}

\author[0000-0002-6221-1829]{Jianwei Lyu}
\affiliation{Steward Observatory, University of Arizona, 933 North Cherry Avenue, Tucson, AZ 85721, USA}

\author[0000-0002-8224-4505]{Sandro Tacchella}
\affiliation{Kavli Institute for Cosmology, University of Cambridge, Madingley Road, Cambridge, CB3 0HA, UK}
\affiliation{Cavendish Laboratory, University of Cambridge, 19 JJ Thomson Avenue, Cambridge, CB3 0HE, UK}

\author[0000-0002-8876-5248]{Zihao Wu}
\affiliation{Center for Astrophysics $|$ Harvard \& Smithsonian, 60 Garden St., Cambridge MA 02138 USA}

\author[0000-0003-3307-7525]{Yongda Zhu}
\affiliation{Steward Observatory, University of Arizona, 933 North Cherry Avenue, Tucson, AZ 85721, USA}

\begin{abstract}
We present a detailed study of HUDFJ0332.4–2746.6 (HUDF46), a $z \approx 1.84$ overdensity in the Hubble Ultra Deep Field, previously identified with HST as a proto-cluster. JWST/NIRISS spectroscopy expands its confirmed membership from 18 to 37 galaxies, while deep HST/ACS, JWST/NIRCam, and JWST/MIRI imaging provide a comprehensive multiwavelength view from the rest-frame UV to the mid-infrared. This dataset probes the population across three dex in stellar mass ($M_\bigstar \approx 10^{7.5\text{--}10.5}\,M_\odot$), delivering the first direct view of a young cluster down to such low-$M_\bigstar$ at $z\gtrsim1$. Assuming virialization, we derive a velocity dispersion of $\sigma \approx 670\pm 91\,\mathrm{km\,s^{-1}}$ and a halo mass of $M_{200} \approx (1.2\pm0.2) \times 10^{14}\,M_\odot$, in agreement with X-ray constraints from deep {\it Chandra} data. Despite residing in a massive halo likely in the hot-halo regime, the population is overwhelmingly star-forming, with no established red sequence and no extended X-ray emission from a hot intracluster medium. HUDF46 members have stellar and structural properties nearly indistinguishable from coeval field galaxies, and the structure hosts only one AGN candidate, found in its brightest galaxy, which lies at the cluster center. Overall, HUDF46 appears to be in a transitional phase prior to the onset of environmental quenching, making its galaxy population a key benchmark for tracing the processes that will later build a passive population and shape the assembly of massive clusters at later cosmic times.

\end{abstract}

\keywords{Galaxy evolution (594); Galaxy formation (595); Spectral energy distribution (2129);  High-redshift galaxies (734); Dwarf galaxies
(416); Galaxy environments (2029); Galaxy quenching (2040)}
 
\section{Introduction}

In the $\Lambda$CDM cosmological framework, structure formation proceeds through the collapse of primordial density fluctuations into dark matter halos \citep{Press_Formation_1974, White_core_1978}. Baryons subsequently fall into these potential wells, cool, and form stars, giving rise to galaxies. As gravity drives the hierarchical growth of structure, halos merge into increasingly massive systems, eventually forming the largest gravitationally bound structures in the Universe: galaxy clusters, which host hundreds to thousands of galaxies \citep{Kravtsov_Formation_2012}.

Given their cosmological importance, galaxy clusters have long been the focus of extensive observational and theoretical studies, revealing a strong connection between galaxy properties and environment. Early evidence for this link was provided by \citet{butcher_evolution_1978} and later on by \citet{dressler_galaxy_1980}, who showed that galaxies in dense regions exhibit distinct morphologies and colors compared to field galaxies at low redshift. Subsequent work has confirmed and extended these results across cosmic time (see \citealt{dressler_evolution_1984, boselli_environmental_2006, de_lucia_build-up_2007, blanton_physical_2009, boselli_galex_2014, alberts_clusters_2022}, and references therein), demonstrating that environmental processes play a fundamental role in driving galaxy evolution (e.g., \citealt{peng_mass_2010, thomas_environment_2010}).

As observations extended to higher redshift, studies showed that environmental effects are already in place at $z \gtrsim 1$ (e.g., \citealt{rettura_early-type_2011, muzzin_gemini_2012, noble_kinematic_2013, balogh_evidence_2016, kawinwanichakij_effect_2017, ji_evidence_2018}), with field galaxies assembling their stellar mass ($M_{\bigstar}$) over longer timescales, while cluster galaxies are predominantly early-type galaxies (ETGs), defining what is now commonly defined as a red sequence in galaxy clusters, with smooth, ellipsoidal structures and old stellar populations out to $z \approx 1.5\text{--}2$ (see \citealt{alberts_clusters_2022}). Together, these results indicate that dense environments accelerate galaxy evolution and quenching (e.g., \citealt{mei_evolution_2009, stanford_idcs_2012, muzzin_discovery_2013}), with such signatures already in place at high redshift (e.g., \citealt{li_epochs_2025}), motivating constraints on their underlying mechanisms and onset epoch.

However, {\it how do these processes take place?} In massive dark matter halos tracing group- and cluster-scale structures, gas is expected to be shock-heated to high virial temperatures, suppressing cold accretion and establishing a hot intracluster medium (ICM; \citealt{birnboim_virial_2003, dekel_galaxy_2006}). The presence of a hot ICM is typically adopted as a signpost of already virialized, massive systems \citep{foley_discovery_2011, mantz_deep_2020}, through its detection in X-ray emission \citep{Sarazin_x-ray_1986} and via the Sunyaev–Zel'dovich effect \citep{birkinshaw_sunyaev-zeldovich_1999}\footnote{Searches for clusters using these methods have historically been biased toward more mature, already virialized systems.}. Galaxies accreting onto these halos may experience ram-pressure stripping and tidal perturbations \citep{gunn_infall_1972, moore_morphological_1998}, potentially even before full cluster infall \citep{berrier_assembly_2009, mcgee_accretion_2009}. In parallel, increasingly efficient feedback from Active Galactic Nuclei (AGNs), both energetic and maintenance modes, can further regulate gas cooling and re-accretion \citep{croton_many_2006, henriques_galaxy_2017}. Although halo assembly is well characterized, the epoch at which these environmental processes begin to significantly affect galaxies remains empirically uncertain.

Addressing {\it when} and {\it how} these transformations occur requires observing clusters during their early assembly phase, before environmental processes have fully reshaped their galaxy populations. At this stage, these systems lack a developed hot ICM, rendering them undetectable through classical methods (e.g., X-ray and Sunyaev–Zel'dovich). At these early stages, structures commonly identified as protoclusters or overdensities are not yet virialized, and at high redshift not all overdensities are expected to collapse into present-day clusters (\citealt{muldrew_what_2015}). As a result, the distinction between groups, clusters, and protoclusters is often applied pragmatically, since observational limitations rarely permit an unambiguous classification (see \citealt{overzier_realm_2016}).

With the advent of JWST \citep{gardner_james_2023}, the identification and characterization of galaxy overdensities and group-scale structures—likely progenitors of present-day clusters—has become increasingly routine, owing to the unprecedented depth and resolution of NIRCam \citep{rieke_performance_2023} and MIRI \citep{wright_mid-infrared_2023}. These capabilities enable the detection of overdense regions even during the Epoch of Reionization \citep{helton_identification_2024, Fudamoto_sapphires_2025, wu_jades_2026}. Crucially, JWST now allows us to probe the low-mass regime within these structures, revealing that low-mass quiescent galaxies may follow evolutionary pathways distinct from their massive counterparts. Recent analyses (see discussion in \citealt{cutler_structure_2025}) suggest that, at $z\gtrsim2$, the initial quenching of low-mass systems is not predominantly environmentally driven, and may instead reflect bursty or temporarily suppressed star formation (e.g. \citealt{tacchella_confinement_2016}), with long-timescale environmental mechanisms becoming dominant only at later epochs. Access to the low-mass population is therefore essential to disentangle the onset of quenching from the subsequent environmental processing that ultimately reshapes galaxies in dense environments.

Therefore, identifying cluster- and protocluster-scale overdensities represents only the first step toward understanding the formation of massive cosmic structures observed at later epochs. A complete picture requires determining their evolutionary stage and fully characterizing their galaxy populations by (1) establishing robust membership across a wide mass range, from the most massive systems to the faintest low-mass galaxies, and (2) constraining their stellar populations and star-formation histories. Only by linking galaxy demographics to their physical properties can we reconstruct the pathway from assembling protoclusters to mature clusters, and quantify how environmental processes shape galaxy evolution across cosmic time.

\vspace{3mm}

In this paper, we study HUDFJ0332.4$-$2746.6 (hereafter HUDF46), an overdensity at $z = 1.84 \pm 0.01$ in the Hubble Ultra Deep Field (HUDF; \citealt{Beckwith_hubble_2006}), identified by \citet{mei_star-forming_2015} as a protocluster with an estimated halo mass of $M_{200} = (2.2 \pm 1.8) \times 10^{14}\,M_{\odot}$, an intrinsic velocity dispersion ($\sigma$) of $730\pm260$~km\,s$^{-1}$ and $R_{200}=0.9\pm0.3$~Mpc\footnote{$R_{200}$ is the radius within which the mean cluster density equals 200 times the critical density of the Universe.}. The halo mass was inferred assuming virialization, based on the Gaussian velocity distribution of member galaxies, the detection of an $18\sigma$ overdensity consistent with a massive node of the cosmic web (\citealt{hahn_evolution_2007, cautun_subhalo_2014}), and upper limits from 4~Ms \emph{Chandra} X-ray observations (\citealt{xue_chandra_2011}).
\citet{mei_star-forming_2015} showed that HUDF46 lacks a well-defined red sequence (i.e., missing quenched galaxies) and suggested the presence of a substantial population of star-forming ETGs in a dense environment at these redshifts. Here, we exploit new ultra-deep JWST observations obtained with NIRCam, MIRI, and, critically, the Near Infrared Imager and Slitless Spectrograph (NIRISS; \citealt{willott_near-infrared_2022, doyon_near_2023}), which provide robust spectroscopic redshifts. This leap in sensitivity allows us to revisit and significantly extend the analysis of \citet{mei_star-forming_2015} by probing its member galaxies across three dex in stellar mass ($M_{\bigstar}\approx10^{7.5\text{--}10.5}\,M_\odot$). By revising its membership and incorporating low-mass members, HUDF46 provides a uniquely powerful laboratory for studying cluster evolution in action, enabling a direct assessment of how environmental mechanisms begin to shape galaxy populations across a wide range in stellar mass.

\vspace{3mm}
This paper is structured as follows. Section 2 presents the dataset used in this work. Section 3 outlines the sample selection and redshift assessment. Section 4 details the methodology adopted to identify HUDF46. Section 5 presents a detailed analysis of HUDF46 and its galaxy population in comparison to coeval field galaxies. Section 6 summarizes and discusses the main results. Throughout this paper, we consider a cosmology with $H_{0} = 70\; \rm km\;s^{-1}\;Mpc^{-1}$, $\Omega_{M} = 0.3$, and $\Omega_{\Lambda} =0.7$. All magnitudes are total and refer to the AB system \citep{oke_secondary_1983}. A \citet{kroupa_variation_2001} initial mass function (IMF) is assumed (0.1--100 M$_{\odot}$).

\section{Dataset}
We relied on HST and JWST data available in HUDF. This section describes all datasets used, with particular focus on the reduction of NIRISS data (imaging and wide field slitless spectroscopy). We emphasize that in NIRISS wide-field slitless spectroscopy observations, the associated NIRISS direct imaging serves as pre-imaging to support the grism data reduction, providing source positions, extraction apertures, and contamination estimates. In contrast, the photometry in the NIRISS wavelength range (F115W, F150W, F200W) used for spectral energy distribution (SED) fitting is derived from the corresponding NIRCam imaging, which is significantly deeper. This approach ensures a more robust photometric characterization of faint galaxies, while all spectroscopic measurements are obtained exclusively from the NIRISS grism observations.

\subsection{HST}
As for the HST data, we used the ACS/WFC and WFC3/IR data from the Hubble Legacy Field (HLF) observations that cover GOODS-S. The HLF provides deep imaging in 9 HST bands covering a wide range of wavelengths (0.4$-$1.6 $\mu$m), from the optical (ACS/WFC F435W, F606W, F775W, F814W, and F850LP filters) to the near-infrared (WFC3/IR F105W, F125W, F140W and F160W filters). In this work, we make use of HST/ACS only for our analysis, since the overlap of HST/WFC3 with NIRCam. For a more detailed description of this dataset, we refer the reader to \citet{whitaker_hubble_2019}.

\subsection{NIRCam}
We made use of NIRCam data from the JWST Advanced Deep Extragalactic Survey (JADES) NIRCam Data Release 2 (JADES DR2 -- PIDs: 1180, 1210; PIs.: D. Eisenstein, N. Luetzgendorf; \citealt{eisenstein_jades_2023, eisenstein_overview_2023}), which includes observations from the JWST Extragalactic Medium-band Survey (JEMS -- PID: 1963; PIs: C. C. Williams, S. Tacchella, M. Maseda; \citealt{williams_jems_2023}) and the First Reionization Epoch Spectroscopically Complete Observations (FRESCO -- PID: 1895; PI: P. Oesch; \citealt{oesch_jwst_2023}).

The combined NIRCam data from all these surveys data cover a broad wavelength range ($\approx1$–$5\,\mu$m): F090W, F115W, F150W, F182M, F200W, F210M, F277W, F335M, F356W, F430M, F444W, F460M, and F480M. For more details, we refer the reader to the JADES DR2 (\citealt{eisenstein_jades_2023, eisenstein_overview_2023}).

\subsection{MIRI}

We also considered MIRI data from the SMILES program (PID: 1207; PI: G. H. Rieke, \citealt{rieke_smiles_2024}), which enable us to cover a wide range of wavelengths, from $5.6\;\mu$m to $25.5\;\mu$m. A more detailed discussion on the MIRI data reduction and overall quality of the data can be found in \citet{alberts_smiles_2024} for SMILES.

\subsection{NIRISS}

We made use of imaging and wide-field slitless spectroscopy (WFSS) data from the Next Generation Deep Extragalactic Exploratory Public (NGDEEP) program (PID 2079, PIs: S. Finkelstein, C. Papovich, N. Pirzkal), which covers HUDF with three filters (F115W, F150W, and F200W) over $\approx5.4$~arcmin$^{2}$ with a 1D line sensitivity (5$\sigma$) of $\approx1.3\times10^{-18}\,\,\rm erg\,s^{-1}\,cm^{-2}$. We refer the reader to \citet{shen_ngdeep_2024}, \citet{forrey_ngdeep_2025}, and \citet{shen_ngdeep_2025} for more details on the NGDEEP dataset. 

The data were reduced with our custom NIRISS pipeline, which incorporates additional steps originally developed for NIRCam and MIRI image processing and already applied in several recent works (e.g., \citealt{rinaldi_midis_2023, caputi_midis_2024, iani_midis_2024, navarro-carrera_constraints_2024, rinaldi_deciphering_2025}). The NIRISS pipeline is built on the official \textsc{jwst} pipeline ({\tt 1.16.1}; {\tt pmap\_1303}) and the {\sc grizli} pipeline ({\tt 1.12.8}; \citealt{brammer_gbrammergrizli_2021}). We refer the reader to the Appendix \ref{appendix_niriss_data_reduction}, which provides a schematic overview of the full reduction workflow.

\section{Sample Selection and redshift evaluation}

Since our goal is to better characterize HUDF46, reported by \citet{mei_star-forming_2015}, we restricted our analysis to galaxies within the redshift range $z_{\rm phot}\approx1.5-2.5$. Accordingly, we pre-selected sources directly from the JADES DR2 catalog based on their photometric redshifts \citep{hainline_cosmos_2024}.

We then made use of {\sc grizli} to extract all pre-selected sources from the JADES DR2 catalog blindly (i.e., without applying any prior from the photometric redshifts during the fit with {\sc grizli}). The main advantage of using {\sc grizli} is its ability to forward-model templates directly onto the 2D grism frames, incorporating source morphology to account for spectral smearing across adjacent pixels in the direct image. Most importantly, {\sc grizli} allows us to also model and remove, during the fitting process, nearby sources that can potentially contaminate the source of interest. We fit the grism data using {\sc grizli}, simultaneously incorporating information from both orthogonal grisms. In this context, for the fitting process and redshift estimation, {\sc grizli} employs templates based on flexible stellar population synthesis models \citep{conroy_fsps_2010}, combined with nebular emission lines. 

When {\sc grizli} extracts and fits each object, it generates multiple output files ({\tt *\_.fits} and {\tt *\_.png}), allowing users to inspect both 1D and 2D spectra (for each grism-filter combination) alongside the extracted spectra and their redshift probability distribution functions (PDFs). If {\sc grizli} fails to fit the extracted spectrum, it still produces all the aforementioned files except for the fitted spectrum and its corresponding probability distribution function (PDF). 

Once {\sc grizli} completed the extraction, we performed multiple tiers of visual inspection. The first tier classified sources into four categories based on their redshift reliability:

\begin{itemize}
    \item {\tt FLAG 1}: Sources with a secure redshift solution based on both the 1D and 2D spectra, supported by the detection of multiple emission lines and characterized by either a single peak or a dominant peak ($\gtrsim90\%$ of the PDF).
    \item {\tt FLAG 2}: Sources where the redshift solution was less certain, resulting in multiple peaks in the PDF without a single dominant solution, either due to faintness or single-line detections, with the quality of the identification assessed using $z_{\rm phot}$ in the second round of inspection.
    \item {\tt FLAG 3}: Sources for which neither the 1D nor 2D spectra yielded a reliable redshift solution, typically due to a chaotic PDF, severe contamination from instrumental artifacts (e.g., lightsaber effects or unflagged snowballs\footnote{See \href{https://jwst-docs.stsci.edu/known-issues-with-jwst-data/niriss-known-issues}{NIRISS Known Issues (JDox)} and \href{https://jwst-docs.stsci.edu/known-issues-with-jwst-data/nircam-known-issues}{NIRCam Known Issues (JDox)}}), or their location at the edge of the grism exposures, resulting in poor spectral coverage.
    \item {\tt FLAG 4}: Sources for which {\sc grizli} failed to converge to any reliable redshift solution\footnote{Some of these objects exhibit clear emission lines but a poorly constrained or unreliable continuum, preventing a robust joint fit of continuum and line emission from {\sc grizli}.}.
\end{itemize}

In the second tier, we reviewed only {\tt FLAG 1} and {\tt FLAG 2} sources to minimize potential outliers from {\sc grizli}. Some were downgraded to {\tt FLAG 3} and discarded due to unreliable redshift solutions (e.g., strong 2D contamination), while a subset of {\tt FLAG 2} sources were upgraded to {\tt FLAG 1} when single-line identifications were supported by the photometric redshifts enabled by the dense HST+NIRCam coverage in HUDF. The third tier focused on {\tt FLAG 4} sources. Since {\sc grizli} does not provide any redshift for this class by default during the extraction process due to poorly constrained or unreliable continuum, we forced the fits using JADES DR2 photometric redshifts as priors, assuming a 10\% uncertainty and adopting a fine redshift grid ($dz = 0.0004$) centered on the prior value (following \citealt{watson_glass-jwst_2025}). This process allowed us to recover several sources (e.g., with clear emission lines), which were then reclassified as either {\tt FLAG 1} or {\tt FLAG 2} based on their PDFs, while others were reassigned to {\tt FLAG 3} and discarded. When possible, we also made use of spectroscopic redshifts available in the field from 3D-HST (\citealt{brammer_3d-hst_2012}), MUSE Hubble Ultra-Deep Field  (\citealt{bacon_muse_2023}), FRESCO (\citealt{oesch_jwst_2023}), ASPECS (\citealt{walter_alma_2016}), and recent Near-Infrared Spectrograph
(NIRSpec; \citealt{ferruit_near-infrared_2022, jakobsen_near-infrared_2022}) observations from JADES (\citealt{deugenio_jades_2024}) and SMILES (\citealt{Zhu_smiles_2025}). At this point, our sample includes only three quality flags: 1, 2, and 3.

Finally, we reviewed all sources classified as {\tt FLAG 1} and {\tt FLAG 2}, leveraging all available information to refine our final assessment, thus including spectral energy distributions (SEDs) from JADES and postage stamps across HST and JWST filters. 

\begin{figure}
    \centering
    \includegraphics[width=1.\linewidth]{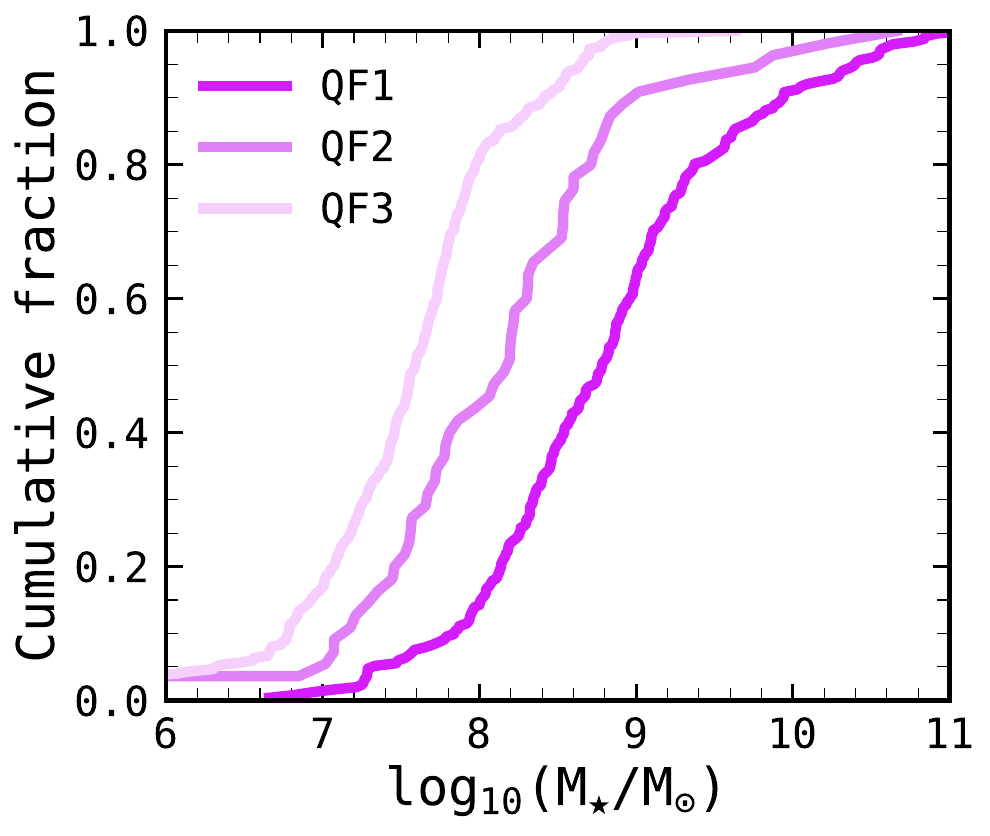}
    \caption{Stellar mass distributions of galaxies grouped by quality flag of their redshift (QF1, QF2, QF3), shown as cumulative fractions to enable direct comparison between the samples. For QF1 sources, stellar masses are derived in this work by fixing the redshift to the spectroscopic value (see Section 5.4.1), while for QF2 and QF3 sources stellar masses are taken from existing field catalogs and are based on photometric redshifts (\citealt{rinaldi_emergence_2024, navarro-carrera_burstiness_2024}). The vast majority of galaxies in QF3 ($\approx80\text{--}85\%$) have stellar masses below $\approx10^{8}\,M_{\odot}$.}
    \label{fig:mass_distribution}
\end{figure}

At the end of this multi-tiered inspection, we obtain 259 sources with {\tt FLAG 1}, 54 with {\tt FLAG 2}, and 303 with {\tt FLAG 3}. We notie that most {\tt FLAG 3} sources are either low-mass galaxies (with 85\% having $\lesssim10^{8}\,M_{\odot}$) or lie in regions of the NIRISS/WFSS footprint where information cannot be reliably extracted due to edge effects or severe contamination from nearby sources (see Figure \ref{fig:mass_distribution}). 

\begin{figure*}
    \includegraphics[width=0.99\linewidth]{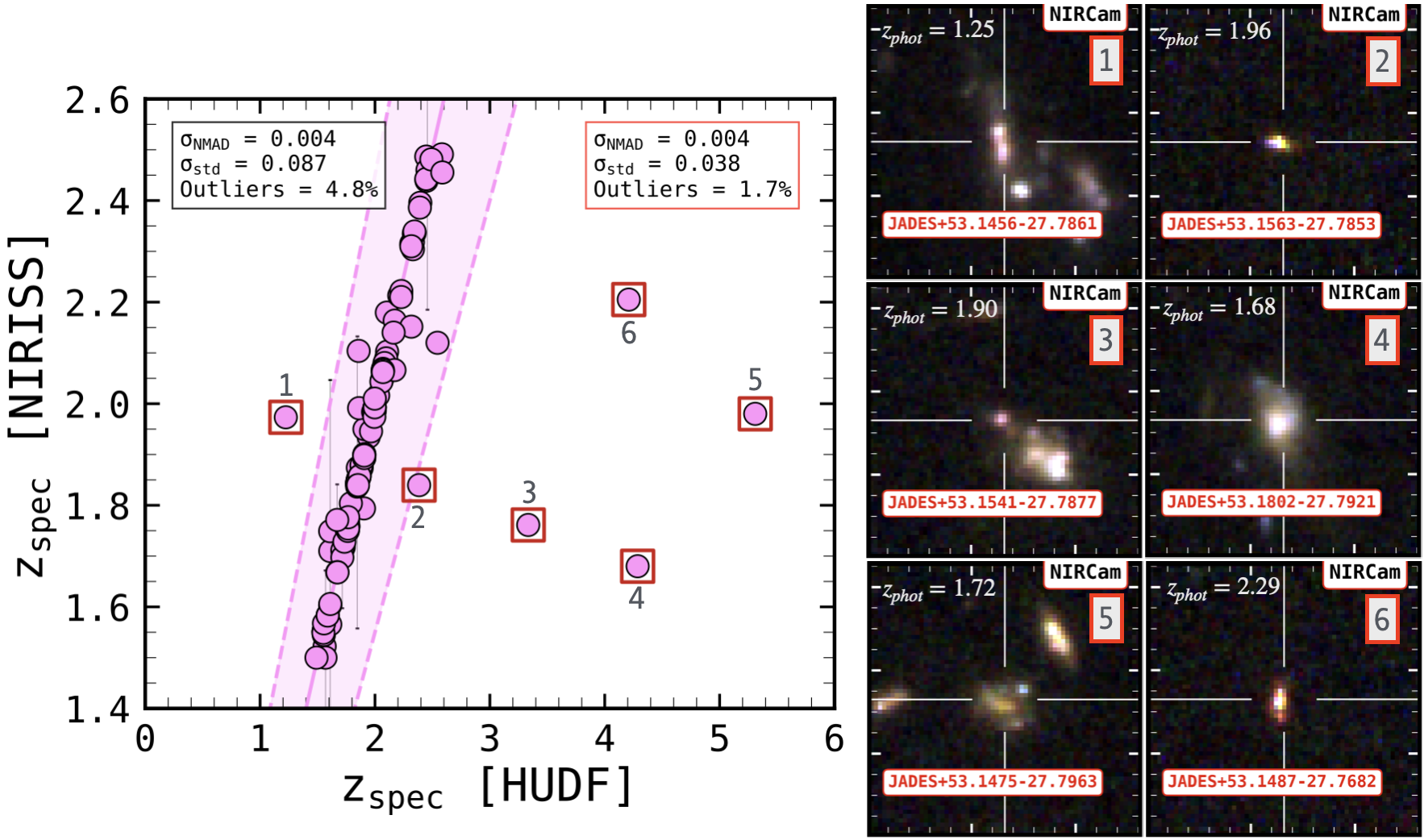}
    \caption{We perform a benchmark comparison between our spectroscopic redshifts from NIRISS/WFSS and those available in the HUDF from 3D-HST, MUSE, FRESCO, ASPECS, and recent NIRSpec observations from JADES and SMILES (only their QF=1). The left panel shows the comparison between the redshifts measured here with previous results. Overall, we find good agreement with the existing literature (black box, top left). The right panel highlights the few outliers (denoted by numbers on the left panel) which we visually inspected. We find that the sources previously spectroscopically classified as high-redshift objects are likely affected by unreliable quality flags in the earlier datasets. Their photometric redshifts (from JADES DR2) agree well with our NIRISS-based spectroscopic redshifts, further supporting the revised lower-redshift interpretation. After removing these sources, the agreement between $z_{\text{NIRISS}}$ and $z_{\text{HUDF}}$ improves further (red box, top right).}
    \label{fig:z_comparison}
\end{figure*}

In Figure \ref{fig:z_comparison}, we show the comparison between our {\tt FLAG 1} (i.e., the sources we will focus on throughout this work) with the available spectroscopic redshifts in HUDF from 3D-HST, MUSE Hubble Ultra-Deep Field, FRESCO, ASPECS, and recent NIRSpec observations from JADES and SMILES. We find a relatively small outlier fraction of 4.8\% (where outliers are defined as $|\Delta z| > 0.15$; $N_{\text{matched}} = 125$). The normalized median absolute deviation is $\sigma_{\rm NMAD} = 0.004$, indicating tight agreement between $z_{\rm NIRISS}$ and $z_{\rm spec}$, while the standard deviation of $\sigma_{\rm std} = 0.084$ reflects a small tail of larger deviations (see black box). As shown in Figure~\ref{fig:z_comparison}, some sources (highlighted with red squares; specifically, sources: 3, 4, 5, and 6) were classified at much higher redshifts than what we infer from NIRISS/WFSS. We visually inspected these outliers (see RGB postage stamps from NIRCam F090W, F115W, and F150W), and find that their photometric redshifts agree well with our NIRISS-based spectroscopic redshifts. Further inspection of their SEDs confirms that these sources are not at high redshift (as initially pointed out in the previous literature), and we conclude that their published $z_{\rm spec}$ values are likely the result of unreliable quality flags (either from MUSE or 3D-HST). After removing these sources, the agreement between $z_{\rm NIRISS}$ and $z_{\rm spec}$ improves further, reducing the outlier fraction to 1.7\% (see red box).

We also assessed the agreement between our $z_{\mathrm{NIRISS}}$ estimates and the photometric redshifts from JADES DR2, finding an outlier fraction of 3.4\%, with $\sigma_{\rm NMAD} = 0.031$ and $\sigma_{\rm std} = 0.161$, indicating overall good consistency despite a broader tail.

\section{Identification of overdensties in HUDF}

Identifying galaxy overdensities is non-trivial. Spectroscopic samples provide robust redshifts but suffer from sparse sampling and complex selection functions, limiting classical density–contrast methods that assume uniform coverage. Slitless spectroscopy (e.g., NIRISS) offers more homogeneous sampling and is less sensitive to edge effects and bright neighbors than imaging, though incompleteness persists due to 1) spectral overlap and 2) ambiguities when only a single line is detected. By contrast, photometric redshifts enable larger, more complete samples but introduce projection effects from their higher uncertainties.
To address these challenges, several approaches have been developed, including Voronoi tessellations (e.g., \citealt{Ramella_finding_2001}), adaptive kernel smoothing (e.g., \citealt{Darvish_comparative_2015}), and the Friends-of-Friends (FoF) algorithm (e.g., \citealt{huchra_groups_1982, berlind_percolation_2006}), with the latter being adopted in this work (see next section).

\subsection{Group Identification via Friends-of-Friends and Evidence Accumulation Clustering}

In this work, we apply a FoF algorithm to the spectroscopic catalog (Section 3) using a multi-tiered grid of projected separations and velocity differences, thereby avoiding reliance on a single linking choice and capturing a range of plausible groupings. The goal of this section is to identify overdensities within the NIRISS field over the redshift range $z \approx 1.5\text{--}2.5$, which includes independent confirmation of HUDF46 and its properties using the near-IR coverage and sensitivity of NIRISS. FoF is a well-tested group finder in both simulations and redshift surveys, linking galaxies in redshift space based on projected and line-of-sight separations and associating those likely residing within a common dark-matter halo. When properly parameterized, FoF minimizes fragmentation and over-merging, recovers unbiased group multiplicity functions, and yields nearly universal halo mass functions accurate to $\approx10\%$ across redshift and cosmology (e.g., \citealt{courtin_imprints_2011}. To assess the robustness of the resulting group catalogs across the FoF parameter space, we also employ a custom framework based on Evidence Accumulation Clustering (EAC; \citealt{fred_combining_2005}). The most secure overdensities identified in this way are then used as a benchmark for applying the method of \citet{Watson_hst_2025}, which builds upon established observational techniques, including the modified \citet{eisenhardt_clusters_2008} criteria introduced by \citet{noirot_hst_2018}.

We explored a $9\times20$ grid in the FoF linking parameters. The projected linking length spanned $r_{\perp}=0.10$--$0.50$\,Mpc in $0.05$\,Mpc steps, and the line-of-sight velocity threshold $\sigma=100$--$2000$\,km\,s$^{-1}$ in $100$\,km\,s$^{-1}$ steps. These values were chosen to reflect the NIRISS field of view at $z\approx2$ and the typical velocity uncertainty of our spectroscopic redshifts. Each of the resulting FoF runs produced an independent partition $P^{(r)}$ of galaxies into groups, effectively sampling the possible ways galaxies can be physically connected under different linking assumptions.

Then, the EAC method aggregates these multiple realizations into a single, consensus grouping by interpreting each FoF run as a piece of evidence on the physical association between galaxies. For each galaxy pair $(i,j)$, we recorded whether it appeared in the same FoF group across the ensemble of runs, defining the \textit{co-association matrix} as:
\begin{equation}
C_{ij} = \frac{1}{N}\sum_{r=1}^{N}\mathbf{1}\!\bigl[i,j\in\text{same group in }P^{(r)}\bigr],
\end{equation}
whose entries represent the empirical probability that galaxies $i$ and $j$ belong to the same structure. In this framework, FoF provides the base clusterings, while EAC serves as a consensus layer that extracts only those associations that are consistently recovered across the entire parameter grid.

We interpret $C$ as a similarity matrix and construct an undirected graph using {\tt networkx}, where galaxies are vertices and edges connect pairs with $C_{ij}\ge\pi_{\min}$\footnote{Where, $\pi_{\min}$ is the minimum required co-association probability for declaring a link between two galaxies in the EAC graph.}. Varying $\pi_{\min}$ from 0.20 to 1.00 in steps of 0.05 yields 17 provisional partitions. An edge is considered \textit{robust} if it appears in at least half of these partitions (i.e.\ in $\geq50\%$ of $\pi_{\min}$ sweeps). The connected components of the resulting \textit{consensus graph} represent the final EAC-refined FoF groups. This ensures that only galaxy associations that persist across both the FoF parameter grid and the EAC threshold space are retained.

For each galaxy $i$, we further quantified the stability of its membership through three metrics:
\begin{itemize}
  \item \textit{Intra-group coherence}, $P_{\mathrm{group}}(i)$, the biweight location of $C_{ij}$ values linking $i$ to the other members of its consensus group.
  \item \textit{Grouping frequency}, $f_{\mathrm{grouped}}(i)=n^{\mathrm{grp}}_i/N_{\mathrm{runs}}$, the fraction of FoF realizations in which $i$ appears in \textit{any} group, measuring its overall tendency to cluster.
  \item \textit{Robustness score}, $P_{\mathrm{score}}(i)=f_{\mathrm{grouped}}(i)\,P_{\mathrm{assoc}}(i)$, where $P_{\mathrm{assoc}}$ is the 75$^{\mathrm{th}}$ percentile of the $C_{ij}$ values with $C_{ij}\ge0.3$, penalizing galaxies that are either rarely grouped or lack consistently strong connections.
\end{itemize}

\begin{figure}
    \centering
    \includegraphics[width=1.\linewidth]{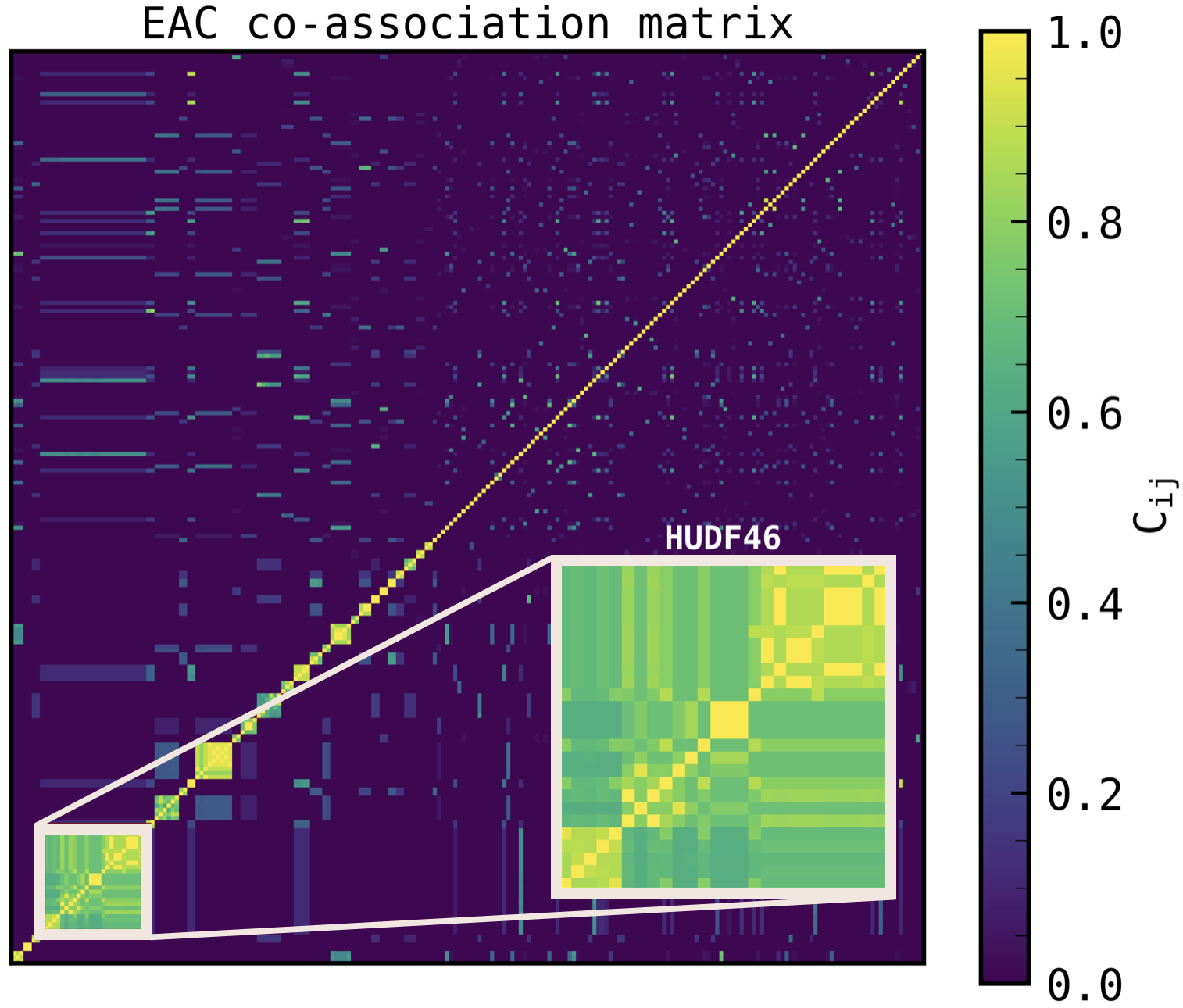}
    \caption{EAC co-association matrix for the HUDF spectroscopic sample. Galaxies are ordered by their final EAC consensus group for visualization only. The color scale shows the co-association value $C_{ij}$, i.e., the fraction of FoF realizations (across a broad grid of projected separations and velocity thresholds) in which galaxies $i$ and $j$ are grouped together. Diagonal elements are unity by construction. Most pairs have low $C_{ij}$, indicating weak association across parameter space. The highlighted region and inset zoom into HUDF46 at $z\approx1.84$, revealing a compact set of galaxies consistently recovered as a group. While the visual appearance of the matrix depends on the chosen ordering, all pairwise co-association values and derived stability metrics are invariant under permutations of the axes. We note that, beyond HUDF46, other overdensities are present within HUDF, which are however less rich in member galaxies.}
    \label{fig:EAC}
\end{figure}

Finally, we classified as \textit{secure members} those galaxies with $P_{\mathrm{group}} \ge 0.5$ and $P_{\mathrm{score}} \ge 0.5$, i.e.\ objects that are both frequently recovered and tightly connected to their neighbours. This majority-rule selection is applied consistently at both the pairwise and per-galaxy level, ensuring that only statistically robust associations are retained. We visually show the result of this methodology in Figure \ref{fig:EAC}.

\subsection{Final identification}

The consensus grouping identifies the two previously reported overdensities HUDF46 at $z \approx 1.84$ and HUDFJ0332.5$-$2747.3 at $z \approx 1.89$; both were originally classified as proto-clusters by \citet{mei_star-forming_2015} and are now expanded with additional low-mass members enabled by the deeper JWST data. We also identify two new overdensities at $z\approx1.75$ and $z\approx1.99$.  In what follows, we restrict our discussion to HUDF46, as it is the most prominent and best-sampled structure in our dataset, allowing a detailed characterization of both its global structure and galaxy population.

\section {A Deep dive into HUDF46}

HUDF46 was originally identified using HST grism spectroscopy and classified as a high-redshift proto-cluster at $z=1.84\pm0.01$, with a halo mass of $M_{200}=(2.2\pm1.8)\times10^{14}\,M_{\odot}$ \citep{mei_star-forming_2015}. Using new JWST/NIRISS spectroscopy, we confirm HUDF46 as the richest overdensity in the NIRISS field of view at $z\approx1.84$, revising its membership from 18 to 37 spectroscopically confirmed galaxies (see next section). This highlights JWST’s ability to probe the faint end of the galaxy population previously inaccessible to HST (\citealt{mei_star-forming_2015}). 

\subsection{The final membership of HUDF46}

Building on the FoF+EAC analysis presented in the previous section, we take advantage of the ultra-deep NIRISS spectroscopy to construct a richer census of HUDF46 member galaxies, down to low $M_{\bigstar}$ ($\approx10^{7.5}\,M_{\odot}$).  To refine the membership of HUDF46, we applied the framework introduced by \citet{Watson_hst_2025}, which builds on established observational techniques; specifically, the modified \citet{eisenhardt_clusters_2008} method proposed by \citet{noirot_hst_2018}. This step provides the final identification of HUDF46 members used throughout the analysis.

Briefly, we made use of the secure members (QF1) to estimate an initial cluster redshift ($z_{cl}$), central position (R.A., DEC), velocity dispersion ($\sigma$), and $R_{200}$ (using the formalism from \citealt{danese_velocity_1980} and \citealt{carlberg_average_1997}). We then applied the method described in \citet{Watson_hst_2025}, with one key modification: instead of adopting fixed thresholds, we implemented dynamic radial and velocity cuts based on $3 \times \sigma$ and $3 \times R_{200}$, assuming virialization (see next section). Cluster properties and memberships are iteratively refined using $3\sigma$ clipping \citep{yahil_velocity_1977, blanton_first_2000} until convergence. During this process, we also updated the central position at each step and cap the velocity dispersion at 2000 km\,s$^{-1}$, while allowing the spatial search to extend across the full NIRISS field of view.

This procedure recovers the same core members identified by the FoF$+$EAC method for HUDF46 and includes additional galaxies. While the core remains unchanged, the \citet{Watson_hst_2025} approach systematically adds sources that were weakly or inconsistently linked in the FoF runs and thus excluded by the consensus classification. This highlights the conservative nature of FoF$+$EAC compared to the more inclusive spatial and kinematic criteria adopted here. In total, we identify 34 members for HUDF46 (QF=1). 

When cross-matching with the 18 members originally reported by \citet{mei_star-forming_2015}, two are excluded from our list: the first one (GMASS220) lies outside our footprint and the second one (UDF-2127) has NIRSpec confirmation at higher redshift with respect to what reported by \citet{mei_star-forming_2015}\footnote{Specifically: $z_{\rm UDF-2127,\, old} = 1.858$ vs. $z_{\rm UDF-2127,\,new} = 1.863$.}. Another object (UDF-2433), instead, lies in a crowded region dominated by very bright nearby sources. In particular, it can be cross-matched with three objects: our HUDF46+53.1472-27.7776, a passive galaxy with NIRSpec spectroscopic confirmation at $z_{\rm NIRSpec}=1.6398$, and a higher-redshift source with $z_{\rm NIRISS}=1.941$. Given the proximity of these systems and the resulting ambiguity in the association, we conservatively adopt HUDF46+53.1472–27.7776 as the most likely counterpart.

\begin{figure*}
    \includegraphics[width=0.99\linewidth]{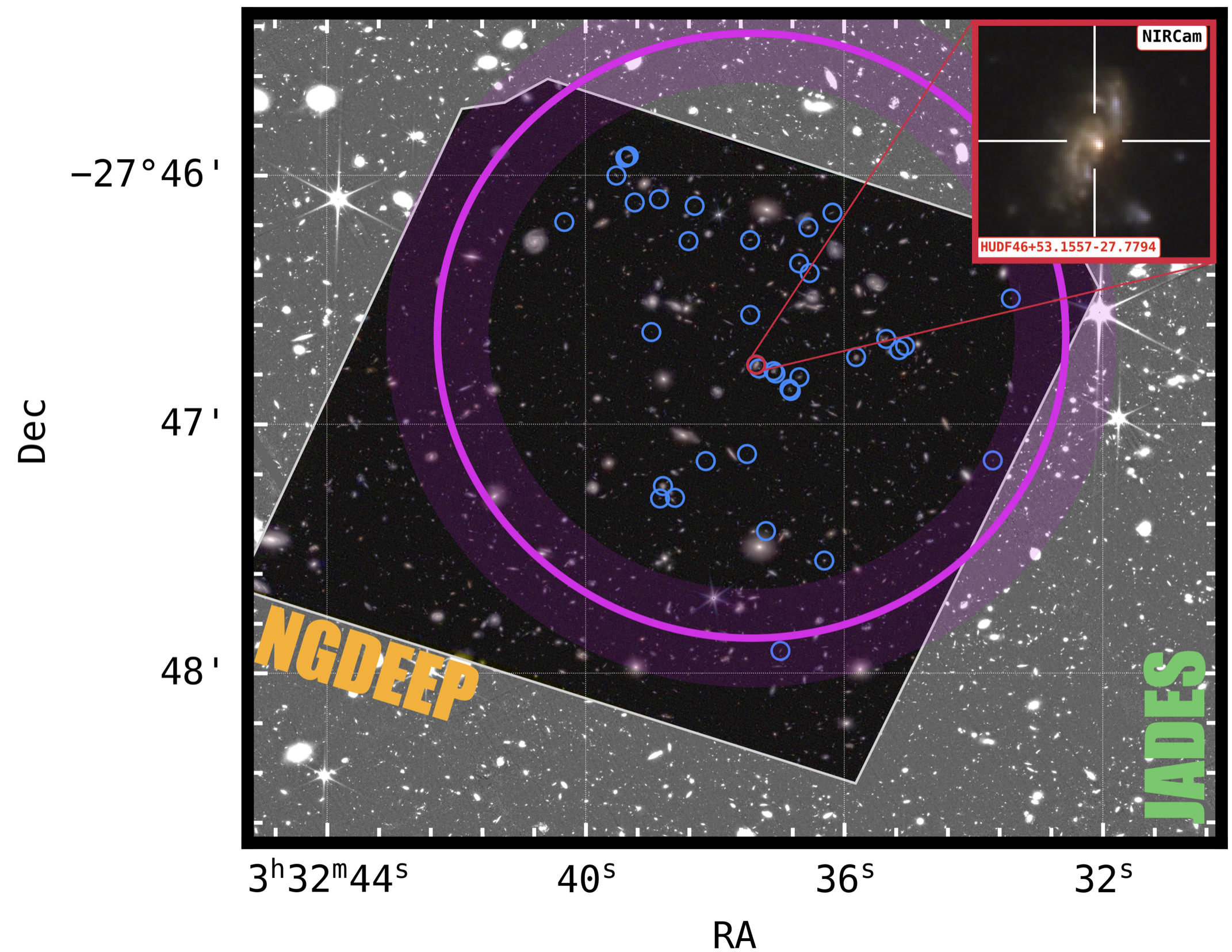}
    \caption{RGB image of the HUDF obtained with JWST/NIRISS (F200W, F150W, F115W), overlaid on the NIRCam/F150W grayscale mosaic from JADES. HUDF46 members are highlighted with blue circles. The virial radius, $R_{200}$, is shown in deep magenta, with its uncertainty indicated by the shaded region. We also show a zoom-in (3\arcsec$\times$\,3\arcsec) on the brightest member (HUDF46+53.1557-27.7794), which is also the object that has been classified as an AGN according to its X-ray-to-radio luminosity ratio in \citet{lyu_agn_2022}.}
    \label{fig:rgb_niriss}
\end{figure*}

We also find that three galaxies previously assigned a quality flag $\rm QF = 3$ in our NIRISS analysis exhibit redshifts (from NIRSpec) fully consistent with the HUDF46 structure; therefore we include these sources in the updated HUDF46 census, bringing the total membership to 37 galaxies. These newly identified sources have low stellar masses, with $M_{\bigstar} \lesssim 10^{8.5\text{--}9}\,M_{\odot}$. We also identified additional potential members located outside the NIRISS footprint having NIRSpec data (for a total of 13)\footnote{One of these corresponds to a source reported by \citet{mei_star-forming_2015} (GMASS220), previously excluded from our cross-match because it lies outside the NIRISS footprint.}, extending out to $\approx 3\times R_{200}$. However, to avoid the strong incompleteness and selection biases inherent to these non-uniform NIRSpec datasets, we restrict our analysis to the uniform NIRISS-based sample. In Figure \ref{fig:rgb_niriss}, we display HUDF46 members. The spectroscopically confirmed members of HUDF46 are listed in Table~\ref{tab:hudf46}.

\begin{deluxetable*}{lccclccc}
\tabletypesize{\scriptsize}
\tablecaption{Spectroscopic Members of HUDF46}
\tablewidth{0pt}
\tablehead{
\colhead{ID} &
\colhead{$z_{\rm NIRISS}$} &
\colhead{$\log_{10}(M_\bigstar/M_\odot)$} &
\colhead{$\mathcal{S}$} &
\colhead{ID} &
\colhead{$z_{\rm NIRISS}$} &
\colhead{$\log_{10}(M_\bigstar/M_\odot)$} &
\colhead{$\mathcal{S}$}
}
\startdata
HUDF46+53.1541--27.7985 & $1.838\pm0.001$ & $8.71_{-0.08}^{+0.07}$ & H$\beta$,[O\,{\sc iii}],H$\alpha$ & HUDF46+53.1513--27.7925* & $1.837\pm0.001$ & $9.40_{-0.03}^{+0.06}$ & H$\beta$,[O\,{\sc iii}],H$\alpha$ \\
HUDF46+53.1619--27.7883* & $1.840\pm0.001$ & $8.82_{-0.07}^{+0.05}$ & H$\beta$,[O\,{\sc iii}],H$\alpha$ & HUDF46+53.1609--27.7883 & $1.846\pm0.001$ & $7.84_{-0.08}^{+0.09}$ & H$\beta$,[O\,{\sc iii}],H$\alpha$ \\
HUDF46+53.1617--27.7875*$^{\Diamond}$ & $1.849\pm0.001$ & $10.45_{-0.08}^{+0.11}$ & H$\beta$,[O\,{\sc iii}],H$\alpha$ & HUDF46+53.1589--27.7858 & $1.843\pm0.001$ & $7.54_{-0.14}^{+0.12}$ & H$\beta$,[O\,{\sc iii}] \\
HUDF46+53.1563--27.7853 & $1.840\pm0.002$ & $7.97_{-0.10}^{+0.12}$ & H$\beta$,[O\,{\sc iii}] & HUDF46+53.1404--27.7858 & $1.828\pm0.001$ & $8.51_{-0.06}^{+0.05}$ & H$\beta$,[O\,{\sc iii}],H$\alpha$ \\
HUDF46+53.1534--27.7811* & $1.845\pm0.001$ & $7.50_{-0.03}^{+0.06}$ & H$\beta$,[O\,{\sc iii}],H$\alpha$ & HUDF46+53.1535--27.7810* & $1.852\pm0.001$ & $8.88_{-0.03}^{+0.05}$ & H$\beta$,[O\,{\sc iii}],H$\alpha$ \\
HUDF46+53.1544--27.7799 & $1.845\pm0.001$ & $8.55_{-0.03}^{+0.04}$ & H$\beta$,[O\,{\sc iii}],H$\alpha$ & HUDF46+53.1545--27.7798* & $1.850\pm0.001$ & $9.57_{-0.05}^{+0.07}$ & H$\alpha$ \\
HUDF46+53.1555--27.7796* & $1.839\pm0.001$ & $8.99_{-0.07}^{+0.10}$ & H$\beta$,[O\,{\sc iii}],H$\alpha$ & HUDF46+53.1557--27.7794*$\dagger\,^{\star}$ & $1.845\pm0.001$ & $10.43_{-0.05}^{+0.07}$ & H$\alpha$ \\
HUDF46+53.1529--27.7802* & $1.854\pm0.001$ & $9.39_{-0.05}^{+0.07}$ & [O\,{\sc ii}],H$\beta$,[O\,{\sc iii}] & HUDF46+53.1492--27.7788*$^{\star}$ & $1.850\pm0.002$ & $10.09_{-0.09}^{+0.09}$ & H$\beta$,[O\,{\sc iii}],H$\alpha$ \\
HUDF46+53.1561--27.7760 & $1.840\pm0.001$ & $7.43_{-0.07}^{+0.08}$ & H$\beta$,[O\,{\sc iii}] & HUDF46+53.1465--27.7784* & $1.857\pm0.000$ & $8.62_{-0.06}^{+0.05}$ & H$\beta$,[O\,{\sc iii}],H$\alpha$ \\
HUDF46+53.1473--27.7776 & $1.857\pm0.001$ & $9.47_{-0.04}^{+0.04}$ & H$\beta$,[O\,{\sc iii}],H$\alpha$ & HUDF46+53.1461--27.7781 & $1.857\pm0.001$ & $8.05_{-0.19}^{+0.12}$ & H$\beta$,[O\,{\sc iii}],H$\alpha$ \\
HUDF46+53.1680--27.7698 & $1.857\pm0.002$ & $8.76_{-0.07}^{+0.11}$ & H$\beta$,[O\,{\sc iii}] & HUDF46+53.1600--27.7711* & $1.840\pm0.001$ & $8.81_{-0.08}^{+0.06}$ & H$\beta$,[O\,{\sc iii}],H$\alpha$ \\
HUDF46+53.1522--27.7732 & $1.844\pm0.001$ & $8.98_{-0.16}^{+0.03}$ & H$\beta$,[O\,{\sc iii}],H$\alpha$ & HUDF46+53.1635--27.7685 & $1.845\pm0.001$ & $7.94_{-0.07}^{+0.05}$ & H$\beta$,[O\,{\sc iii}] \\
HUDF46+53.1529--27.7726* & $1.843\pm0.001$ & $9.11_{-0.05}^{+0.08}$ & H$\beta$,[O\,{\sc iii}],H$\alpha$ & HUDF46+53.1561--27.7710* & $1.843\pm0.001$ & $9.42_{-0.04}^{+0.04}$ & H$\beta$,[O\,{\sc iii}],H$\alpha$ \\
HUDF46+53.1596--27.7687 & $1.843\pm0.001$ & $7.74_{-0.11}^{+0.15}$ & H$\beta$,[O\,{\sc iii}] & HUDF46+53.1523--27.7702*$^{\star}$ & $1.840\pm0.005$ & $9.75_{-0.04}^{+0.06}$ & H$\beta$,[O\,{\sc iii}],H$\alpha$ \\
HUDF46+53.1647--27.7667* & $1.845\pm0.001$ & $9.14_{-0.04}^{+0.04}$ &  [O\,\sc{ii}],H$\beta$,[O\,{\sc iii}] & HUDF46+53.1393--27.7749 & $1.838\pm0.001$ & $8.55_{-0.08}^{+0.06}$ & H$\beta$,[O\,{\sc iii}],H$\alpha$ \\
HUDF46+53.1640--27.7654* & $1.844\pm0.001$ & $8.77_{-0.09}^{+0.07}$ & H$\beta$,[O\,{\sc iii}],H$\alpha$ & HUDF46+53.1641--27.7655 & $1.846\pm0.001$ & $7.90_{-0.14}^{+0.10}$ & H$\beta$,[O\,{\sc iii}],H$\alpha$ \\
HUDF46+53.1508--27.7692* & $1.844\pm0.004$ & $8.57_{-0.07}^{+0.07}$ & H$\beta$,[O\,{\sc iii}],H$\alpha$ & HUDF46+53.1639--27.7654* & $1.846\pm0.001$ & $9.14_{-0.07}^{+0.05}$ & H$\beta$,[O\,{\sc iii}],H$\alpha$ \\
HUDF46+53.1624--27.7772 & $1.850\pm0.001$ & $8.45_{-0.07}^{+0.02}$ & $\ddagger$ & HUDF46+53.1550--27.7905 & $1.849\pm0.001$ & $8.99_{-0.07}^{+0.06}$ & $\ddagger$ \\
HUDF46+53.1619--27.7683 & $1.850\pm0.001$ & $8.07_{-0.07}^{+0.06}$ & $\ddagger$ &  &  &  &  \\
\enddata
\label{tab:hudf46}
\tablecomments{$\mathcal{S}$: spectral features identified in NIRISS/WFSS. An asterisk (*) refers to sources with a literature spectroscopic redshift consistent with our $z_{\rm NIRISS}$. A dagger ($\dagger$) marks the brightest member, located at the center of HUDF46. A double dagger ($\ddagger$) identifies sources with $\mathrm{QF}=3$ (from NIRISS) confirmed by NIRSpec, typically exhibiting H$\beta$, [O\,\textsc{iii}], H$\alpha$, and additional emission lines. Sources with $^{\star}$ have been recently studied in \citet{forrey_ngdeep_2025}, which also have $M_{\bigstar}$ consistent with our measurements. Finally, $\diamond$ marks the brightest member of HUDF46.}
\end{deluxetable*}

\subsection{The updated properties for HUDF46}

In \citet{mei_star-forming_2015}, the halo mass was estimated under the assumption of virialization based on multiple lines of evidence including 1) the velocity distribution of the member galaxies is gaussian, 2) an $18\sigma$ overdensity has a high probability of being a massive node in the cosmic web rather than a filament (\citealt{hahn_evolution_2007, cautun_subhalo_2014}), and 3) an upper limit derived using 4 Ms Chandra X-ray imaging (\citealt{xue_chandra_2011}) was consistent with halo masses of $M_{200}\lesssim10^{14}\,M_{\odot}$. Such halos are expected to be largely virialized by $z\approx1.5$ \citep{evrard_virial_2008}, with recent Millenium N-body simulations coupled with semi-analytic models of \citet{henriques_galaxy_2015} predicting their  gravitational collapse is largely complete by $z\gtrsim1.5$, transitioning from the protocluster to the dynamically relaxed cluster phase \citep{chiang_galaxy_2017, lim_flamingo_2024}.

Given our new, expanded membership for HUDF46, we assessed the significance of HUDF46 using the Nth-nearest neighbor technique (\citealt{papovich_spitzer-selected_2010, gobat_wfc3_2013}), as also adopted by \citet{mei_star-forming_2015} (see our Appendix \ref{appendix1} for details). As in their study, due to the limited field of view of the NIRISS/NGDEEP observations ($\approx5.4$\,arcmin$^2$), we compute statistics within redshift slices over the HUDF region covered by NIRISS/WFSS, which corresponds to a co-moving size of $\approx 1$~Mpc at these redshifts. 

Our analysis confirms, in agreement with \citet{mei_star-forming_2015}, a significant overdensity at $z\approx1.84$ (HUDF46), with a contrast significance of $\approx17\pm1\sigma$ relative to the field when applying the Nth–nearest–neighbor estimator with $N=5\text{--}7$. We also find a modest increase in significance at $N=3$ and $N=4$, consistent with the densest core of the structure being probed at smaller scales. This trend likely indicates a compact and centrally concentrated galaxy distribution.

Since deeper observations have become available over time with respect to what used in \citet{mei_star-forming_2015}, we next inspected the \textit{Chandra} data (now updated to 7 Ms). Consistent with earlier results, one galaxy in HUDF46 (HUDF46+53.1680$-$27.7697) coincides with X-ray source ID $=$ 512, classified as a “galaxy without AGN” by \citet{xue_chandra_2011}. As it was the case in \citet{mei_star-forming_2015}, no X-ray extended emission is observed even with the latest \textit{Chandra} data, ruling out the presence of a strong/hot ICM. Nonetheless, the increased depth of the 7 Ms data allows a tighter constraint on the total halo mass, yielding an upper limit of $M_{200} \approx (1.1$–$1.8)\times10^{14}\,M_{\odot}$, depending on the assumed ICM gas temperature (1--3 keV)\footnote{These are $3\sigma$ upper limits.}. These values are in good agreement with those reported by \citet{mei_star-forming_2015}. The details of our X-ray analysis are provided in Appendix~\ref{x-ray-chandra}. In particular, we highlight that such values are consistent with the observed and simulated $L_{X}$–$M_{h}$ relations within the associated uncertainties (e.g., \citealt{zheng_measuring_2023, nelson_introducing_2024}).

Given these confirmations, we likewise assume HUDF46 is virialized and compute the line-of-sight velocity dispersion using the formalism of \citet{danese_velocity_1980}\footnote{Which assumes that the observed redshift distribution directly traces the line-of-sight velocity spread, and thus no additional correction for redshift-space distortions is applied (e.g., \citealt{ma_soptics_2025}).}, specifically their Equations (5) and (10), and obtain $\sigma = 670 \pm 91$~km\,s$^{-1}$. This value is slightly lower, though consistent within uncertainties, with the earlier estimate of $\sigma = 730\pm260$~km\,s$^{-1}$ reported by \citet{mei_star-forming_2015}. The reduced dispersion reflects both the increased number of members (37 vs. 18) and the improved redshift precision provided by NIRISS/WFSS.

To remain consistent with \citet{mei_star-forming_2015}, we first estimate the halo mass using their Equation (1), based on the calibration from \citet{munari_relation_2013}, obtaining $M_{200} = (1.2 \pm 0.2) \times 10^{14}\ M_\odot$. The corresponding virial radius, derived using the relation from \citet{carlberg_average_1997}, is $R_{200} = 0.61 \pm 0.10$~Mpc. Alternatively, applying the $\sigma$--$M_{200}$ scaling relation from \citet{evrard_virial_2008} and adopting the spherical overdensity definition, $R_{200} = \left(3 M_{200} / 4\pi \cdot 200 \cdot \rho_{\mathrm{crit}}(z)\right)^{1/3}$, we find a slightly smaller value of $R_{200} = 0.52 \pm 0.10$~Mpc, consistent with the higher critical density at this redshift. Results do not depend on using different values for $A_{1D}$, as pointed out also in \citet{mei_star-forming_2015}. In this work, we will refer to $R_{200} = 0.61 \pm 0.10$~Mpc and $M_{200} = (1.2 \pm 0.2) \times 10^{14}\ M_\odot$ to be consistent with the methods adopted in \citet{mei_star-forming_2015}. We warn the reader that in the case that HUDF46 is only partially or not virialized, the halo mass retrieved in this work must be considered as an upper limit.  

Throughout this paper, we will refer to HUDF46 as a {\it young cluster}\footnote{We remind the reader that distinctions between groups, clusters, and protoclusters are often applied pragmatically, as observational constraints rarely allow an unambiguous classification \citep{overzier_realm_2016}.}, as it has a virial mass consistent with a gravitationally bound, cluster-scale system, while its redshift ($z \approx 1.84$) and galaxy population indicate an early evolutionary stage, close to the epoch when mature clusters first emerge ($z \lesssim 1.5$; e.g., \citealt{kawinwanichakij_effect_2017, tomczak_glimpsing_2017, papovich_effects_2018}). Therefore, the term young cluster reflects both its dynamical state and formative phase.

\begin{figure}
    \centering
    \includegraphics[width=1.\linewidth]{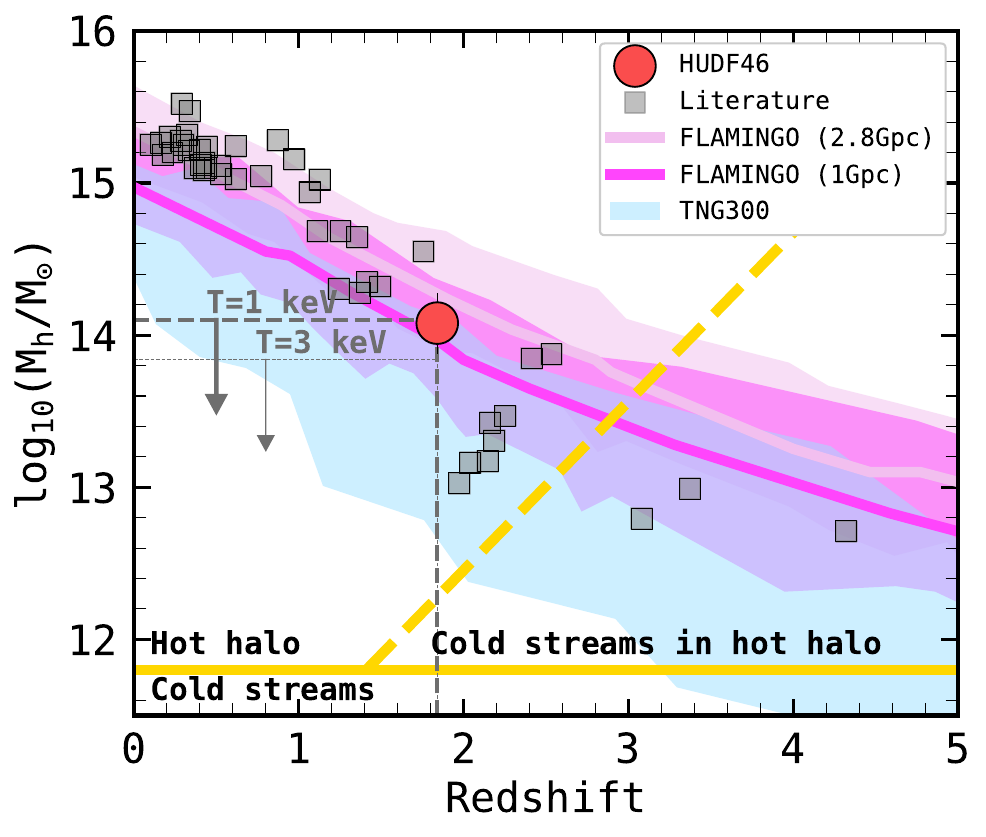}
    \caption{The evolution of the virial mass of haloes as a function of redshift for HUDF46 is shown alongside recent measurements at both low and high redshift (up to $z\approx5$). The observational compilation includes clusters and proto--clusters spanning $0 \lesssim z \lesssim 5$, such as GOODS--N \citep{chapman_submillimeter_2009}, MRC~1138--262 \citep{dannerbauer_excess_2014}, SSA22 \citep{umehata_alma_2015}, CL--J1449+0856 \citep{gobat_mature_2011}, CL--J1001+0220 \citep{wang_revealing_2018}, PHz~G237.01+42.50 \citep{polletta_spectroscopic_2021}, CC2.2 \citep{darvish_spectroscopic_2020}, PCL1002 \citep{casey_massive_2015}, SPT2349--56 \citep{miller_massive_2018}, the ISCS and SPT samples from \citet{brodwin_x-ray_2011}, \citet{jee_scaling_2011}, and \citet{williamson_sunyaev-zeldovich-selected_2011}, as well as the massive system ACT--CL~J0102--4915 \citep{menanteau_atacama_2012}. Predictions for the evolution of $M_{200}$ as a function of cosmic time from the IllustrisTNG (\citealt{nelson_illustristng_2019}) and FLAMINGO simulations (\citealt{schaye_flamingo_2023, lim_flamingo_2024}) are included for comparison. The derived $M_{200}$ aligns well with the expected halo growth at $z\approx1.84$. For comparison, we also show the upper limits from {\it Chandra} 7 Ms, assuming a blackbody ICM model with $T=3\rm \,KeV$ and $T=1\rm \,KeV$. Finally, for reference, the gold dashed lines mark the different circumgalactic--gas phases defined by \citet{dekel_galaxy_2006}.\label{fig:halo_redshift}
}
    \label{fig:halo_redshift}
\end{figure}

In Figure \ref{fig:halo_redshift}, we present our updated $M_{200}$ measurement for HUDF46 compared with cluster and proto-cluster measurements across cosmic time ($z\approx0\text{--}5$) and predictions from cosmological simulations. The simulation tracks are taken from \citet{lim_is_2021, lim_flamingo_2024}, who tracked the progenitors of the 25 most massive $z=0$ clusters in IllustrisTNG (\citealt{nelson_illustristng_2019}) and the 100 highest-SFR haloes in FLAMINGO (\citealt{schaye_flamingo_2023}) at each redshift snapshot.  Our measured $M_{200}$ places HUDF46 is on the simulated evolutionary tracks and firmly in the hot halo regime (\citealt{dekel_galaxy_2006}).

\subsection{The phase-space diagram of HUDF46}

\begin{figure}
    \centering
    \includegraphics[width=1\linewidth]{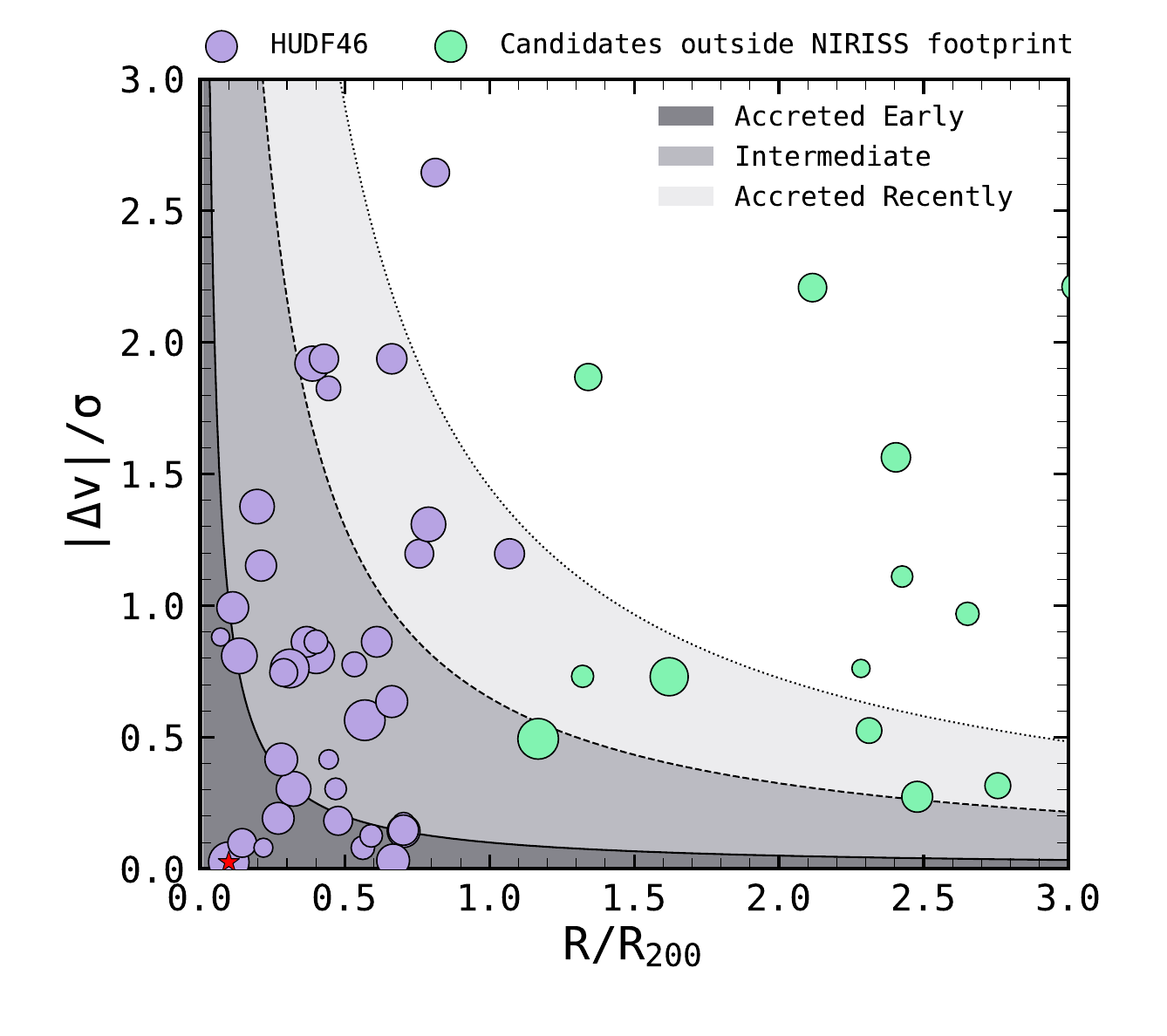}
    \caption{Phase–space diagram for the HUDF46 member galaxies (purple) and 13 candidate members outside the NGDEEP footprint (light gray). Point sizes scale with $M_{\bigstar}$. Caustic lines follow \citet{noble_phase_2016}, adopting constant values of $(r/r_{200})\times(\Delta v/\sigma_v)=0.1$, 0.65, and 1.45 to delineate ancient, intermediate, and recent infallers . 
    Nearly all confirmed HUDF46 members lie within the caustic boundaries, except for one object at $R/R_{200}<1$ but high $|\Delta v|/\sigma_v$. We retain this galaxy in the sample, noting its uncertain dynamical state. Among the candidate members, roughly half lie below the caustics at $R/R_{200}>1$, with the remainder above them. The red star marks the brightest galaxy in HUDF46.
    }
    \label{fig:phase_space}
\end{figure}

Galaxies that experience environmental effects in clusters can do so on both a local or global level via multiple mechanisms.  The effectiveness and timescale of these mechanisms depends on the galaxy's infall history, which is roughly caputred by probing the accretion state and orbital histories of its members in the phase-space diagram \citep[e.g.,][]{noble_kinematic_2013, jaffe_budhies_2015, rhee_phase-space_2017}. By combining projected clustercentric radius and line-of-sight velocity, this diagnostic retains information on time since infall. Virialized galaxies occupy the inner, low-velocity regions, while recently accreted systems lie at larger radii and higher velocities, producing the characteristic trumpet-shaped distribution \citep{jaffe_budhies_2015, noble_phase_2016}\footnote{Although infall-time calibrations are uncertain and largely based on $z\approx0$ systems, phase--space diagrams remain a useful tool for interpreting the dynamical state and assembly of overdensities \citep[e.g.,][]{dou_estimating_2025}.}.

In Figure~\ref{fig:phase_space}, we present the projected phase–space diagram following \citet{noble_kinematic_2013}, where a galaxy’s position relative to the caustics traces its average infall time (commonly distinguishing ancient infallers, intermediate/backsplash galaxies, and infalling systems). We note that at this early stage, a backsplash population (systems that have passed through the core and moved back to larger radii; e.g., \citealt{borrow_there_2023}) is unlikely (\citealt{haggar_three_2020}); we therefore classify galaxies as ancient infallers, recent infallers, or ongoing infallers, the latter not necessarily bound to HUDF46.

All confirmed HUDF46 members lie within the caustic boundaries defined by \citet{noble_phase_2016}, consistent with a population dominated by ancient and recent infallers, with only one galaxy still in an active accretion phase. The brightest member is located at small projected radius and low relative velocity, consistent with a dynamically settled position near the cluster center. These objects therefore constitute a complete core sample of HUDF46, suitable for investigating early environmental processing.

We also expanded our search beyond the NIRISS footprint and identified 13 galaxies spectroscopically confirmed with NIRSpec at the redshift of HUDF46 out to $3\times R_{200}$. Of these, $46\%$ lie within the HUDF46 caustics and are thus likely gravitationally bound members. The remaining systems have less certain dynamical fates and may represent future infallers, either drawn from the surrounding field or associated with a more extended proto-cluster remnant (\citealt{mei_star-forming_2015}). We emphasize, however, that the phase-space diagram is incomplete beyond the NIRISS footprint; given this incompleteness, the present analysis focuses on the core population, deferring a detailed assessment of extended infall to future work.

In what follows, we revisit the phase-space diagram using $\log_{10}((R/R_{200}) \times (|\Delta v|/\sigma_v))$ ($\equiv \log_{10}(r_{\rm norm}\times v_{\rm norm})$) as an approximate proxy for infall time and relate this metric to the member-galaxy properties examined throughout the remainder of the analysis.

\subsection{Demographics of HUDF46}
In this section, we closely examine the galaxy population within HUDF46 and compare its properties to those of field galaxies in the HUDF, in order to assess whether systematic differences are already in place in this young cluster.

\subsubsection{Stellar properties of member galaxies} 

We used \textsc{bagpipes} \citep{carnall_jwst_2024} to derive the stellar properties of both HUDF46 member galaxies and field galaxies. Briefly, \textsc{bagpipes} relies on synthetic templates from \citet{bruzual_stellar_2003} with a \citet{kroupa_variation_2001} IMF, adopting a cut-off mass of 100 $M_{\odot}$, and nebular emission modeled with \textsc{cloudy} \citep{ferland_2013_2013}. We adopted the continuity non-parametric star formation history (SFH) prior from \citet{leja_how_2019}. For each galaxy, we defined the age bin edges (in look-back time from the redshift of observation) following a logarithmic spacing from $z=30$ to the age of the Universe at the corresponding redshift; the same approach was used for the age prior. Stellar masses were allowed to range from $10^5$ to $10^{13}\ M_{\odot}$, using a uniform prior in $\log_{10}( M_{\bigstar}/M_{\odot})$. We adopted a Calzetti dust attenuation law \citep{calzetti_dust_2000}, allowing $A_{V}$ to vary between 0 and 6, and set metallicity ($Z/Z_{\odot}$) to vary between 0 and 2.5. The ionization parameter ($\log_{10} (U)$) was allowed to vary freely between $-4$ and $-0.001$. Redshifts were fixed to the spectroscopic values from NIRISS/WFSS. No AGN component was considered during the SED fitting\footnote{We tested whether including an AGN component for the brightest member galaxy (confirmed AGN; see next section) affects the derived stellar properties, particularly $M_{\bigstar}$, and find no significant differences, with all estimates consistent within $1\sigma$.}. In particular, the SED fitting is based on Kron photometry from HST/ACS, JWST/NIRCam (JADES DR2; \citealt{eisenstein_jades_2023}), and JWST/MIRI (SMILES; \citealt{alberts_smiles_2024}).

The SED fitting analysis shows that the HUDF46 galaxies span nearly three orders of magnitude in $M_{\bigstar}$, from $10^{7.5}$ to $10^{10.5}\,M_{\odot}$, with the majority ($\approx68\%$) lying below $10^{9}\,M_{\odot}$. In contrast, only $\approx30\%$ of the 18 sources reported by \citet{mei_star-forming_2015} had $M_{\bigstar}\lesssim10^{9}\,M_{\odot}$. This underscores the leap forward enabled by deep NIRISS/WFSS observations in probing the low-mass end of the galaxy population in HUDF46. Compared to \citet{mei_star-forming_2015}, the present dataset also enables a more robust spectroscopic redshift determination, owing to the simultaneous detection of multiple emission lines, including H$\alpha$. We infer the stellar properties of the field galaxies ($\rm QF=1$) using the same {\sc Bagpipes} configuration and find a similarly broad distribution in $M_{\bigstar}$ among the field population.

The HUDF46 galaxies show a wide range of dust attenuation, with $A_V$ values spanning $0\text{--}1.5$~mag, similarly to the field galaxies. The most massive systems in HUDF46 ($M_{\bigstar}\gtrsim10^{10}\,M_{\odot}$), which represent only $\approx8\%$ of the total sample, are also the most dust-obscured, typically exhibiting $A_V\gtrsim1$~mag. We also find that the HUDF46 galaxies have predominantly sub-solar metallicities, consistent with the results of \citet{shen_ngdeep_2025}, whose analysis of the same NGDEEP dataset spans a similar redshift range to that of our sample. Yet, since we do not perform spectrophotometric fitting in this work, no further constraints can be drawn.

\subsubsection{Does HUDF46 host AGNs?}

Observationally, studies have shown that the fraction of galaxies hosting AGN increases with redshift, both in the field and in dense environments. In galaxy groups and clusters, this increase is particularly pronounced, with the AGN fraction rising rapidly toward $z\gtrsim1$ and in some cases exceeding field values (e.g.,  \citealt{martini_cluster_2013, alberts_star_2016, hashiguchi_agn_2023, shah_enhanced_2025}). For this reason, we investigated the presence of AGNs among the HUDF46 members by cross-matching with the existing pre-JWST AGN catalog of \citet{lyu_agn_2022} (which is partly based on \citealt{luo_chandra_2017}), which identifies AGNs based the multi-wavelength data from the X-ray to the radio with eight independent selection methods We found a single match: HUDF46+53.1556–27.7793, classified as an AGN because its X-ray–to–radio luminosity ratio exceeds that expected from stellar processes. Interestingly, this source is also the brightest member of HUDF46 and lies close to its center (see zoom-in in Figure \ref{fig:rgb_niriss}). 

We also cross-matched our sources with the recent MIRI-AGN catalog of \citet{lyu_active_2024}, based on SMILES data, which is sensitive to lower-luminosity AGN than previous selection methods and extends to systems of lower stellar mass compared to pre-JWST catalogs. The MIRI-AGN catalog provides one additional match, but while their work lists the source at $z=1.98$, our spectroscopic data place it at $z_{\mathrm{spec}}=1.8574$ based on the detection of [O\,{\sc iii}]$\lambda\lambda4959,5007$ and H$\beta$ emission. Given this discrepancy and the potential biases introduced by adopting an incorrect redshift during their SED fitting, we classify this source as non-AGN in our analysis\footnote{Our visual inspection suggests that their conclusion may be driven by strong Paschen-$\alpha$ entering MIRI/F560W at that redshift, likely biasing their SED fitting results.}.  

It is important to note that the SMILES data are relatively shallow (\citealt{alberts_smiles_2024}), reaching only HUDF46 members with $M_{\bigstar} \gtrsim 10^{9}\,M_\odot$. As a result, 17 galaxies ($\approx45\%$ of the sample) are detected, while the remainder lack MIRI detection. The current MIRI-AGN catalog therefore does not allow us to fully probe the low-mass regime and assess the incidence of AGN activity in the least massive systems. However, we warn the reader that AGN identification in low-mass galaxies remains uncertain even with MIRI data (Iani et al., in prep.), as low-metallicity star-forming systems can mimic AGN-like mid-IR colors (e.g., \citealt{hainline_mid-infrared_2016}). In summary, while existing catalogs rule out AGNs for all but possibly one HUDF46 member in the high-mass end, the presence of AGNs among the low-mass population cannot be fully excluded.

\subsubsection{Yet to quench: HUDF46 members and field galaxies aligned along the Star-Forming Main Sequence}

A defining feature of galaxy clusters is the buildup of a red sequence (i.e., an excess of quenched galaxies relative to the field; e.g., \citealt{de_lucia_build-up_2007}). While quiescent galaxies are observed in some high-$z$ protoclusters and in clusters at $z\approx2$ (e.g., \citealt{sun_jwsts_2024, witten_not_2025}), widespread quenching is typically established only by $z\lesssim1.5$ \citep{noordeh_quiescent_2021}, with significant diversity during the collapse and virialization phase \citep{chiang_galaxy_2017}. 

In the case of HUDF46, \citet{mei_star-forming_2015} found no evidence of a red sequence in their HST-selected membership. Using the catalog from \citet{guo_candels_2013}, they identified only six additional red galaxies, the brightest with spectroscopic redshifts outside the structure, confirming them as non-members. However, they cautioned, however, that some passive galaxies may lie beyond the reach of their available data, as spectroscopic confirmation is often more challenging for quiescent systems in the absence of sufficiently deep observations. 

The NIRISS spectroscopy adopted in this work now doubles the number of confirmed members, enabling a reassessment of potentially missed passive systems. We reach a $5\sigma$ 1D line-flux sensitivity of $\approx 1.3\times10^{-18}\,\rm erg\,s^{-1}\,cm^{-2}$, consistent with NGDEEP studies (\citealt{shen_ngdeep_2024, forrey_ngdeep_2025, shen_ngdeep_2025})\footnote{Almost $40\times$ deeper than HST grism observations adopted in \citet{mei_star-forming_2015} in the same field.}, corresponding to an H$\alpha$ star-formation rate (SFR) of $\approx 0.17\,M_{\odot}\,\rm yr^{-1}$ (\citealt{kennicutt_star_2012})\footnote{Note that we do not rely solely on H$\alpha$ to identify membership; therefore, H$\alpha$ only provides an approximate estimate of our completeness with respect to passive galaxies.}. Adopting a redshift-dependent specific SFR (sSFR) threshold of the form $\mathrm{sSFR}{\rm crit} \propto 0.2/t_{\mathrm{H}}(z)$ \citep{pacifici_evolution_2016}, we consider our membership to be highly complete for red galaxies down to $M_{\bigstar} \approx 10^{9.5}\,M_{\odot}$\footnote{Although the emission-line threshold drives our membership identification, the SFRs used throughout this paper are derived from SED fitting, not from H$\alpha$. Therefore, even for sources identified via [O\,{\sc iii}] with H$\alpha$ below the detection limit, we still obtain an SFR estimate.}.

\begin{figure}
    \centering
    \includegraphics[width=1\linewidth]{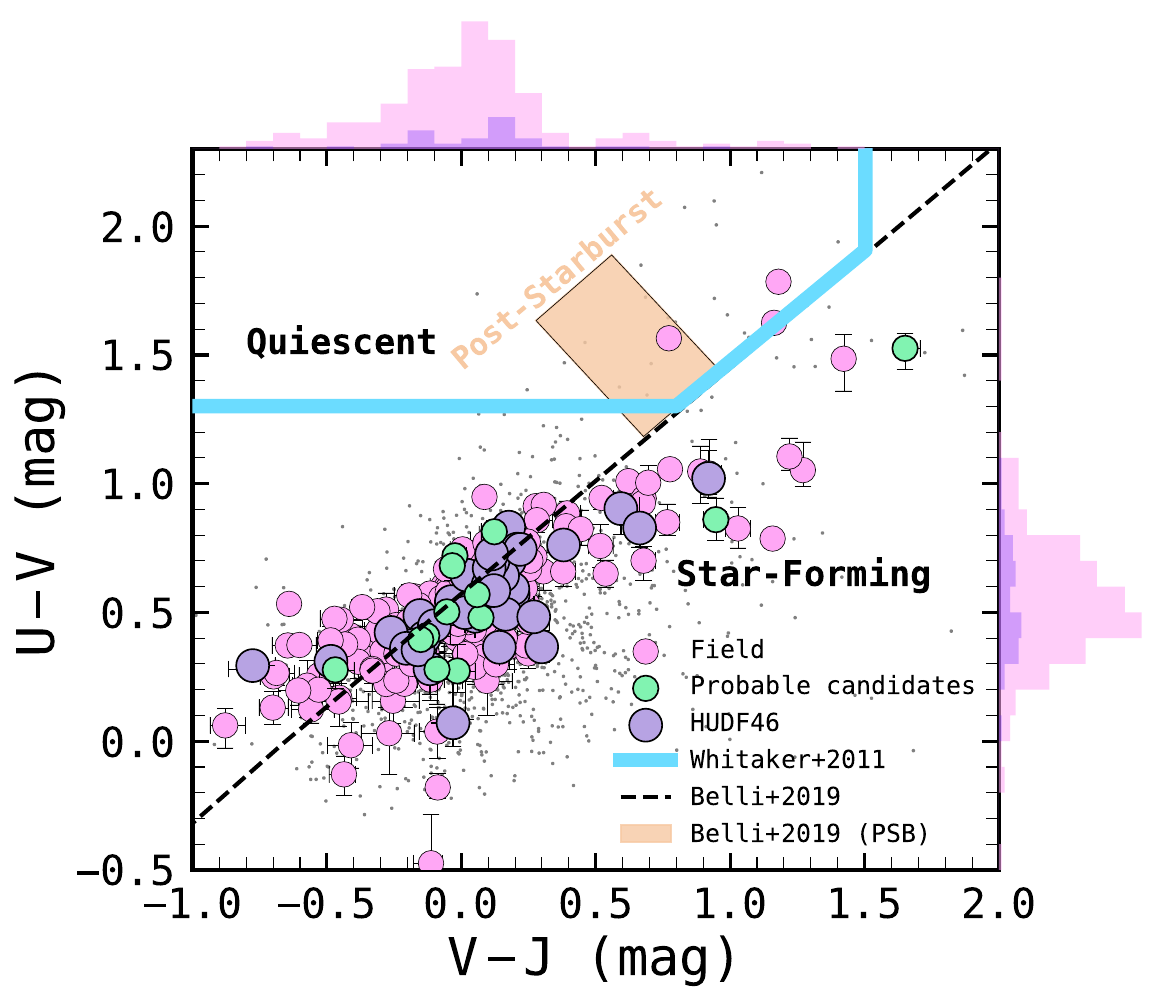}
    \caption{Rest-frame UVJ diagram for both field galaxies at $z\approx1.5\text{--}2.5$ (pink), HUDF46 (purple) members, and the JADES DR2 photometric catalog (comparable redshifts) in the HUDF (gray points; same area of NIRISS). We also show the 13 possible candidates that might belong to HUDF46 located outside the NGDEEP footprint (within $3\times R_{200}$). The solid line (turquoise blue) shows the quiescent selection of \citet{whitaker_newfirm_2011}, while the dashed line indicates the modified criterion of \citet{belli_mosfire_2019}. HUDF46 members are all consistent with the star-forming region, confirming the absence of a red sequence and in agreement with their late-type morphologies. A few members lie above the modified criterion from \citet{belli_mosfire_2019}, yet they show signatures of ongoing star-formation activity (through $H\beta$, [O\,{\sc iii}], and H$\alpha$).
}
    \label{fig:UVJ}
\end{figure}

As a first attempt to look for a red sequence in HUDF46, we adopt the UVJ diagram\footnote{UVJ magnitudes were computed with both {\sc eazy-py} (\citealt{brammer_eazy_2008}) and {\sc bagpipes}, yielding consistent results; Figure~\ref{fig:UVJ} shows the {\sc bagpipes}-based colors.} \citep{williams_detection_2009, whitaker_newfirm_2011} to search for passive and post-starburst (PSB) galaxies, along with the modified criteria of \citet{belli_mosfire_2019} to recover recently quenched, lower-mass systems (see Figure \ref{fig:UVJ}).  We note that this selection has caveats: 1) by relaxing the color cuts, it can introduce contaminants and 2) it may miss some spectroscopically confirmed high-$z$ quiescent galaxies \citep{antwi-danso_beyond_2023, baker_abundance_2025, baker_exploring_2025}; however, these effects are expected to be less significant in the redshift range considered here.

We find that a few field galaxies occupy the classical quiescent region, and all are spectroscopically inconsistent with HUDF46. Some lower-mass cluster members formally satisfy the modified PSB criteria; however, their spectra show H$\alpha$, H$\beta$, and [O\,{\sc iii}] emission, placing them on the star-forming main sequence (see next paragraphs). Given the known limitations of color-based selections and the mixed spectroscopic properties of PSBs \citep{nielsen_evidence_2025}, we conservatively flag these as PSB candidates pending further analysis.

Even with the expanded membership, HUDF46 shows no evidence for a red sequence within $R_{200}$, consistent with \citet{mei_star-forming_2015}. A search for candidates outside the NIRISS footprint (out to $3\times R_{200}$) yields no photometric candidates consistent with HUDF46 and located in the classical UVJ quenching region (\citealt{whitaker_star_2012}). This result remains tentative, as photometric redshifts can be uncertain for passive systems, and spectroscopic constraints are severely incomplete outside the NIRISS footprint.

\begin{figure*}
    \centering
    \includegraphics[width=0.8\linewidth]{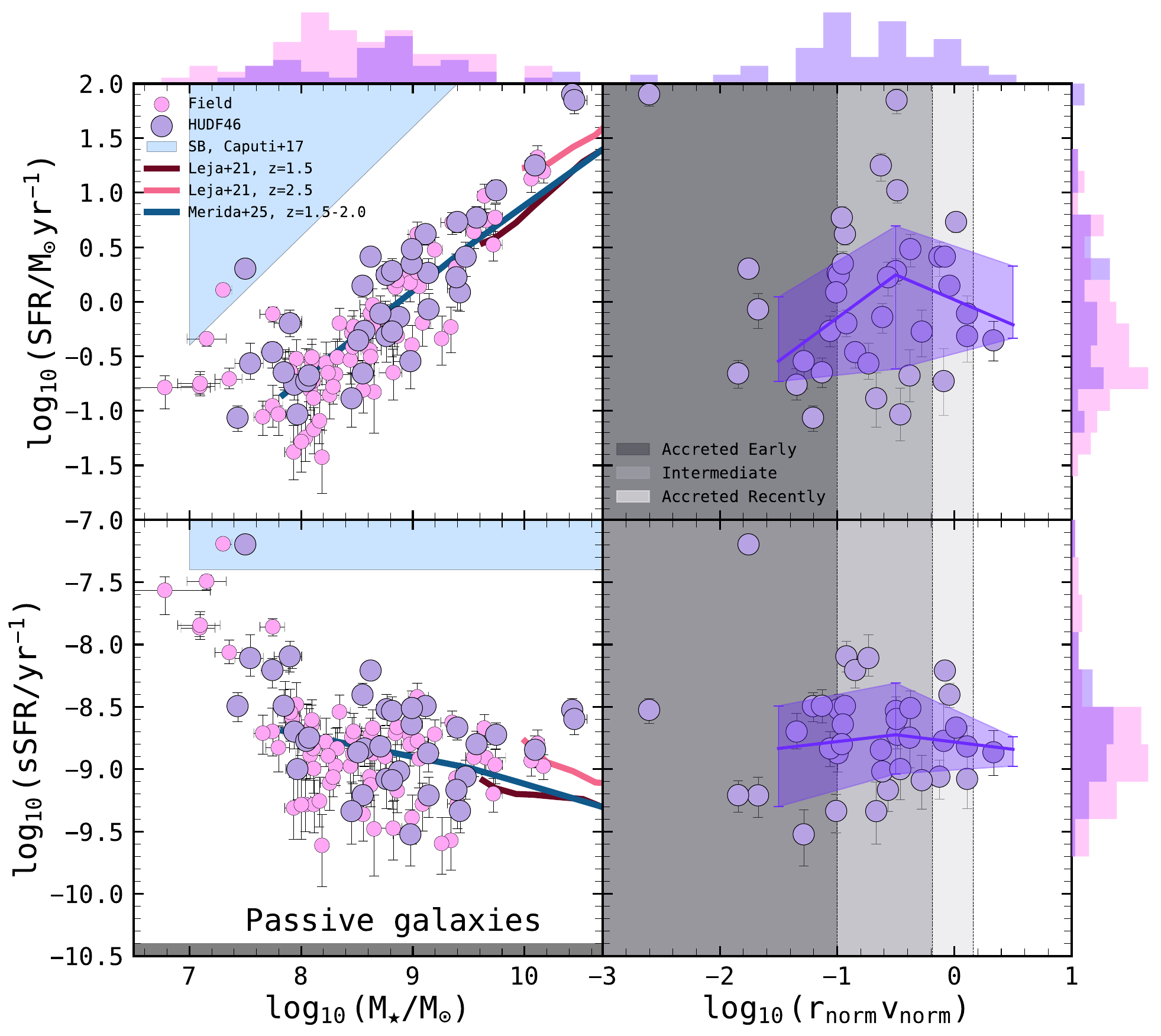}
    \caption{{\bf Left}: The SFR--$M_{\bigstar}$ and sSFR--$M_{\bigstar}$ for both field galaxies at $z\approx1.7\text{--}2$ (pink) and HUDF46 members (purple). For reference, we include the starburst region from \citet{caputi_star_2017} and the main-sequence parameterization at similar redshifts from \citet{leja_new_2022} and \citet{merida_probing_2025}. In the sSFR--$M_{\bigstar}$, we also show in gray the region below which galaxies are defined as quiescent systems based on the redshift-dependent formalism from \citet{pacifici_evolution_2016}, where $\rm sSFR_{\rm crict}\propto0.2/t_{\rm H(z)}$.  On average, field and HUDF46 galaxies occupy the same locus in the SFR–$M_{\bigstar}$ plane, indicating no systematic environmental differences. A few low-mass galaxies, including one HUDF46 object (HUDF46+53.1534326–27.7811176), lie significantly above the main sequence, possibly signaling a starburst episode at the time of observation. {\bf Right}: SFR and sSFR as a function of $\log_{10}(r_{\rm norm}\times v_{\rm norm})$ (following the separation adopted in Figure \ref{fig:phase_space}) along with the median trend. Contrary to what is commonly observed in low-redshift, more mature clusters (e.g., \citealt{noble_kinematic_2013, noble_phase_2016}), we find no substantial evolution across the accretion histories probed in HUDF46, with all members appearing overwhelmingly star-forming regardless of their orbital history.
 }
    \label{fig:main-sequence}
\end{figure*}

\begin{figure*}
    \includegraphics[width=0.99\linewidth]{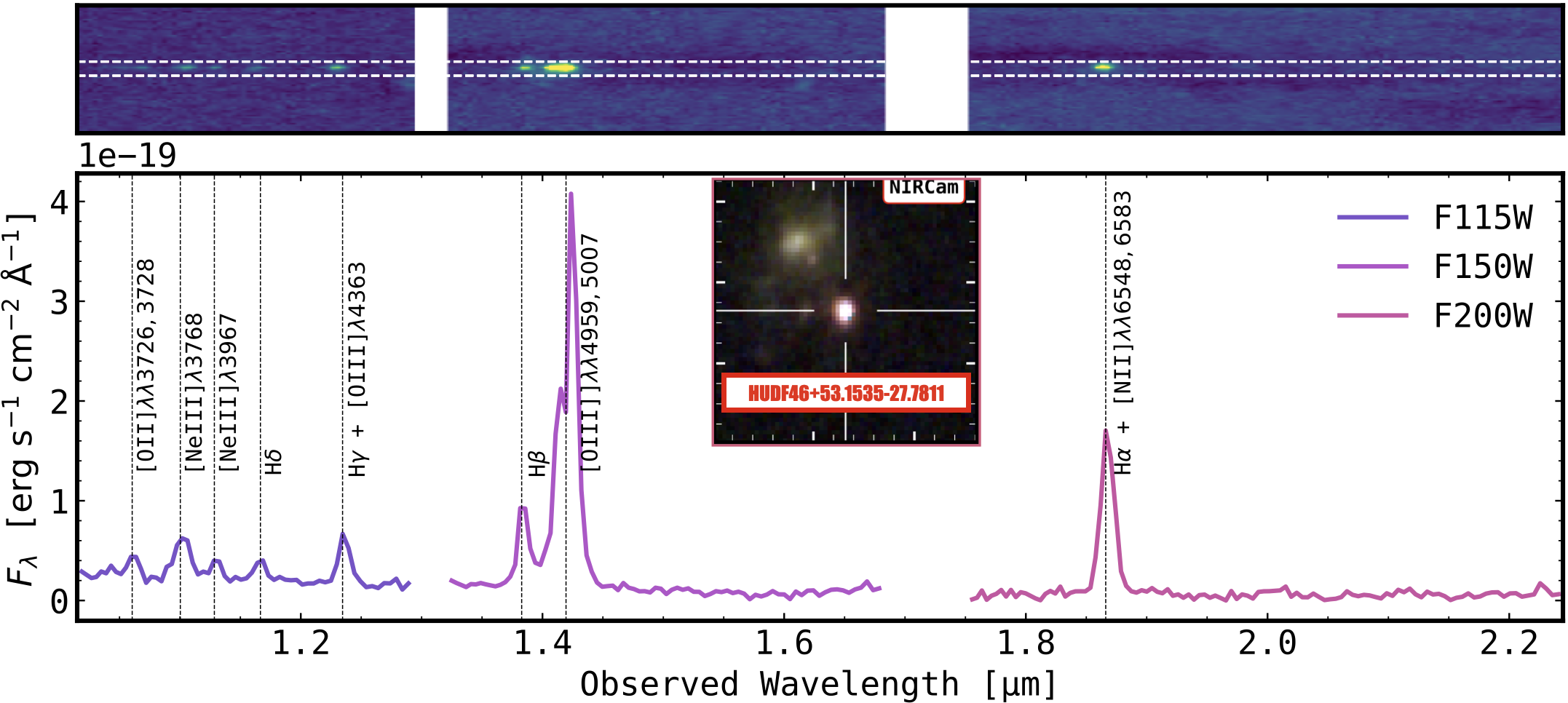}
    \caption{NIRISS spectrum of HUDF46+53.1534–27.7811. {\bf Top:} continuum-subtracted 2D spectrum. {\bf Bottom:} 1D spectrum. This is the only galaxy located in the starburst cloud (\citealt{caputi_star_2017}) of the SFR$\text{--}M_{\bigstar}$ plane, showing multiple strong emission lines. This galaxy has $\log_{10}(M_{\bigstar}/M_{\odot}) = 7.89\pm0.15$, $A_{V} = 0.31\pm0.05$~mag, a UV slope of $\beta=-2.06\pm0.03$, and $M_{UV}=-18.55\pm0.05$. Its red appearance in the NIRCam RGB cutout (F090W, F115W, and F200W) is likely due to strong H$\alpha$ emission falling within the NIRCam/F200W filter.}
    \label{fig:NIRISS_spectrum_4641}
\end{figure*}

A classical UVJ selection alone may be insufficient to robustly identify passive galaxies, though. It is often complemented by a cut in specific SFR (sSFR), adopting a redshift-dependent threshold of the form $\rm sSFR_{\rm crit} \propto 0.2/t_{\rm H(z)}$ (\citealt{pacifici_evolution_2016}), which corresponds to $\approx 10^{-10.40}\,\rm{yr^{-1}}$ at $z \approx 1.84$. Therefore, as a complementary approach to the UVJ diagram, we examine the $\mathrm{SFR}\text{--}M_{\bigstar}$ and $\mathrm{sSFR}\text{--}M_{\bigstar}$ relations (Figure \ref{fig:main-sequence}). From Figure \ref{fig:main-sequence} (left panels), we notice that nearly all HUDF46 members follow the star-forming main sequence (SFMS;  \citealt{brinchmann_physical_2004, speagle_highly_2014, rinaldi_emergence_2024}), with only one clear starburst galaxy (\citealt{caputi_star_2017, rinaldi_galaxy_2022}; specifically, HUDF46+53.1534–27.7811; Figure~\ref{fig:NIRISS_spectrum_4641}).

Adopting the SFMS parameterization of \citet{leja_new_2022}, we find consistency at the high-mass end with their relation, and at low masses with its extrapolation recently derived by \citet{merida_probing_2025}. In both the SFR–$M_{\bigstar}$ and sSFR–$M_{\bigstar}$ planes, HUDF46 and field galaxies overlap, consistent with results for massive clusters at $z>1$ \citep{alberts_star_2016}, lower-mass systems \citep{williams_alma_2022}, and simulations from IllustrisTNG \citep{nelson_illustristng_2019, andrews_galaxy_2025}, which find no clear environmental dependence of the MS. 

Motivated by the possibility of identifying distinct accretion histories within our sample using the projected phase--space diagram, in Figure~\ref{fig:main-sequence} (right panels) we show the SFR and sSFR as a function of $(r/r_{200}) \times (\Delta v/\sigma_v)$\footnote{$(r/r_{200}) \times (\Delta v/\sigma_v)$ is the dimensionless phase–space coordinate tracing projected distance and velocity offset from the cluster center.} to look for for slowly varying SFR and sSFR, as observed in lower-redshift clusters.   Unlike low-$z$ clusters (where early infallers show suppressed star formation; \citealt{noble_kinematic_2013, noble_phase_2016}), we find no significant trend: both SFR and sSFR remain broadly flat, similar to field results \citep{noble_phase_2016}. However, previous studies focused on more massive galaxies ($M_{\star}\gtrsim10^{9.5}\,M_{\odot}$) and more evolved clusters at $z\lesssim1.5$, where quenching is already advanced \citep{dressler_evolution_1984, balogh_bimodal_2004, kauffmann_environmental_2004, baldry_galaxy_2006, muzzin_discovery_2013, kawinwanichakij_effect_2017, tomczak_glimpsing_2017, papovich_effects_2018}. Our results extend this into the low-mass end of the galaxy population within a cluster and it may therefore reflect either weak environmental effects in HUDF46 at Cosmic Noon or mass-dependent environmental mechanisms, as suggested by \citet{hamadouche_jwst_2025}.

\subsubsection{Morphology of HUDF46 members}

\begin{figure*}
    \includegraphics[width=0.99\linewidth]{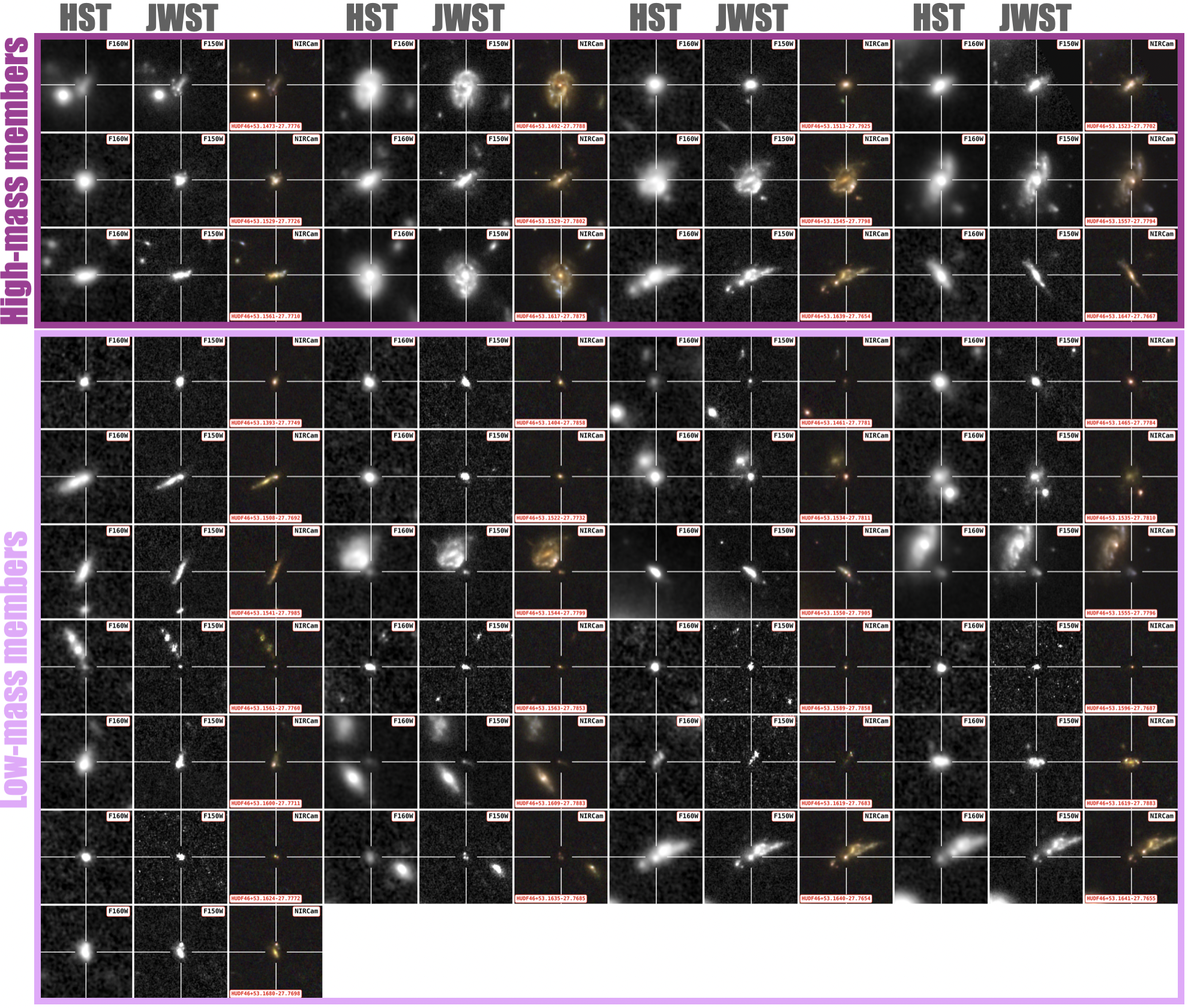}
    \caption{The 37 galaxies identified as HUDF46 members. For each object, we show HST/WFC3 F160W, JWST/NIRCam F150W, and an RGB composite built from NIRCam F090W, F115W, and F150W. Each cutout spans 1.5\arcsec$\times$1.5\arcsec. The sample is divided into high-mass ($>10^{9}\,M_{\odot}$) and low-mass ($\lesssim10^{9}\,M_{\odot}$) galaxies. The direct comparison between HST and NIRCam highlights how JWST’s superior sensitivity and angular resolution allow a far more reliable characterization of galaxy morphology. Several sources that appeared smooth or spheroidal in HST (potentially leading \citet{mei_star-forming_2015} to classify them as ETGs) reveal clear substructure in the NIRCam data.}
    \label{fig:morphology}
\end{figure*}

At $z\gtrsim1.5$, some galaxy populations in clusters and protoclusters already show a clear morphological mix, with both ETGs and late-type galaxies (LTGs) present. For instance, \citet{mei_morphology-density_2023} analyzed overdensities in the Clusters Around Radio-Loud AGN (CARLA) survey, which revealed that a significant fraction of galaxies display early-type morphologies even at $z\approx2$, while late-type and irregular systems remain common, reflecting the ongoing assembly of these environments.

Interestingly, \citet{mei_star-forming_2015} suggested that some HUDF46 members could be \textit{star-forming} ETGs based on visual inspection. While ETGs are classically defined as morphologically early-type systems hosting old stellar populations and occupying the passive locus of the SFR–$M_{\bigstar}$ plane in the local Universe (e.g., \citealt{george_revealing_2015}), several studies have identified star-forming ETGs both locally and at intermediate redshift. In SDSS, such systems show SFRs of $0.5\text{--}50\,M_{\odot}\,\rm{yr}^{-1}$ and often exhibit extended star formation and merger signatures \citep{kollmeier_sloan_2025, nolan_young_2007, george_structural_2017, george_deep_2023}. Similarly, in $z\approx1\text{--}1.5$ clusters, \citet{wagner_star_2015} found ETGs with enhanced specific SFRs relative to local counterparts, though still below those of coeval field SFGs.

Although these findings demonstrate that an early-type morphology does not preclude ongoing star formation, recent JWST results show that deeper, higher-resolution imaging can revise morphological classifications (\citealt{ferreira_jwst_2023}). Using HST data, \citet{mei_star-forming_2015} classified several HUDF46 members as star-forming ETGs ($\approx44\%$) of their sample). We therefore visually re-examined all 37 confirmed members, with particular focus on those identified as star-forming ETGs in their work.

For each source, we inspected HST and JWST imaging at comparable observed wavelengths ($\approx0.8\text{--}1.6\,\mu$m) to assess potential biases arising from HST’s coarser resolution (0.13\arcsec/px for WFC3) relative to NIRCam (0.031\arcsec/px in the short-wavelength channel). We classify galaxies as LTGs when exhibiting disks, irregular features, or clumps, and as ETGs when showing smooth, centrally concentrated light profiles. Figure~\ref{fig:morphology} presents the HUDF46 members split into two stellar-mass bins, low-mass ($M_{\bigstar}\lesssim10^{9}\,M_{\odot}$) and high-mass ($M_{\bigstar}>10^{9}\,M_{\odot}$), displaying HST/WFC3 F160W and NIRCam/F150W images, along with RGB composites from F090W, F115W, and F150W.
 
Our visual inspection shows that nearly all high-mass galaxies exhibit disturbed morphologies, with only one system appearing centrally concentrated. In that case, the bulge-to-total ratio ($B/T$) in the NIRCam bands is $\approx0.3$ \citep{genin_dawn_2025}, consistent with systems transitioning from disk- to bulge-dominated morphologies \citep{dimauro_coincidence_2022}; by comparison, ETGs are typically defined by $B/T\gtrsim0.6$ \citep{dimauro_coincidence_2022}. Among low-mass members, $\approx60\%$ display disturbed structures, while the remainder appear compact. Several galaxies also show clumpy or fragmented morphologies, indicative of gas-rich, gravitationally unstable disks \citep{zhu_clump-like_2026}.

Among the member galaxies, the eight systems reported as star-forming ETGs by \citet{mei_star-forming_2015} warrant closer examination\footnote{Notably, the object UDF-2383, previously reported at $z=1.865$, is now confirmed by MUSE spectroscopy to lie at much lower redshift (\citealt{bacon_muse_2023}).}. We complement our reclassification by using the morphological parameter estimates provided by \citet{genin_dawn_2025} (i.e., the bulge-to-total ratio). 

Our JWST-based visual classifications show that none of the galaxies previously identified as star-forming ETGs display the defining morphologies of bona fide early-type systems. In all cases, the higher spatial resolution of NIRCam reveals structures that differ markedly from the smoother HST appearance: although often compact, these galaxies are disk-dominated, irregular, asymmetric, or clumpy, and in some instances show disturbed or tidal features. Their $B/T$ values span $\lesssim0.1$ to $\approx0.4$, well below the $\gtrsim0.6$ threshold typically adopted for ETGs (e.g., \citealt{dimauro_coincidence_2022}), confirming the absence of strongly centrally concentrated light profiles. We therefore conclude that none of these systems can be classified as genuine ETGs. Overall, HUDF46 members show on average $B/T\approx0.2$ in F090W, F115W, and F150W\footnote{Only three very compact objects, with $M_{\bigstar} < 10^{9}\,M_{\odot}$, show values above 0.4 from \citet{genin_dawn_2025}; however, given their extremely small sizes, we consider these measurements likely spurious.}.

Although based on a limited sample, this comparison illustrates the step forward provided by JWST’s superior angular resolution and suggests that some HST-based classifications were affected by resolution and possibly depth limitations (see also \citealt{kuhn_jwst_2024}). Similar reassessments with JWST data may therefore be warranted in other structures. We conclude that, at present, there is no clear evidence for an established morphology–density relation in HUDF46.

\subsubsection{The size-mass relation in HUDF46}

\begin{figure*}
    \centering
    \includegraphics[width=1.\linewidth]{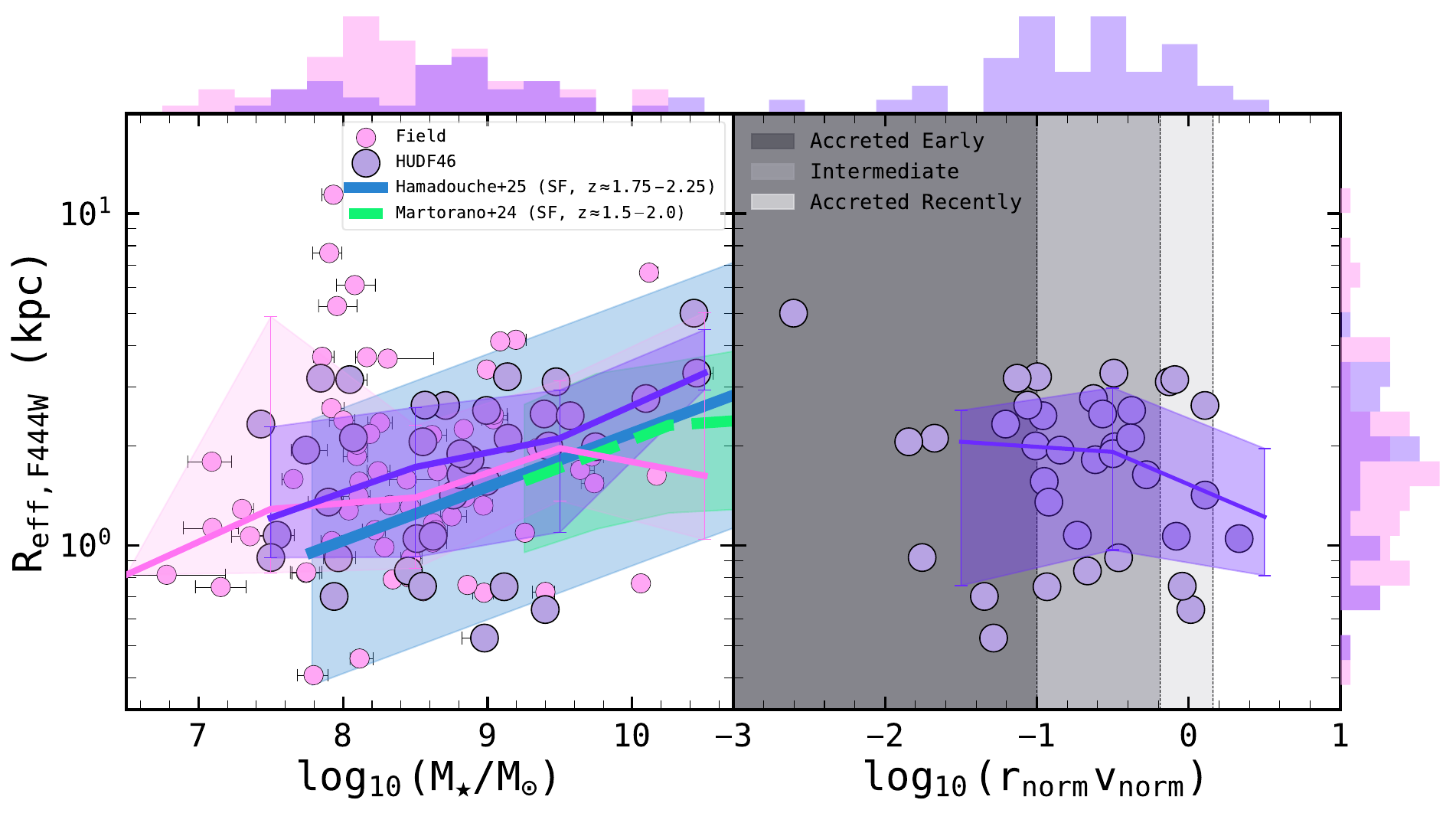}
    \caption{{\bf Left:} Mass–size relation for both field galaxies at $z\approx1.7\text{--}2$ (pink) and HUDF46 (purple) members measured with JWST/NIRCam F444W. Median trends in stellar mass bins are shown, along with the recent size-mass relations for star-forming at comparable redshifts from \citet{hamadouche_jwst_2025} and \citet{martorano_sizemass_2024}. Both field and HUDF46 galaxies follow the mild evolution expected for star-forming systems, and, critically, no significant offset is detected between HUDF46 and field populations. {\bf Right:} $R_{\rm eff, F444W}$ as a function of $\log_{10}(r_{\rm norm}\times v_{\rm norm})$ (following the separation adopted in Figure \ref{fig:phase_space}) along with the median trend. No substantial evolution is observed between early and recently accreted galaxies, with the median trend remaining broadly flat, although a significant diversity is present at fixed $\log_{10}(r_{\rm norm}\times v_{\rm norm})$.
}
    \label{fig:size_mass}
\end{figure*}

Over the past decades, several studies have investigated how environment can influence the structural properties of galaxies (e.g., \citealt{kuchner_effects_2017, carlsten_structures_2021}). In dense environments, environmental quenching mechanisms are expected to act preferentially on the outer regions of galaxy disks, leading to an outside-in suppression of star formation \citep{bluck_how_2020}. As a consequence, galaxies undergoing a transition from star-forming to quiescent phases may appear systematically more compact at fixed stellar mass, not because of an intrinsic structural transformation, but due to the fading of their outer disks (e.g., \citealt{kuchner_effects_2017}). However, observational results remain mixed, with reported trends strongly dependent on sample selection, wavelength, surface-brightness limits, and morphological classification, highlighting the role of significant observational biases. For instance, recent results from, e.g., \citet{pan_uncovermegascience_2025} challenge this view, showing no significant environmental dependence in a protocluster at $z\approx2.6$.

Motivated by this, we perform a similar analysis for our sample. We estimate the sizes of both field and HUDF46 galaxies using the F444W filter. In our case, the sample spans $z\approx1.5\text{--}2.5$, with HUDF46 members located at $z\approx1.84$. At these redshifts, F444W traces rest-frame near-infrared wavelengths, providing a reliable measure of the $M_{\bigstar}$ distribution (\citealt{meidt_reconstructing_2014}).

Structural parameters for both field and HUDF46 galaxies are computed using \textsc{pysersic} \citep{pasha_pysersic_2023}, a Bayesian inference framework for fitting surface brightness profiles to galaxies with robust uncertainty estimation made possible by implementing Markov Chain Monte Carlo (MCMC) methods. The values reported here are derived from the posterior distributions of the structural parameters for single-Sérsic component fits, obtained using the MCMC sampling mode. A detailed description of the procedure for fitting Sérsic profiles to galaxies in JADES, and the resulting structural parameters, can be found in \citet{carreira_jwst_2026}.

In Figure \ref{fig:size_mass} (left panel), we compare the sizes of field and HUDF46 galaxies and find no significant offset between the two populations. However, three confirmed HUDF46 members lie along the lower envelope of the recent size–mass relations reported by \citet{hamadouche_jwst_2025} and show bulge-to-total ratios consistent with transitional systems ($\approx 0.3\text{--}0.4$) found by \citet{genin_dawn_2025}. The brightest member instead lies on the upper bound of the relation from \citet{hamadouche_jwst_2025}, with $R_{\rm eff} \approx 5$ kpc, and is the largest galaxy in the sample. The median trends in each mass bin are consistent with the results for star-forming systems reported by \citet{hamadouche_combined_2022} and \citet{martorano_sizemass_2024}. When examining the sizes of HUDF46 members as a function of $(r/r_{200}) \times (\Delta v/\sigma_v)$ (Figure~\ref{fig:size_mass}, right panel), we find no strong trend in $R_{\rm eff,,F444W}$ across the range of inferred accretion histories, with sizes remaining approximately flat. The largest galaxy in the sample (i.e., the brightest member), however, lies in the region associated with very ancient infall, which may be consistent with its more extended structure compared to more recently accreted members.

\subsubsection{No difference in Balmer/4000\,\AA\, between field and HUDF46 galaxies}

Star formation in galaxies is not eternal; it is a transient phase that can cease temporarily or permanently. When suppressed, short-lived OB stars fade and the light becomes dominated by A-type and older stars, producing a sharp feature at the Balmer limit (e.g., \citealt{bruzual_a_spectral_1983, poggianti_indicators_1997}). It is weak in young systems and prominent in evolved or quenched galaxies, increasing with mass-to-light ratio and decreasing with specific SFR (e.g., \citealt{worthey_old_1994}). In massive galaxies, the Balmer break is therefore commonly used as a proxy for recent quenching and assembly history. In cluster environments, its strength has been shown to depend on accretion history in massive populations, linking stellar population age to environmental processing out to intermediate redshift (e.g., \citealt{kim_gradual_2023}). We can now extend this analysis across the full mass range, from massive to low-mass galaxies in a young cluster, and directly compare their Balmer break strengths to those of field galaxies at similar redshift.

Following the discussion in \citet{wilkins_first_2024}, we quantify the break strength in field and HUDF46 member galaxies following \citet{binggeli_balmer_2019} as the more classic definition introduced by \citet{bruzual_a_spectral_1983}\footnote{Defined as $D_{4000}$, which relies on using windows at [3750, 3950]\,\AA\, and [4050, 4250]\,\AA.} is not useful for quantifying the break in younger galaxies. Leveraging the multi-wavelength photometry and SED-fitting results, we measure the continuum in $F_{\nu}$ at 3500\,\AA\ and 4200\,\AA\ and define $B = F_{\nu,4200\text{\AA}} / F_{\nu,3500\text{\AA}}$. This definition encompasses both the Balmer and classical 4000\,\AA\ breaks; we therefore refer to $B$ as the Balmer/D4000 index\footnote{We note that resolving the two components would require high-resolution spectroscopy, beyond the capabilities of the available NIRISS data given their limited spectral resolution ($R \approx 150$) and wavelength coverage, which at $z\approx1.84$ does not fully sample the rest-frame region of interest.}.

\begin{figure*}
    \centering
    \includegraphics[width=1.\linewidth]{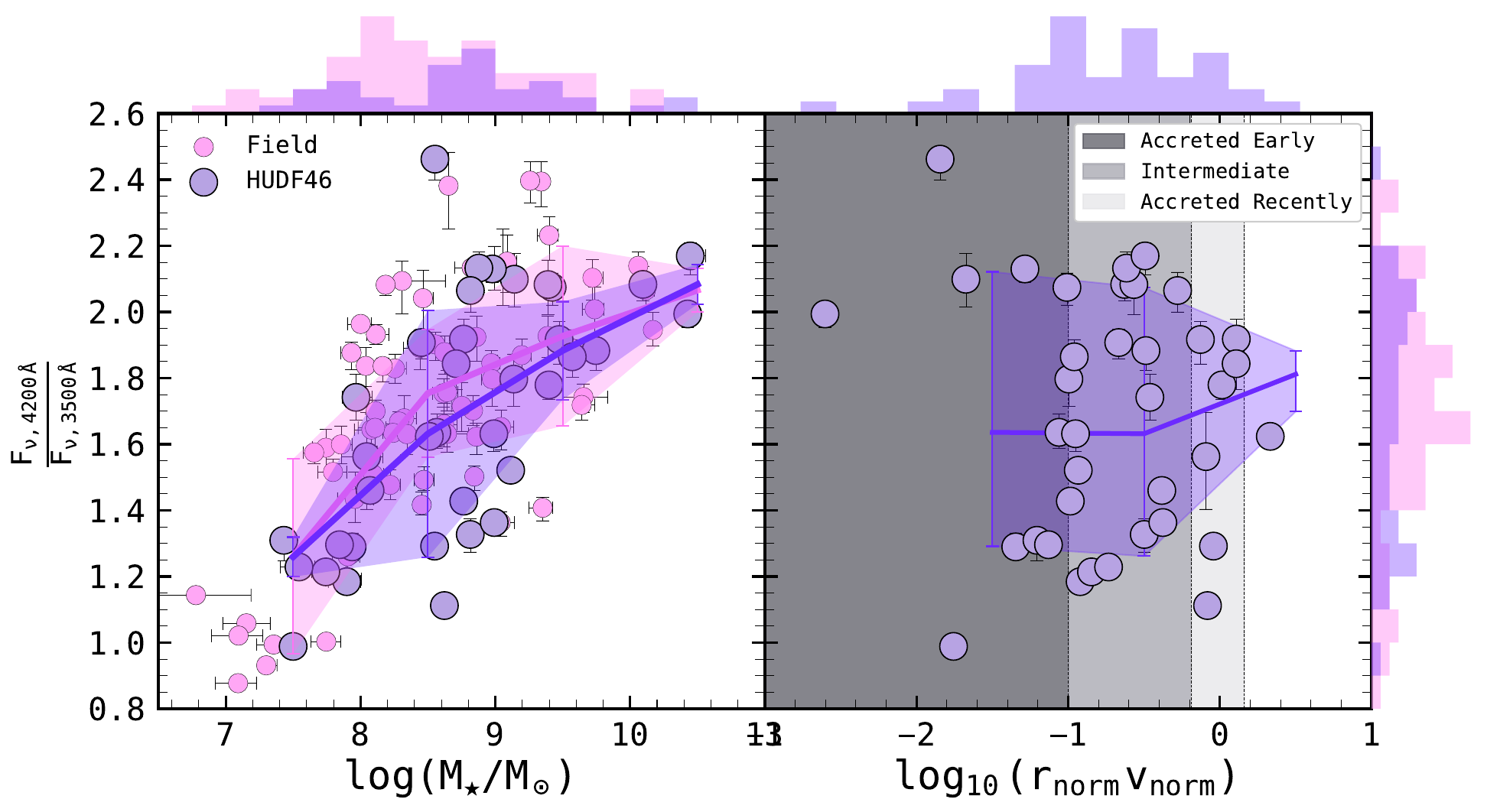}
    \caption{{\bf Left:} The Balmer/4000\,\AA\;strength ($B$) as a function of $M_{\bigstar}$ for both field galaxies at $z\approx1.7\text{--}2$ (pink) and HUDF46 (purple) members. A clear mass dependence is present, with more massive galaxies exhibiting higher $B$ values, in line with the downsizing (\citealt{cowie_new_1996}). Notably, no systematic differences are observed between field and HUDF46 members. {\bf Right:} $B$ as a function of $\log_{10}(r_{\rm norm}\times v_{\rm norm})$ along with the median trend. No strong variation is observed across the range of accretion histories probed in HUDF46, in clear contrast to what is typically found in more mature clusters at lower redshifts (e.g., \citealt{noble_phase_2016}). On the other hand, HUDF46 members exhibit a wide spread in $B$ at fixed $\log_{10}(r_{\rm norm}\times v_{\rm norm})$, indicating substantial diversity in their stellar populations.
}
    \label{fig:balmer_break_mass}
\end{figure*}

In Figure~\ref{fig:balmer_break_mass}, we compare the Balmer/4000\,\AA\; ($B$) between field and HUDF46 member galaxies. A clear stellar-mass dependence of $B$ is observed in both environments.  Figure \ref{fig:balmer_break_mass} shows that there is no significant difference in $B$ between field galaxies and HUDF46 galaxies, especially in the low-mass regime ($M_{\bigstar}\lesssim10^{9}\,M_{\odot}$). One object from HUDF46 appears to exhibit a large Balmer break ($B\approx2.5$). After inspecting this source, we found that its photometry is likely contaminated by a nearby, much more massive HUDF46 member. We therefore regard this feature as spurious. 

Overall, the HUDF46 members span a broad range of $B$ values (particularly at low stellar masses) that fully overlaps with the parameter space of field galaxies. Nonetheless, as shown in Figure~\ref{fig:main-sequence}, HUDF46 galaxies lie on the star-forming main sequence, indicating ongoing star formation at the epoch of observation (with one object experiencing a pronounced starburst).

Importantly, the coexistence of relatively strong Balmer/D4000 breaks and ongoing star formation does not imply that these systems are globally quiescent. Main-sequence galaxies, particularly at low stellar masses ($M_{\bigstar}\lesssim10^{8\text{--}10}\,M_{\odot}$), are expected to exhibit intrinsically bursty star formation, with variability increasing toward lower masses \citep{atek_star_2022, navarro-carrera_burstiness_2024, simmonds_bursting_2025}. In this framework, the observed $B$ values may trace temporary lull phases between bursts rather than sustained quenching \citep{looser_recently_2024}. At higher stellar masses, however, enhanced Balmer break/4000\,\AA\; in systems that remain consistent with the main sequence are more plausibly linked to rejuvenation episodes, frequently associated with merger-driven events at cosmic noon, which play a significant role in the buildup of the quiescent population at later times \citep{harrold_role_2026}. 

Finally, we investigate whether $B$ shows any dependence on accretion history by examining its behavior as a function of $(r/r_{200}) \times (\Delta v/\sigma_v)$ in the right panel of Figure~\ref{fig:balmer_break_mass}. In more evolved clusters at lower redshifts, $B$ has been observed to vary systematically across phase space, decreasing monotonically toward more recent infallers \citep[e.g.,][]{noble_phase_2016}, a trend that has also been reported in more recent work \citep{kim_gradual_2023}. In the case of HUDF46, however, we find no evidence for such a trend. Instead, $B$ spans a wide range of values at fixed $\log_{10}[(r/r_{200}) \times (\Delta v/\sigma_v)]$, indicating substantial diversity among the member galaxies. As discussed earlier, this discrepancy may reflect distinct environmental quenching processes operating in low- and high-mass galaxies (e.g., \citealt{hamadouche_jwst_2025}).

\section{Discussion and Conclusions}

Studies of clusters and protoclusters often have to contend with biases imposed by observational limitations. (Proto)cluster selection techniques, inhomogeneous datasets, lack of spectroscopy, and restricted areal coverage can all complicate our understanding of the role of environment in galaxy evolution.  Our concept and identification of galaxy clusters often hinges on the detection of an established ICM and evolved, massive populations.  It is well established that the widespread quenching and morphological transformation of cluster populations is largely in place by $z\approx1$ (e.g., \citealt{muzzin_gemini_2012, balogh_evidence_2016, webb_evolution_2013, liu_morphological_2021}) and early studies projected the environmental processing of these populations back to the pre-virialized, proto-cluster phase (e.g., \citealt{thomas_epochs_2005, de_lucia_hierarchical_2007}). Cluster studies at $z\approx1\text{--}2$ have now, however, revealed substantial system-to-system variation at fixed halo mass:  high-redshift clusters host both vigorously star-forming and quenched massive galaxies (see \citealt{overzier_realm_2016, alberts_clusters_2022}).  A similar diversity is found in proto-clusters, which contain both active (e.g., \citealt{wang_discovery_2016}) and quenched populations (e.g., \citealt{baker_double_2025, pan_uncovermegascience_2025}).  Interpreting the protocluster to cluster transition requires a deeper look into this epoch and, particularly, the parameter space opened by JWST's deep spectroscopic and imaging capabilities in the infrared.

\subsection{HUDF46: a cluster in transition?}

In this work, we have used ultra-deep NIRISS spectroscopy to confirm that HUDF46 is a significant overdensity ($\delta\approx17\sigma$) and measured a halo mass of $M_{200}=(1.2\pm0.2)\times10^{14}\,M_{\odot}$, given the velocity distribution of its member galaxies and the assumption of virialization. This indicates that HUDF46 is a young cluster at $z=1.84$, however, we have also shown that it generally lacks the hallmarks of clusters observed at later times.  HUDF46 shows no signs of a red sequence, as first found in \citet{mei_star-forming_2015}, and the lack of X-ray detection, though consistent with the measured halo mass given its upper limit, implies a relatively weak or early stage ICM.  Additionally, and puzzlingly, HUDF46 exhibits a deficit of massive ($\gtrsim10^{11}\,M_{\odot}$) galaxies. 

At $\log_{10} (M_{\rm h}/M_{\odot})\approx14$, the expected stellar mass fractions are $M_{\bigstar,\mathrm{central}}/M_{\rm halo}\approx0.002$ and $M_{\bigstar,\mathrm{total}}/M_{\rm halo}\approx0.01$ (e.g., \citealt{behroozi_average_2013, shuntov_cosmos2020_2022, zhang_halo_2024}), albeit with substantial scatter driven by galaxy properties (e.g., SFR and morphology; \citealt{zhang_halo_2024}) and number density of satellites (e.g., \citealt{lu_galaxy_2016, zhang_halo_2024}). HUDF46 falls well below these expectations; indeed, its most massive (central) galaxy is star-forming with $\log_{10}(M_{\bigstar}/M_{\odot})\approx10.5$, and the total stellar mass within the virial radius is only $\log_{10}(M_{\bigstar}/M_{\odot})\approx11$, comparable to that of a typical Bright Cluster Galaxy (BCG) in a $\log_{10}(M_{\rm h}/M_{\odot})\approx14$ halo at much later times. 

One possibility is that the assumption of full virialization is incorrect, at least in part, which would lower the inferred halo mass. However, if one were to invoke simple abundance-matching arguments, the stellar mass of HUDF46’s central galaxy would correspond to a halo mass of only $\log_{10}(M_{\rm h}/M_{\odot})\approx12\text{--}12.5$. This is difficult to reconcile with its location in a $17\sigma$ overdensity and likely fails to capture the accelerated growth expected for BCGs in dense environments at later times (e.g., \citealt{demaio_growth_2020}).  Furthermore, studies of the GOGREEN clusters at $z\approx1\text{--}1.5$ show that the number of massive galaxies ($\log_{10}(M_{\bigstar}/M_{\odot})>10.2$) within 1000~kpc spans more than an order of magnitude at nearly fixed halo mass (\citealt{van_der_burg_gogreen_2020}), with the low end of this distribution consistent with HUDF46.

This deficit of massive galaxies in HUDF46 may explain its lack of a quenched population, which sets HUDF46 apart from most of the clusters studied to date at $z>1.5$.  Often these very early clusters are reported to be evolved systems with high QG fractions\footnote{This likely reflects selection biases rather than intrinsic properties, as cluster selection that relies on detection of the ICM or identifying galaxy overdensities has historically become harder with increasing redshift.  This is changing in the JWST era.} \citep[e.g.][]{cooke_mature_2016, lee-brown_ages_2017, newman_spectroscopic_2014, strazzullo_galaxy_2019}; however this excess is almost exclusively confined to the most massive galaxies ($\gtrsim10^{10.5\text{--}11}\,M_{\odot}$). At lower stellar masses ($\lesssim10^{10}\,M_{\odot}$), early cluster QG fractions are generally consistent with those of the field (e.g., \citealt{lee-brown_ages_2017, kiyota_cluster_2025}). This is consistent with the stellar-mass dependence of quenching in clusters generally seen at $z>1$ (e.g., \citealt{pintos-castro_evolution_2019, van_der_burg_gogreen_2020}).  For the most massive cluster galaxies, multiple lines of evidence indicate they may quench predominantly through secular processes, prior to infall: their stellar ages are indistinguishable between cluster and field environments \citep{webb_gogreen_2020, ahad_environment-dependent_2024}, quenching is already evident in infall regions \citep{werner_satellite_2022}, and stellar mass functions imply early shutdown \citep{van_der_burg_gogreen_2020}.  As the HUDF itself is known to be underdense in bright sources, even compared to the broader GOODS-S field \citep{Beckwith_hubble_2006}, we posit that very massive, pre-quenched galaxies were too rare in this field to fall into HUDF46 by the epoch of observation. Projecting whether HUDF46 can potentially accrete sufficient mass to approach the expected $M_{\bigstar}$--$M_{h}$ relation at later times, however, requires spectroscopic coverage over a {\it much} larger area in order to characterize the full infall population and to confirm or rule out the presence of an extended structure.

So, {\it is HUDF46 a cluster in transition?} With the exception of very massive galaxies, widespread quenching appears to begin in earnest around $z\approx1.5$ in massive clusters \citep{dressler_evolution_1984, balogh_bimodal_2004, kauffmann_environmental_2004, baldry_galaxy_2006, muzzin_discovery_2013, alberts_evolution_2014, alberts_star_2016, tomczak_glimpsing_2017, papovich_effects_2018}.  What triggers this transition is not well understood.  Recently, \citet{mei_morphology-density_2023} showed that, for radio-selected clusters at $z\approx1.3$–2 from the CARLA survey \citealp{wylezalek_galaxy_2013, wylezalek_galaxy_2014, noirot_hst_2018} with $\log_{10}(M_h/M_\odot)\approx13.5$–14.5, the fractions of massive quenched and early-type galaxies correlate with the {\it local} environment rather than with global halo mass. Local studies of the effectiveness of ram pressure stripping in clusters alternatively point to a threshold in the density of the ICM \citep{roberts_quenching_2019, roberts_lotss_2022}, which could mean either globally or in terms of an individual galaxy's trajectory after infall. 

\begin{figure}
    \includegraphics[width=0.99\linewidth]{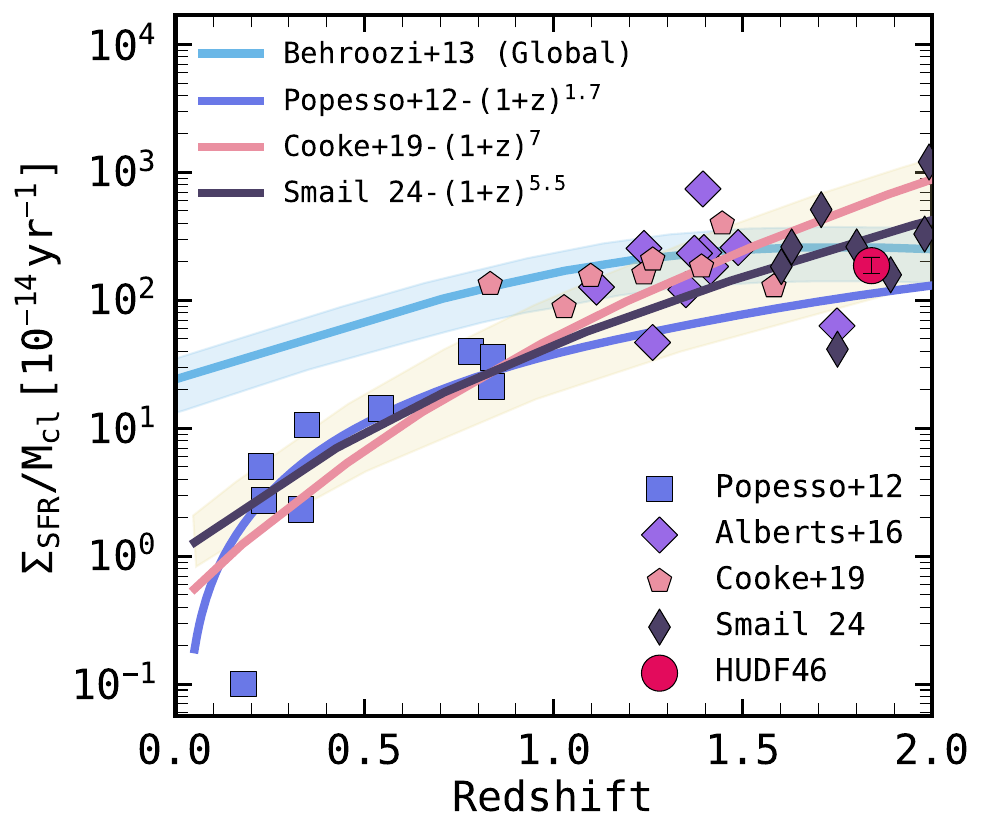}
    \caption{Halo mass-normalized integrated SFR as a function of redshift. We show our measurement for HUDF46 along with recent results from the literature and the derived evolution as a function of redshift from \citet{popesso_evolution_2012, alberts_star_2016, cooke_submillimetre_2019, smail_obscured_2024}. In addition, we compare to the global evolution of the mass-normalized SFR for all galaxies (sky blue line; \citealt{behroozi_average_2013})}.
    \label{fig:sigma_sfr_vs_z}
\end{figure}

At $z=1.84$, HUDF46 has been caught before this epoch of widespread quenching, potentially prior the conditions necessary to quench its population, despite its halo mass. Indeed, the system is dominated by star-forming galaxies: members lie on the star-forming main sequence, largely lack signs of post-starburst activity, and are predominantly disk-like/disturbed in morphology. With the exception of the most massive central galaxy, there is no indication of widespread AGN activity, despite some evidence that AGN fractions rise sharply with redshift in dense environments (e.g., \citealt{galametz_cosmic_2009, alberts_star_2016}). The star-forming galaxies in HUDF46 are not starbursts (with the exception of one member), but instead occupy the main sequence, similarly to field galaxies at similar redshifts. Indeed, its mass-normalized cluster SFR is $\Sigma_{\rm SFR}=222_{-22}^{+33}\,M_{\odot}\,\mathrm{yr}^{-1}$, which corresponds to $\Sigma_{\rm SFR}/M_{\rm cl}=187_{-24}^{+32}\times10^{-14}\,\mathrm{yr}^{-1}$ (see Figure~\ref{fig:sigma_sfr_vs_z}); this value is consistent with theoretical expectations for protoclusters at $z\approx1.5\text{--}2$ (e.g., \citealt{lim_is_2021}) and there is no clear evidence for a systematic offset from the field, consistent with the global relation of \citet{behroozi_average_2013}.

Interestingly, recent works have shown, however, that merger-driven activity in overdensities can elevate star formation {\it along} the main sequence without triggering a starburst phase (e.g., \citealt{andrews_galaxy_2025}), a scenario that is potentially consistent with the high incidence of disturbed morphologies observed among HUDF46 members, particularly at $M_{\ast}\gtrsim10^{9}\,M_{\odot}$.  
More detailed, spatially resolved analysis is needed, however, to determine if the morphologies of HUDF46 member galaxies signal the beginnings of environmentally-driven quenching.

\subsection{Low mass galaxies in overdense environments}

Owing to observational limitations, past studies of environmentally driven galaxy evolution have largely focused on high-mass galaxies. This complicates interpretation at high redshift, as massive galaxies are rapidly quenched by secular processes, making it difficult to disentangle environmental effects.  Mass and environmental quenching have been found to be largely separable down to $\log_{10}( M_\bigstar/M_{\odot} \approx 9$) and out to $z \approx 1$ \citep{peng_mass_2010, peng_mass_2012}; however, at $z>1$ the quenching of massive cluster galaxies becomes strongly mass dependent (e.g., \citealt{balogh_evidence_2016, kawinwanichakij_effect_2017, papovich_effects_2018, pintos-castro_evolution_2019, van_der_burg_gogreen_2020}).  Dwarf galaxies, by contrast, may be intrinsically sensitive to environment. A survey of local dwarfs ($\log_{10} (M_\bigstar/M_{\odot})<9$) found that the quenched fraction in the field is extremely low, implying that nearly all quenching in dwarf galaxies occurs in overdense environments \citep{geha_stellar_2012}. This would make dwarfs the perfect laboratory for isolating environmental quenching mechanisms. Support for this picture comes from spatially resolved studies reporting outside-in quenching in dwarf galaxies \citep{bluck_how_2020}, although it is challenged by observational evidence for strong winds and other signatures of secular feedback \citep{mcquinn_galactic_2019}. 

JWST has opened a new window by pushing the stellar-mass frontier to much lower masses at all redshifts (e.g., \citealt{rinaldi_emergence_2024}), enabling comprehensive studies in both typical and overdense environments out to high redshift. In this work, we have used ultra-deep NIRISS spectroscopy to assemble a highly complete cluster membership spanning nearly three dex in stellar mass ($10^{7.5\text{--}10.5}\,M_{\odot}$). Our analysis shows that essentially all confirmed members are star-forming, with a small subset potentially caught in transition toward a post-starburst phase based on their UVJ classification. We note, however, that despite pushing the membership census to very low stellar masses, the sample remains spectroscopically incomplete for passive systems below $\log_{10}(M_{\bigstar}/M_{\odot}) \lesssim 9.5$. The population spans a wide range of Balmer-break strengths, comparable to those observed in the field, and likely reflecting bursty star-formation histories (e.g., \citealt{navarro-carrera_constraints_2024, simmonds_bursting_2025}). Unlike the field, however, living in the hot halo regime (Figure~\ref{fig:halo_redshift}) is expected to begin the process of starvation; a delay period is expected after infall with gas stripping becoming effective at high ICM density (\citealt{boselli_origin_2008, roberts_quenching_2019}).  Both cluster and local environment studies find a sharp increase in PSB and quenched low-mass galaxies at $z<1.5$ (\citealt{kawinwanichakij_effect_2017, mcnab_gogreen_2021}, ), which again is potentially related to reaching a threshold in the density of the ICM.

In addition to being highly susceptible to gas stripping due to their shallow potential wells (\citealt{boselli_virgo_2020}), low mass galaxies in overdense environments may also more strongly feel the effects of interactions and fly-bys.  This again suggests an environmental role in the disturbed/irregular morphologies we observe for the HUDF46 members.  Moreover, cluster members broadly follow the size–mass relation at similar redshift, with the exception of three systems ($\log_{10}(M_{\bigstar}/M_{\odot})>9$) that lie along its lower envelope, possibly indicating outside-in quenching and fading of their outer disks (e.g., \citealt{ito_dynamical_2026}). As we go to lower and lower mass, however, identifying signs of interactions and spatially resolved clues like color gradients become increasingly difficult.  Statistical samples of both star-forming and quenched dwarf galaxies in the extreme environments of clusters will be needed to identify trends versus the field.  This input will be invaluable for dwarf galaxy models and simulations, which tend to overpredict environmental quenching at low mass (e.g., \citealt{baker_double_2025}).

\subsection{Summary}

In this paper, we use ultra-deep JWST imaging and slitless spectroscopy to reassess the nature of the HUDF46 overdensity at $z \approx 1.84$. Exploiting NIRISS/WFSS spectroscopy together with NIRCam and MIRI data, we construct a highly complete spectroscopic census down to low stellar masses ($10^{7.5}\,M_{\odot}$), almost doubling the confirmed membership relative to earlier HST-based studies (\citealt{mei_star-forming_2015}). We derive updated dynamical properties, and characterize the stellar, structural, and star-forming properties of its galaxy population in comparison with the field. This places HUDF46 in a critical transitional phase between proto-clusters and mature clusters, prior to the onset of widespread environmental quenching. We summarize our finding below.

\vspace{5mm}
\begin{itemize}
    \item New JWST data expand the HUDF46 census from 18 to 37 confirmed members (excluding two outside NGDEEP and one low-$z$ interloper), doubling the sample relative to \citet{mei_star-forming_2015} while preserving a high overdensity significance of $17\pm1\sigma$.

    \item The expanded census allows us to revise the HUDF46 properties: we measure $\sigma=670\pm91\,\mathrm{km\,s^{-1}}$, $M_{200}=(1.2\pm0.2)\times10^{14}\,M_{\odot}$, and $R_{200}=0.61\pm0.1\,\mathrm{Mpc}$, all consistent with \citet{mei_star-forming_2015} within uncertainties; no extended ICM emission is detected with the new {\it Chandra} 7~Ms data, which however set an upper limit ($3\sigma$) of $\approx(1.1$–$1.8)\times10^{14}\,M_{\odot}$, in line with expectations at Cosmic Noon.

    \item The projected phase–space distribution indicates that HUDF46 hosts a compact, gravitationally bound core dominated by already accreted galaxies. Objects beyond the NGDEEP footprint populate both the accreted and infalling regions of the phase-space diagram, consistent with ongoing assembly within the cluster-scale radii probed by the NIRISS data. We note, however, that our coverage does not extend beyond this scale, preventing a direct assessment of any larger surrounding structure.

    \item Using {\sc bagpipes}, we derive stellar properties for HUDF46 and field galaxies and find no significant differences: HUDF46 spans $M_\bigstar\approx10^{7.5}\text{--}10^{10.5}\,M_\odot$, with 68\% below $10^{9}\,M_\odot$ (a major gain over HST), shows $A_V\approx0\text{--}1.5$ mag and mostly sub-solar metallicities, and hosts at most one AGN.

    \item The UVJ diagram places all confirmed HUDF46 members in the star-forming region, with only a small fraction formally overlapping with the modified selection of \citet{belli_mosfire_2019}  to recover recently quenched, lower-mass systems, despite exhibiting clear emission lines. This classification is independently supported by their location on the SFR–$M_{\bigstar}$ and sSFR–$M_{\bigstar}$ main sequence. Moreover, neither SFR nor sSFR shows a systematic dependence on projected phase space, in contrast to the well-established trends observed in more mature clusters at lower redshift, indicating that environmental quenching has not yet developed in this young system.

    \item Revisiting morphologies with ultra-deep NIRCam, we find that the majority of HUDF46 members exhibit disky and/or disturbed morphologies—particularly at high stellar masses—while only a small fraction appears compact. The eight star-forming ETGs previously identified by \citet{mei_star-forming_2015} (with HST) instead show low bulge-to-total ratios ($\lesssim0.6$; \citealt{genin_dawn_2025}), consistent with disk-dominated or transitional systems (\citealt{dimauro_coincidence_2022}). This indicates that earlier ETG classifications from \citet{mei_star-forming_2015} were largely driven by resolution and depth limitations.

    \item We inspected the rest-frame near-IR sizes of HUDF46 and field galaxies using F444W. HUDF46 members follow a similar size–mass relation as field galaxies and are in broad agreement with the expected evolution of the relation for star-forming galaxies at similar epochs, indicating that environmental effects have not yet produced measurable structural differences at $z \approx 1.84$. Galaxy sizes show no systematic variation with respect to their inferred accretion history. However, the brightest member, which hosts the only AGN candidate, has $R_{\rm eff} \approx 5$ kpc and lies in the region associated with very ancient infall, making it a mild outlier in phase space relative to the rest of the sample.

    \item The Balmer/4000\,\AA\, index exhibits a clear stellar-mass dependence but shows no systematic difference between HUDF46 members and field galaxies, nor any significant trend with projected phase--space location, further indicating that environmental effects on stellar populations are not yet established in this young cluster.

\end{itemize}

This work highlights the power of deep JWST surveys to uncover structures that fall outside standard selection frameworks. At $z\gtrsim1.5\text{--}2$, clusters have traditionally been identified and studied through their massive galaxy populations and/or clear signatures of maturity, such as a detectable ICM, which naturally biases the census toward already evolved systems. By contrast, studies of galaxy clusters at low redshift probe only the final stages of structure formation, after the dominant physical processes have largely concluded. The progenitors of these systems at earlier epochs must therefore exist in a much less evolved, {\it youthful} state; these transitional clusters can now be more readily identified and characterized through JWST’s deep near-infrared spectroscopy. In this context, HUDF46 may represent the missing link, providing a rare view of the early phases of structure assembly that eventually give rise to the massive clusters observed at much lower redshifts. In future work, we will conduct a more in-depth analysis of its population.

\acknowledgments
The authors thank Cristian Vignali for useful discussions.

This work is based on observations made with the NASA/ESA/CSA JWST. The data were obtained from the Mikulski Archive for Space Telescopes at the Space Telescope Science Institute, which is operated by the Association of Universities for Research in Astronomy, Inc., under NASA contract NAS 5-03127 for JWST. These observations are associated with JWST programs GTO \#1180, GTO \#1207, GO \#1210, GO \#1963, GO \#1895, GO \#2079, and \# 3215. The authors acknowledge the FRESCO, JEMS, and \# 3215 teams led by coPIs P. Oesch, C. C. Williams, M. Maseda, D. Eisenstein, R. Maiolino, S. Finkelstein, C. Papovich, and N. Pirzkal for developing their observing program with a zero-exclusive-access period. 

Processing for the JADES NIRCam data release was performed on the lux cluster at the University of California, Santa Cruz, funded by NSF MRI grant AST 1828315. Also based on observations made with the NASA/ESA Hubble Space Telescope obtained from the Space Telescope Science Institute, which is operated by the Association of Universities for Research in Astronomy, Inc., under NASA contract NAS 526555.

PR, CNAW, ZY and YZ  acknowledge support from JWST/NIRCam contract to the University of Arizona NAS5-02105.

PR and SA acknowledges JWST program \#3432.

SA acknowledges support from the JWST Mid-Infrared Instrument (MIRI) Science Team Lead, grant 80NSSC18K0555, from NASA Goddard Space Flight Center to the University of Arizona. 

WMB gratefully acknowledges support from DARK via the DARK fellowship. This work was supported by a research grant (VIL54489) from VILLUM FONDEN. GN acknowledges support by the Canadian Space Agency under a contract with NRC Herzberg Astronomy and Astrophysics. 

AJB acknowledges funding from the "FirstGalaxies" Advanced Grant from the European Research Council (ERC) under the European Union’s Horizon 2020 research and innovation programme (Grant agreement No. 789056). 

CC acknowledges support from the JWST/NIRCam Science Team contract to the University of Arizona, NAS5-02105, and JWST Program 3215. JWST/NIRCam contract to the University of Arizona, NAS5-02105. 

ST acknowledges support by the Royal Society Research Grant G125142. 

CJEG and LB acknowledge financial support from the Inter-University Institute for Data Intensive Astronomy (IDIA), a partnership of the University of Cape Town, the University of Pretoria and the University of the Western Cape, and Innovation’s National Research Foundation under the ISARP RADIOMAP+ Joint Research Scheme (DSI-NRF Grant Number 150551).

LB acknowledges the financial support of the South African Department of Science and the CPRR HIPPO Project (DSI-NRF Grant Number SRUG22031677). 

CJEG acknowledges the financial assistance of the South African Radio Astronomy Observatory (SARAO) (www.sarao.ac.za), and the CPRR Projects (DSI-NRF Grant Number SRUG2204254729).

\appendix

\section{NIRISS data reduction}\label{appendix_niriss_data_reduction}

We begin from the {\tt *\_uncal.fits} exposures and run the standard {\tt stage 1} of the {\tt jwst pipeline} to produce {\tt *\_rate.fits} products. On these {\tt *\_rate.fits} files, we apply an additional custom striping mitigation (similar to what it is usually done with NIRCam; \citealt{bagley_next_2024}), generating corrected rate products that are used as input to subsequent processing. During {\tt stage 1}, we also include an extra {\tt column\_jump} step\footnote{Following the algorithm presented here: \url{https://github.com/chriswillott/jwst}} applied after dark current subtraction and before jump detection. This correction mitigates random column jumps (typically $\approx50$ per $N_g=100$ NISRAPID ramp) that would otherwise increase the noise. Here, “columns” refer to detector coordinates (rows in the DMS orientation), distinct from large-scale 1/f stripes.

Starting from the corrected {\tt *\_rate.fits}, we run a dedicated {\tt stage 2} preparation to generate {\tt *\_cal.fits} products for imaging and {\it pseudo}-{\tt *\_cal.fits} products for WFSS. In the latter case, we use a modified version of the {\sc grizli} initialization routines to apply the WCS, perform flat-fielding and WFSS background corrections, and ensure that all required {\sc grizli} header keywords are present for downstream spectral extraction.

The resulting {\tt *\_(pseudo)cal.fits} files are used to produce mosaics for both imaging and WFSS. These mosaics serve as a baseline to construct an initial source mask, generated using {\sc Source Extractor} \citep{bertin_sextractor_1996} and later refined using {\sc NoiseChisel} \citep{akhlaghi_noise-based_2017}, which is particularly effective at detecting faint and extended emission. This mask is reprojected onto each individual exposure and used to refine background subtraction and large-scale corrections.

For imaging, the exposures are aligned to the JADES DR2 astrometric reference catalog using a modified version of the {\tt tweakreg\_step} within the {\tt jwst} pipeline. We then run the standard {\tt outlier\_step} and {\tt resampling\_step} to produce the final mosaics, iterating the masking and background subtraction to minimize residual large-scale structures. The estimated $5\sigma$ depths (measured in 0.2\arcsec\ radius apertures) are 29.4 mag in F115W, 28.7 mag in F150W, and 28.7 mag in F200W, corresponding to $\approx0.4$–$0.5$ mag deeper than ETC predictions.
The final, background-subtracted WFSS exposures are converted into a {\sc grizli}-compatible {\tt *\_rate.fits} format, including a cleanup step that replaces NaNs with zeros, as expected by {\sc grizli}. At this stage, the imaging reduction is complete, and the WFSS products are ready for spectral extraction and analysis. Because the imaging and WFSS data have already been pre-processed with our custom NIRISS pipeline, we bypass {\sc grizli}’s default pre-processing and use {\sc grizli} directly to (i) register grism exposures to the aligned imaging, (ii) model contamination from neighboring sources, (iii) extract 2D and 1D spectra, and (iv) fit continua and emission lines. We refer the reader to \citet{noirot_first_2023} for more details (Section 3).

\section{Significance of the overdensities in HUDF}\label{appendix1}

As described in \citet{mei_star-forming_2015} (Section 4.1), the redshift range we focused on this work ($z \approx1.5-2.5$) spans different rest-frame emission lines entering and exiting the grism+filter coverage. Specifically, at $z \lesssim 1.75$, spectra typically contain H$\alpha$ $\lambda$6563, [O\,\textsc{iii}]$\lambda\lambda$4959,5007, and H$\beta$ $\lambda$4861. By $z \approx 2.3$, however, H$\alpha$ is redshifted to the red edge of the GR150+F200W range, where the throughput declines, reducing sensitivity. As a result, detections at higher redshift rely increasingly on [O\,\textsc{iii}] and H$\beta$, with a higher effective line-flux limit due to both weaker lines and reduced instrument sensitivity\footnote{We note that [O\,\textsc{iii}] $\lambda\lambda$4959,5007 is not necessarily weaker than H$\alpha$ $\lambda$6563, especially at high-z (e.g., Fig. 6 in \citealt{noirot_hst_2018})}. As noted by \citet{mei_star-forming_2015}, this leads to systematically lower background counts at high redshift, rendering our significance estimates at the high-$z$ end conservative.

For each overdensity, we compute projected galaxy densities using the Nth-nearest neighbor method, with $\Sigma_N = N / (\pi D_N^2)$, where $N$ is the number of neighbors and $D_{N}$ is the projected distance in Mpc to the Nth nearest neighbor (e.g., \citealt{dressler_galaxy_1980}). We evaluate $\Sigma_N$ in redshift slices of $\Delta z = 0.02$ between $z=1.5$ and $z=2.5$, which balances sensitivity to real structures with minimal smoothing. Following \citet{mei_star-forming_2015}, the overdensity significance is defined as $\mathrm{SNR} = (\Sigma_N - \Sigma_N^{\mathrm{bkg}})/\sigma_N^{\mathrm{bkg}}$, where $\Sigma_N^{\mathrm{bkg}}$ and $\sigma_N^{\mathrm{bkg}}$ are the mean and standard deviation of the field distribution at a given redshift.

\section{Updated X-ray measurement from Chandra 7MS}\label{x-ray-chandra}

To determine an upper limit on the X-ray luminosity of the ICM, we used the all available {\it Chandra} data covering the cluster, with a total exposure time of $\approx$ 7Ms. We re-reduced and extracted the flux from all the {\it Chandra} observations using the Chandra Interactive Analysis of Observations (CIAO; v4.17 and CALDB 4.11.0; \citealt{fruscione_ciao_2006}) tools \texttt{chandra\_repro} and \texttt{srcflux}. We extracted the source counts from a $\approx$ 17\arcsec\, radius circular region within the previously-determined $R_{200}$ of the cluster, in between the known X-ray sources. The background counts were extracted from a nearby source-free 35\arcsec\, radius region. Assuming a Galactically-absorbed blackbody model for the ICM emission, with temperatures ranging between $T=1-3$\,keV, we found a $3\sigma$ upper luminosity limit of  $1.3-2.8 \times 10^{43}\,\rm erg\, s^{-1}$, depending on the temperature assumed. Finally, we adopted the cluster mass-luminosity relation from \citet{rykoff_lx-m_2008} to retrieve $M_{200}$.

\newpage

\bibliography{references}{}

@article{alberts_evolution_2014,
	title = {The evolution of dust-obscured star formation activity in galaxy clusters relative to the field over the last 9 billion years},
	volume = {437},
	issn = {0035-8711},
	url = {https://ui.adsabs.harvard.edu/abs/2014MNRAS.437..437A/abstract},
	doi = {10.1093/mnras/stt1897},
	language = {en},
	number = {1},
	urldate = {2026-02-25},
	journal = {Monthly Notices of the Royal Astronomical Society},
	author = {Alberts, Stacey and Pope, Alexandra and Brodwin, Mark and Atlee, David W. and Lin, Yen-Ting and Dey, Arjun and Eisenhardt, Peter R. M. and Gettings, Daniel P. and Gonzalez, Anthony H. and Jannuzi, Buell T. and Mancone, Conor L. and Moustakas, John and Snyder, Gregory F. and Stanford, S. Adam and Stern, Daniel and Weiner, Benjamin J. and Zeimann, Gregory R.},
	month = jan,
	year = {2014},
	pages = {437--457},
}

@article{shah_enhanced_2025,
	title = {Enhanced active galactic nucleus activity in overdense galactic environments at 2 \&lt; z \&lt; 4},
	volume = {704},
	issn = {0004-6361},
	url = {https://ui.adsabs.harvard.edu/abs/2025A&A...704A.101S/abstract},
	doi = {10.1051/0004-6361/202452055},
	language = {en},
	urldate = {2026-02-24},
	journal = {Astronomy and Astrophysics},
	author = {Shah, Ekta A. and Lemaux, Brian C. and Forrest, Ben and Hathi, Nimish and Shen, Lu and Cucciati, Olga and Hung, Denise and Giddings, Finn and Sikorski, Derek and Lubin, Lori and Gal, Roy R. and Zamorani, Giovanni and Golden-Marx, Emmet and Bardelli, Sandro and Cassarà, Letizia Pasqua and Garilli, Bianca and Gururajan, Gayathri and Suh, Hyewon and Vergani, Daniela and Zucca, Elena},
	month = dec,
	year = {2025},
	pages = {A101},
}

@article{martini_cluster_2013,
	title = {The {Cluster} and {Field} {Galaxy} {Active} {Galactic} {Nucleus} {Fraction} at z = 1-1.5: {Evidence} for a {Reversal} of the {Local} {Anticorrelation} between {Environment} and {AGN} {Fraction}},
	volume = {768},
	issn = {0004-637X},
	shorttitle = {The {Cluster} and {Field} {Galaxy} {Active} {Galactic} {Nucleus} {Fraction} at z = 1-1.5},
	url = {https://ui.adsabs.harvard.edu/abs/2013ApJ...768....1M/abstract},
	doi = {10.1088/0004-637X/768/1/1},
	language = {en},
	number = {1},
	urldate = {2026-02-24},
	journal = {The Astrophysical Journal},
	author = {Martini, Paul and Miller, E. D. and Brodwin, M. and Stanford, S. A. and Gonzalez, Anthony H. and Bautz, M. and Hickox, R. C. and Stern, D. and Eisenhardt, P. R. and Galametz, A. and Norman, D. and Jannuzi, B. T. and Dey, A. and Murray, S. and Jones, C. and Brown, M. J. I.},
	month = may,
	year = {2013},
	pages = {1},
}

@article{haggar_three_2020,
	title = {The {Three} {Hundred} project: backsplash galaxies in simulations of clusters},
	volume = {492},
	issn = {0035-8711},
	shorttitle = {The {Three} {Hundred} project},
	url = {https://ui.adsabs.harvard.edu/abs/2020MNRAS.492.6074H/abstract},
	doi = {10.1093/mnras/staa273},
	language = {en},
	number = {4},
	urldate = {2026-02-24},
	journal = {Monthly Notices of the Royal Astronomical Society},
	author = {Haggar, Roan and Gray, Meghan E. and Pearce, Frazer R. and Knebe, Alexander and Cui, Weiguang and Mostoghiu, Robert and Yepes, Gustavo},
	month = mar,
	year = {2020},
	pages = {6074--6085},
}

@article{tacchella_confinement_2016,
	title = {The confinement of star-forming galaxies into a main sequence through episodes of gas compaction, depletion and replenishment},
	volume = {457},
	issn = {0035-8711},
	url = {https://ui.adsabs.harvard.edu/abs/2016MNRAS.457.2790T/abstract},
	doi = {10.1093/mnras/stw131},
	language = {en},
	number = {3},
	urldate = {2026-02-20},
	journal = {Monthly Notices of the Royal Astronomical Society},
	author = {Tacchella, Sandro and Dekel, Avishai and Carollo, C. Marcella and Ceverino, Daniel and DeGraf, Colin and Lapiner, Sharon and Mandelker, Nir and Primack Joel, R.},
	month = apr,
	year = {2016},
	pages = {2790--2813},
}

@article{cutler_structure_2025,
	title = {The {Structure} and {Formation} {Histories} of {Low}-mass {Quiescent} {Galaxies} in the {A2744} {Cluster} {Environment}},
	volume = {993},
	issn = {0004-637X},
	url = {https://ui.adsabs.harvard.edu/abs/2025ApJ...993..169C/abstract},
	doi = {10.3847/1538-4357/ae0629},
	language = {en},
	number = {2},
	urldate = {2026-02-20},
	journal = {The Astrophysical Journal},
	author = {Cutler, Sam E. and Weaver, John R. and Whitaker, Katherine E. and Greene, Jenny E. and Setton, David J. and Webb, Zach J. and Abdullah, Ayesha and Medrano, Aubrey and Bezanson, Rachel and Brammer, Gabriel and Feldmann, Robert and Furtak, Lukas J. and Glazebrook, Karl and Labbe, Ivo and Leja, Joel and Marchesini, Danilo and Miller, Tim B. and Mitsuhashi, Ikki and Nanayakkara, Themiya and Nelson, Erica J. and Pan, Richard and Price, Sedona H. and Suess, Katherine A. and Wang, Bingjie},
	month = nov,
	year = {2025},
	pages = {169},
}

@article{butcher_evolution_1978,
	title = {The evolution of galaxies in clusters. {I}. {ISIT} photometry of {Cl} 0024+1654 and {3C} 295.},
	volume = {219},
	issn = {0004-637X},
	url = {https://ui.adsabs.harvard.edu/abs/1978ApJ...219...18B/abstract},
	doi = {10.1086/155751},
	language = {en},
	urldate = {2026-02-15},
	journal = {The Astrophysical Journal},
	author = {Butcher, H. and Oemler, A.},
	month = jan,
	year = {1978},
	pages = {18--30},
}

@article{galametz_cosmic_2009,
	title = {The {Cosmic} {Evolution} of {Active} {Galactic} {Nuclei} in {Galaxy} {Clusters}},
	volume = {694},
	issn = {0004-637X},
	url = {https://ui.adsabs.harvard.edu/abs/2009ApJ...694.1309G/abstract},
	doi = {10.1088/0004-637X/694/2/1309},
	language = {en},
	number = {2},
	urldate = {2026-02-15},
	journal = {The Astrophysical Journal},
	author = {Galametz, Audrey and Stern, Daniel and Eisenhardt, Peter R. M. and Brodwin, Mark and Brown, Michael J. I. and Dey, Arjun and Gonzalez, Anthony H. and Jannuzi, Buell T. and Moustakas, Leonidas A. and Stanford, S. Adam},
	month = apr,
	year = {2009},
	pages = {1309--1316},
}

@article{roberts_lotss_2022,
	title = {{LoTSS} jellyfish galaxies. {III}. {The} first identification of jellyfish galaxies in the {Perseus} cluster},
	volume = {658},
	issn = {0004-6361},
	url = {https://ui.adsabs.harvard.edu/abs/2022A&A...658A..44R/abstract},
	doi = {10.1051/0004-6361/202142294},
	language = {en},
	urldate = {2026-02-15},
	journal = {Astronomy and Astrophysics},
	author = {Roberts, I. D. and van Weeren, R. J. and Timmerman, R. and Botteon, A. and Gendron-Marsolais, M. and Ignesti, A. and Rottgering, H. J. A.},
	month = feb,
	year = {2022},
	pages = {A44},
}

@article{roberts_quenching_2019,
	title = {Quenching {Low}-mass {Satellite} {Galaxies}: {Evidence} for a {Threshold} {ICM} {Density}},
	volume = {873},
	issn = {0004-637X},
	shorttitle = {Quenching {Low}-mass {Satellite} {Galaxies}},
	url = {https://ui.adsabs.harvard.edu/abs/2019ApJ...873...42R/abstract},
	doi = {10.3847/1538-4357/ab04f7},
	language = {en},
	number = {1},
	urldate = {2026-02-15},
	journal = {The Astrophysical Journal},
	author = {Roberts, Ian D. and Parker, Laura C. and Brown, Toby and Joshi, Gandhali D. and Hlavacek-Larrondo, Julie and Wadsley, James},
	month = mar,
	year = {2019},
	pages = {42},
}

@article{merida_probing_2025,
	title = {Probing the {Star} {Formation} {Main} {Sequence} down to 10\${\textasciicircum}\{7\} {M}\_{\textbackslash}odot\$ at \$1 \&lt; z \&lt; 9\$},
	url = {https://ui.adsabs.harvard.edu/abs/2025arXiv250922871M/abstract},
	doi = {10.48550/arXiv.2509.22871},
	language = {en},
	urldate = {2026-02-15},
	journal = {arXiv e-prints},
	author = {Mérida, Rosa M. and Sawicki, Marcin and Iyer, Kartheik G. and Noirot, Gaël and Willott, Chris J. and Bradač, Maruša and Desprez, Guillaume and Martis, Nicholas S. and Muzzin, Adam and Rihtaršič, Gregor and Sarrouh, Ghassan T. E. and Favaro, Jeremy and Gaspar, Gaia and Harshan, Anishya and Judež, Jon},
	month = sep,
	year = {2025},
	pages = {arXiv:2509.22871},
}

@article{hashiguchi_agn_2023,
	title = {{AGN} number fraction in galaxy groups and clusters at z \&lt; 1.4 from the {Subaru} {Hyper} {Suprime}-{Cam} survey},
	volume = {75},
	issn = {0004-6264},
	url = {https://ui.adsabs.harvard.edu/abs/2023PASJ...75.1246H/abstract},
	doi = {10.1093/pasj/psad066},
	language = {en},
	number = {6},
	urldate = {2026-02-13},
	journal = {Publications of the Astronomical Society of Japan},
	author = {Hashiguchi, Aoi and Toba, Yoshiki and Ota, Naomi and Oguri, Masamune and Okabe, Nobuhiro and Ueda, Yoshihiro and Imanishi, Masatoshi and Yamada, Satoshi and Goto, Tomotsugu and Koyama, Shuhei and Lee, Kianhong and Mitsuishi, Ikuyuki and Nagao, Tohru and Nishizawa, Atsushi J. and Noboriguchi, Akatoki and Oogi, Taira and Sakuta, Koki and Schramm, Malte and Shibata, Mio and Terashima, Yuichi and Yamashita, Takuji and Yanagawa, Anri and Yoshimoto, Anje},
	month = dec,
	year = {2023},
	pages = {1246--1261},
}

@article{noirot_first_2023,
	title = {The first large catalogue of spectroscopic redshifts in {Webb}'s first deep field, {SMACS} {J0723}.3-7327},
	volume = {525},
	issn = {0035-8711},
	url = {https://ui.adsabs.harvard.edu/abs/2023MNRAS.525.1867N/abstract},
	doi = {10.1093/mnras/stad1019},
	language = {en},
	number = {2},
	urldate = {2026-02-13},
	journal = {Monthly Notices of the Royal Astronomical Society},
	author = {Noirot, Gaël and Desprez, Guillaume and Asada, Yoshihisa and Sawicki, Marcin and Estrada-Carpenter, Vicente and Martis, Nicholas and Sarrouh, Ghassan and Strait, Victoria and Abraham, Roberto and Bradač, Maruša and Brammer, Gabriel and Iyer, Kartheik and MacFarland, Shannon and Matharu, Jasleen and Mowla, Lamiya and Muzzin, Adam and Pacifici, Camilla and Ravindranath, Swara and Willott, Chris J. and Albert, Loïc and Doyon, René and Hutchings, John B. and Rowlands, Neil},
	month = oct,
	year = {2023},
	pages = {1867--1884},
}

@article{willott_near-infrared_2022,
	title = {The {Near}-infrared {Imager} and {Slitless} {Spectrograph} for the {James} {Webb} {Space} {Telescope}. {II}. {Wide} {Field} {Slitless} {Spectroscopy}},
	volume = {134},
	issn = {0004-6280},
	url = {https://ui.adsabs.harvard.edu/abs/2022PASP..134b5002W/abstract},
	doi = {10.1088/1538-3873/ac5158},
	language = {en},
	number = {1032},
	urldate = {2026-02-13},
	journal = {Publications of the Astronomical Society of the Pacific},
	author = {Willott, Chris J. and Doyon, René and Albert, Loic and Brammer, Gabriel B. and Dixon, William V. and Muzic, Koraljka and Ravindranath, Swara and Scholz, Aleks and Abraham, Roberto and Artigau, Étienne and Bradač, Maruša and Goudfrooij, Paul and Hutchings, John B. and Iyer, Kartheik G. and Jayawardhana, Ray and LaMassa, Stephanie and Martis, Nicholas and Meyer, Michael R. and Morishita, Takahiro and Mowla, Lamiya and Muzzin, Adam and Noirot, Gaël and Pacifici, Camilla and Rowlands, Neil and Sarrouh, Ghassan and Sawicki, Marcin and Taylor, Joanna M. and Volk, Kevin and Zabl, Johannes},
	month = feb,
	year = {2022},
	pages = {025002},
}

@article{wylezalek_galaxy_2014,
	title = {The {Galaxy} {Cluster} {Mid}-infrared {Luminosity} {Function} at 1.3 \&lt; z \&lt; 3.2},
	volume = {786},
	issn = {0004-637X},
	url = {https://ui.adsabs.harvard.edu/abs/2014ApJ...786...17W/abstract},
	doi = {10.1088/0004-637X/786/1/17},
	language = {en},
	number = {1},
	urldate = {2026-02-13},
	journal = {The Astrophysical Journal},
	author = {Wylezalek, Dominika and Vernet, Joël and De Breuck, Carlos and Stern, Daniel and Brodwin, Mark and Galametz, Audrey and Gonzalez, Anthony H. and Jarvis, Matt and Hatch, Nina and Seymour, Nick and Stanford, Spencer A.},
	month = may,
	year = {2014},
	pages = {17},
}

@article{li_epochs_2025,
	title = {{EPOCHS} paper – {X}. {Environmental} effects on {Galaxy} formation and protocluster {Galaxy} candidates at 4.5 \&lt; z \&lt; 10 from {JWST} observations},
	volume = {539},
	issn = {0035-8711},
	url = {https://ui.adsabs.harvard.edu/abs/2025MNRAS.539.1796L/abstract},
	doi = {10.1093/mnras/staf543},
	language = {en},
	number = {2},
	urldate = {2026-02-13},
	journal = {Monthly Notices of the Royal Astronomical Society},
	author = {Li, Qiong and Conselice, Christopher J. and Sarron, Florian and Harvey, Thomas and Austin, Duncan and Adams, Nathan and Trussler, James A. A. and Duan, Qiao and Ferreira, Leonardo and Westcott, Lewi and Harris, Honor and Dole, Hervé and Grogin, Norman A. and Frye, Brenda and Koekemoer, Anton M. and Robertson, Clayton and Windhorst, Rogier A. and Polletta, Maria del Carmen and Hathi, Nimish P. and Jansen, Rolf A.},
	month = may,
	year = {2025},
	pages = {1796--1819},
}

@article{rykoff_lx-m_2008,
	title = {The {L}{\textless}{SUB}{\textgreater}{X}{\textless}/{SUB}{\textgreater}-{M} relation of clusters of galaxies},
	volume = {387},
	issn = {0035-8711},
	url = {https://ui.adsabs.harvard.edu/abs/2008MNRAS.387L..28R/abstract},
	doi = {10.1111/j.1745-3933.2008.00476.x},
	language = {en},
	number = {1},
	urldate = {2026-02-13},
	journal = {Monthly Notices of the Royal Astronomical Society},
	author = {Rykoff, E. S. and Evrard, A. E. and McKay, T. A. and Becker, M. R. and Johnston, D. E. and Koester, B. P. and Nord, B. and Rozo, E. and Sheldon, E. S. and Stanek, R. and Wechsler, R. H.},
	month = jun,
	year = {2008},
	pages = {L28--L32},
}

@article{ito_dynamical_2026,
	title = {Dynamical properties and star formation history of a low-mass quenched galaxy at {Cosmic} {Noon}},
	url = {https://ui.adsabs.harvard.edu/abs/2026arXiv260101722I/abstract},
	doi = {10.48550/arXiv.2601.01722},
	language = {en},
	urldate = {2026-02-13},
	journal = {arXiv e-prints},
	author = {Ito, K. and Valentino, F. and Baker, W. M. and Brammer, G. and Gottumukkala, R. and Kakimoto, T. and Lagos, C. D. P. and Onodera, M. and Pensabene, A. and Scarpe, G. and Tanaka, M. and Whitaker, K. E. and Reddy, N. A. and Sanders, R. L. and Shapley, A. E.},
	month = jan,
	year = {2026},
	pages = {arXiv:2601.01722},
}

@article{strazzullo_galaxy_2019,
	title = {Galaxy populations in the most distant {SPT}-{SZ} clusters. {I}. {Environmental} quenching in massive clusters at 1.4 ≲ z ≲ 1.7},
	volume = {622},
	issn = {0004-6361},
	url = {https://ui.adsabs.harvard.edu/abs/2019A&A...622A.117S/abstract},
	doi = {10.1051/0004-6361/201833944},
	language = {en},
	urldate = {2026-02-13},
	journal = {Astronomy and Astrophysics},
	author = {Strazzullo, V. and Pannella, M. and Mohr, J. J. and Saro, A. and Ashby, M. L. N. and Bayliss, M. B. and Bocquet, S. and Bulbul, E. and Khullar, G. and Mantz, A. B. and Stanford, S. A. and Benson, B. A. and Bleem, L. E. and Brodwin, M. and Canning, R. E. A. and Capasso, R. and Chiu, I. and Gonzalez, A. H. and Gupta, N. and Hlavacek-Larrondo, J. and Klein, M. and McDonald, M. and Noordeh, E. and Rapetti, D. and Reichardt, C. L. and Schrabback, T. and Sharon, K. and Stalder, B.},
	month = feb,
	year = {2019},
	pages = {A117},
}

@article{lee-brown_ages_2017,
	title = {The {Ages} of {Passive} {Galaxies} in a z = 1.62 {Protocluster}},
	volume = {844},
	issn = {0004-637X},
	url = {https://ui.adsabs.harvard.edu/abs/2017ApJ...844...43L/abstract},
	doi = {10.3847/1538-4357/aa7948},
	language = {en},
	number = {1},
	urldate = {2026-02-13},
	journal = {The Astrophysical Journal},
	author = {Lee-Brown, Donald B. and Rudnick, Gregory H. and Momcheva, Ivelina G. and Papovich, Casey and Lotz, Jennifer M. and Tran, Kim-Vy H. and Henke, Brittany and Willmer, Christopher N. A. and Brammer, Gabriel B. and Brodwin, Mark and Dunlop, James and Farrah, Duncan},
	month = jul,
	year = {2017},
	pages = {43},
}

@article{lu_galaxy_2016,
	title = {Galaxy {Groups} in the {2Mass} {Redshift} {Survey}},
	volume = {832},
	issn = {0004-637X},
	url = {https://ui.adsabs.harvard.edu/abs/2016ApJ...832...39L/abstract},
	doi = {10.3847/0004-637X/832/1/39},
	language = {en},
	number = {1},
	urldate = {2026-02-13},
	journal = {The Astrophysical Journal},
	author = {Lu, Yi and Yang, Xiaohu and Shi, Feng and Mo, H. J. and Tweed, Dylan and Wang, Huiyuan and Zhang, Youcai and Li, Shijie and Lim, S. H.},
	month = nov,
	year = {2016},
	pages = {39},
}

@article{webb_evolution_2013,
	title = {The {Evolution} of {Dusty} {Star} formation in {Galaxy} {Clusters} to z = 1: {Spitzer} {Infrared} {Observations} of the {First} {Red}-{Sequence} {Cluster} {Survey}},
	volume = {146},
	issn = {0004-6256},
	shorttitle = {The {Evolution} of {Dusty} {Star} formation in {Galaxy} {Clusters} to z = 1},
	url = {https://ui.adsabs.harvard.edu/abs/2013AJ....146...84W/abstract},
	doi = {10.1088/0004-6256/146/4/84},
	language = {en},
	number = {4},
	urldate = {2026-02-13},
	journal = {The Astronomical Journal},
	author = {Webb, T. M. A. and O'Donnell, D. and Yee, H. K. C. and Gilbank, David and Coppin, Kristen and Ellingson, Erica and Faloon, Ashley and Geach, James E. and Gladders, Mike and Noble, Allison and Muzzin, Adam and Wilson, Gillian and Yan, Renbin},
	month = oct,
	year = {2013},
	pages = {84},
}

@article{cowie_new_1996,
	title = {New {Insight} on {Galaxy} {Formation} and {Evolution} {From} {Keck} {Spectroscopy} of the {Hawaii} {Deep} {Fields}},
	volume = {112},
	issn = {0004-6256},
	url = {https://ui.adsabs.harvard.edu/abs/1996AJ....112..839C/abstract},
	doi = {10.1086/118058},
	language = {en},
	urldate = {2026-02-13},
	journal = {The Astronomical Journal},
	author = {Cowie, Lennox L. and Songaila, Antoinette and Hu, Esther M. and Cohen, J. G.},
	month = sep,
	year = {1996},
	pages = {839},
}

@article{poggianti_indicators_1997,
	title = {Indicators of star formation: 4000 Å break and {Balmer} lines.},
	volume = {325},
	issn = {0004-6361},
	shorttitle = {Indicators of star formation},
	url = {https://ui.adsabs.harvard.edu/abs/1997A&A...325.1025P/abstract},
	doi = {10.48550/arXiv.astro-ph/9703067},
	language = {en},
	urldate = {2026-02-13},
	journal = {Astronomy and Astrophysics},
	author = {Poggianti, B. M. and Barbaro, G.},
	month = sep,
	year = {1997},
	pages = {1025--1030},
}

@article{worthey_old_1994,
	title = {Old {Stellar} {Populations}. {V}. {Absorption} {Feature} {Indices} for the {Complete} {Lick}/{IDS} {Sample} of {Stars}},
	volume = {94},
	issn = {0067-0049},
	url = {https://ui.adsabs.harvard.edu/abs/1994ApJS...94..687W/abstract},
	doi = {10.1086/192087},
	language = {en},
	urldate = {2026-02-13},
	journal = {The Astrophysical Journal Supplement Series},
	author = {Worthey, Guy and Faber, S. M. and Gonzalez, J. Jesus and Burstein, D.},
	month = oct,
	year = {1994},
	pages = {687},
}

@article{bluck_how_2020,
	title = {How do central and satellite galaxies quench? - {Insights} from spatially resolved spectroscopy in the {MaNGA} survey},
	volume = {499},
	issn = {0035-8711},
	shorttitle = {How do central and satellite galaxies quench?},
	url = {https://ui.adsabs.harvard.edu/abs/2020MNRAS.499..230B/abstract},
	doi = {10.1093/mnras/staa2806},
	language = {en},
	number = {1},
	urldate = {2026-02-12},
	journal = {Monthly Notices of the Royal Astronomical Society},
	author = {Bluck, Asa F. L. and Maiolino, Roberto and Piotrowska, Joanna M. and Trussler, James and Ellison, Sara L. and Sánchez, Sebastian F. and Thorp, Mallory D. and Teimoorinia, Hossen and Moreno, Jorge and Conselice, Christopher J.},
	month = nov,
	year = {2020},
	pages = {230--268},
}

@article{kuhn_jwst_2024,
	title = {{JWST} {Reveals} a {Surprisingly} {High} {Fraction} of {Galaxies} {Being} {Spiral}-like at 0.5 ≤ z ≤ 4},
	volume = {968},
	issn = {0004-637X},
	url = {https://ui.adsabs.harvard.edu/abs/2024ApJ...968L..15K/abstract},
	doi = {10.3847/2041-8213/ad43eb},
	language = {en},
	number = {2},
	urldate = {2026-02-12},
	journal = {The Astrophysical Journal},
	author = {Kuhn, Vicki and Guo, Yicheng and Martin, Alec and Bayless, Julianna and Gates, Ellie and Puleo, A. J.},
	month = jun,
	year = {2024},
	pages = {L15},
}

@article{zhu_clump-like_2026,
	title = {Clump-like {Structures} in {High}-{Redshift} {Galaxies}: {Mass} {Scaling} and {Radial} {Trends} from {JADES}},
	shorttitle = {Clump-like {Structures} in {High}-{Redshift} {Galaxies}},
	url = {https://ui.adsabs.harvard.edu/abs/2026arXiv260115965Z/abstract},
	doi = {10.48550/arXiv.2601.15965},
	language = {en},
	urldate = {2026-02-12},
	journal = {arXiv e-prints},
	author = {Zhu, Yongda and Rieke, Marcia J. and Ji, Zhiyuan and Bunker, Andrew J. and Carreira, Courtney and Danhaive, A. Lola and Duan, Qiao and Egami, Eiichi and Eisenstein, Daniel J. and Hainline, Kevin and Johnson, Benjamin D. and Ma, Zheng and Puskás, Dávid and Rieke, George H. and Rinaldi, Pierluigi and Robertson, Brant and Tacchella, Sandro and Übler, Hannah and Villanueva, Natalia C. and Williams, Christina C. and Willmer, Christopher N. A. and Wu, Zihao and Zhang, Junyu},
	month = jan,
	year = {2026},
	pages = {arXiv:2601.15965},
}

@article{williams_alma_2022,
	title = {{ALMA} {Measures} {Molecular} {Gas} {Reservoirs} {Comparable} to {Field} {Galaxies} in a {Low}-mass {Galaxy} {Cluster} at z = 1.3},
	volume = {929},
	issn = {0004-637X},
	url = {https://ui.adsabs.harvard.edu/abs/2022ApJ...929...35W/abstract},
	doi = {10.3847/1538-4357/ac58fa},
	language = {en},
	number = {1},
	urldate = {2026-02-12},
	journal = {The Astrophysical Journal},
	author = {Williams, Christina C. and Alberts, Stacey and Spilker, Justin S. and Noble, Allison G. and Stefanon, Mauro and Willmer, Christopher N. A. and Bezanson, Rachel and Narayanan, Desika and Whitaker, Katherine E.},
	month = apr,
	year = {2022},
	pages = {35},
}

@article{nielsen_evidence_2025,
	title = {Evidence for multiple types of post-starburst galaxies},
	volume = {700},
	issn = {0004-6361},
	url = {https://ui.adsabs.harvard.edu/abs/2025A&A...700A.116N/abstract},
	doi = {10.1051/0004-6361/202554507},
	language = {en},
	urldate = {2026-02-11},
	journal = {Astronomy and Astrophysics},
	author = {Nielsen, Emma W. and Steinhardt, Charles L. and Harper, Mathieux and McPartland, Conor and Sedgewick, Aidan},
	month = aug,
	year = {2025},
	pages = {A116},
}

@article{antwi-danso_beyond_2023,
	title = {Beyond {UVJ}: {Color} {Selection} of {Galaxies} in the {JWST} {Era}},
	volume = {943},
	issn = {0004-637X},
	shorttitle = {Beyond {UVJ}},
	url = {https://ui.adsabs.harvard.edu/abs/2023ApJ...943..166A/abstract},
	doi = {10.3847/1538-4357/aca294},
	language = {en},
	number = {2},
	urldate = {2026-02-11},
	journal = {The Astrophysical Journal},
	author = {Antwi-Danso, Jacqueline and Papovich, Casey and Leja, Joel and Marchesini, Danilo and Marsan, Z. Cemile and Martis, Nicholas S. and Labbé, Ivo and Muzzin, Adam and Glazebrook, Karl and Straatman, Caroline M. S. and Tran, Kim-Vy H.},
	month = feb,
	year = {2023},
	pages = {166},
}

@article{williams_detection_2009,
	title = {Detection of {Quiescent} {Galaxies} in a {Bicolor} {Sequence} from {Z} = 0-2},
	volume = {691},
	issn = {0004-637X},
	url = {https://ui.adsabs.harvard.edu/abs/2009ApJ...691.1879W/abstract},
	doi = {10.1088/0004-637X/691/2/1879},
	language = {en},
	number = {2},
	urldate = {2026-02-11},
	journal = {The Astrophysical Journal},
	author = {Williams, Rik J. and Quadri, Ryan F. and Franx, Marijn and van Dokkum, Pieter and Labbé, Ivo},
	month = feb,
	year = {2009},
	pages = {1879--1895},
}

@article{pacifici_evolution_2016,
	title = {The {Evolution} of {Star} {Formation} {Histories} of {Quiescent} {Galaxies}},
	volume = {832},
	issn = {0004-637X},
	url = {https://ui.adsabs.harvard.edu/abs/2016ApJ...832...79P/abstract},
	doi = {10.3847/0004-637X/832/1/79},
	language = {en},
	number = {1},
	urldate = {2026-02-11},
	journal = {The Astrophysical Journal},
	author = {Pacifici, Camilla and Kassin, Susan A. and Weiner, Benjamin J. and Holden, Bradford and Gardner, Jonathan P. and Faber, Sandra M. and Ferguson, Henry C. and Koo, David C. and Primack, Joel R. and Bell, Eric F. and Dekel, Avishai and Gawiser, Eric and Giavalisco, Mauro and Rafelski, Marc and Simons, Raymond C. and Barro, Guillermo and Croton, Darren J. and Davé, Romeel and Fontana, Adriano and Grogin, Norman A. and Koekemoer, Anton M. and Lee, Seong-Kook and Salmon, Brett and Somerville, Rachel and Behroozi, Peter},
	month = nov,
	year = {2016},
	pages = {79},
}

@article{forrey_ngdeep_2025,
	title = {{NGDEEP}: {A} {New} {Non}-{Parametric} {Measure} of {Local} {Star}-{Formation} and {Attenuation} at {Cosmic} {Noon}},
	shorttitle = {{NGDEEP}},
	url = {https://ui.adsabs.harvard.edu/abs/2025arXiv251211989F/abstract},
	doi = {10.48550/arXiv.2512.11989},
	language = {en},
	urldate = {2026-02-11},
	journal = {arXiv e-prints},
	author = {Forrey, Grace M. and Simons, Raymond C. and Trump, Jonathan R. and Shen, Lu and Koekemoer, Anton M. and Bagley, Micaela B. and Finkelstein, Steven L. and Papovich, Casey and Pirzkal, Nor},
	month = dec,
	year = {2025},
	pages = {arXiv:2512.11989},
}

@article{witten_not_2025,
	title = {Not all protoclusters host evolved galaxies: {Evidence} for reduced environmental effects in a lower halo mass protocluster at \$z = 7.66\$},
	shorttitle = {Not all protoclusters host evolved galaxies},
	url = {https://ui.adsabs.harvard.edu/abs/2025arXiv251105647W/abstract},
	doi = {10.48550/arXiv.2511.05647},
	language = {en},
	urldate = {2026-02-10},
	journal = {arXiv e-prints},
	author = {Witten, Callum and Oesch, Pascal A. and Bennett, Jake S. and Meyer, Romain A. and Giovinazzo, Emma and Covelo-Paz, Alba and Baker, William M. and Ivey, Lucy R.},
	month = nov,
	year = {2025},
	pages = {arXiv:2511.05647},
}

@article{ma_soptics_2025,
	title = {{sOPTICS}: a modified density-based algorithm for identifying galaxy groups/clusters and brightest cluster galaxies},
	volume = {537},
	issn = {0035-8711},
	shorttitle = {{sOPTICS}},
	url = {https://ui.adsabs.harvard.edu/abs/2025MNRAS.537.1504M/abstract},
	doi = {10.1093/mnras/staf115},
	language = {en},
	number = {2},
	urldate = {2026-02-10},
	journal = {Monthly Notices of the Royal Astronomical Society},
	author = {Ma, Hai-Xia and Takeuchi, Tsutomu T. and Cooray, Suchetha and Zhu, Yongda},
	month = feb,
	year = {2025},
	pages = {1504--1517},
}

@article{lim_flamingo_2024,
	title = {The {FLAMINGO} simulation view of cluster progenitors observed in the epoch of reionization with {JWST}},
	volume = {532},
	issn = {0035-8711},
	url = {https://ui.adsabs.harvard.edu/abs/2024MNRAS.532.4551L/abstract},
	doi = {10.1093/mnras/stae1790},
	language = {en},
	number = {4},
	urldate = {2026-02-10},
	journal = {Monthly Notices of the Royal Astronomical Society},
	author = {Lim, Seunghwan and Tacchella, Sandro and Schaye, Joop and Schaller, Matthieu and Helton, Jakob M. and Kugel, Roi and Maiolino, Roberto},
	month = aug,
	year = {2024},
	pages = {4551--4569},
}

@article{courtin_imprints_2011,
	title = {Imprints of dark energy on cosmic structure formation - {II}. {Non}-universality of the halo mass function},
	volume = {410},
	issn = {0035-8711},
	url = {https://ui.adsabs.harvard.edu/abs/2011MNRAS.410.1911C/abstract},
	doi = {10.1111/j.1365-2966.2010.17573.x},
	language = {en},
	number = {3},
	urldate = {2026-02-08},
	journal = {Monthly Notices of the Royal Astronomical Society},
	author = {Courtin, J. and Rasera, Y. and Alimi, J.-M. and Corasaniti, P.-S. and Boucher, V. and Füzfa, A.},
	month = jan,
	year = {2011},
	pages = {1911--1931},
}

@article{wu_jades_2026,
	title = {{JADES}: {A} {Prominent} {Galaxy} {Overdensity} {Candidate} within the {First} 500 {Myr}},
	shorttitle = {{JADES}},
	url = {https://ui.adsabs.harvard.edu/abs/2026arXiv260115960W/abstract},
	doi = {10.48550/arXiv.2601.15960},
	language = {en},
	urldate = {2026-02-08},
	journal = {arXiv e-prints},
	author = {Wu, Zihao and Eisenstein, Daniel J. and Johnson, Benjamin D. and Hainline, Kevin and Baker, William M. and Bunker, Andrew J. and Cameron, Alex J. and Curtis-Lake, Emma and Danhaive, A. Lola and Hausen, Ryan and Helton, Jakob M. and Ji, Zhiyuan and Looser, Tobias J. and Maiolino, Roberto and Mengistu, Petra and Rinaldi, Pierluigi and Robertson, Brant E. and Sun, Fengwu and Tacchella, Sandro and Trussler, James A. A. and Williams, Christina C. and Willmer, Christopher N. A. and Witstok, Joris},
	month = jan,
	year = {2026},
	pages = {arXiv:2601.15960},
}

@article{mantz_deep_2020,
	title = {Deep {XMM}-{Newton} observations of the most distant {SPT}-{SZ} galaxy cluster},
	volume = {496},
	issn = {0035-8711},
	url = {https://ui.adsabs.harvard.edu/abs/2020MNRAS.496.1554M/abstract},
	doi = {10.1093/mnras/staa1581},
	language = {en},
	number = {2},
	urldate = {2026-02-08},
	journal = {Monthly Notices of the Royal Astronomical Society},
	author = {Mantz, Adam B. and Allen, Steven W. and Morris, R. Glenn and Canning, Rebecca E. A. and Bayliss, Matthew and Bleem, Lindsey E. and Floyd, Benjamin T. and McDonald, Michael},
	month = aug,
	year = {2020},
	pages = {1554--1564},
}

@article{foley_discovery_2011,
	title = {Discovery and {Cosmological} {Implications} of {SPT}-{CL} {J2106}-5844, the {Most} {Massive} {Known} {Cluster} at z\&gt;1},
	volume = {731},
	issn = {0004-637X},
	url = {https://ui.adsabs.harvard.edu/abs/2011ApJ...731...86F/abstract},
	doi = {10.1088/0004-637X/731/2/86},
	language = {en},
	number = {2},
	urldate = {2026-02-08},
	journal = {The Astrophysical Journal},
	author = {Foley, R. J. and Andersson, K. and Bazin, G. and de Haan, T. and Ruel, J. and Ade, P. a. R. and Aird, K. A. and Armstrong, R. and Ashby, M. L. N. and Bautz, M. and Benson, B. A. and Bleem, L. E. and Bonamente, M. and Brodwin, M. and Carlstrom, J. E. and Chang, C. L. and Clocchiatti, A. and Crawford, T. M. and Crites, A. T. and Desai, S. and Dobbs, M. A. and Dudley, J. P. and Fazio, G. G. and Forman, W. R. and Garmire, G. and George, E. M. and Gladders, M. D. and Gonzalez, A. H. and Halverson, N. W. and High, F. W. and Holder, G. P. and Holzapfel, W. L. and Hoover, S. and Hrubes, J. D. and Jones, C. and Joy, M. and Keisler, R. and Knox, L. and Lee, A. T. and Leitch, E. M. and Lueker, M. and Luong-Van, D. and Marrone, D. P. and McMahon, J. J. and Mehl, J. and Meyer, S. S. and Mohr, J. J. and Montroy, T. E. and Murray, S. S. and Padin, S. and Plagge, T. and Pryke, C. and Reichardt, C. L. and Rest, A. and Ruhl, J. E. and Saliwanchik, B. R. and Saro, A. and Schaffer, K. K. and Shaw, L. and Shirokoff, E. and Song, J. and Spieler, H. G. and Stalder, B. and Stanford, S. A. and Staniszewski, Z. and Stark, A. A. and Story, K. and Stubbs, C. W. and Vanderlinde, K. and Vieira, J. D. and Vikhlinin, A. and Williamson, R. and Zenteno, A.},
	month = apr,
	year = {2011},
	pages = {86},
}

@article{croton_many_2006,
	title = {The many lives of active galactic nuclei: cooling flows, black holes and the luminosities and colours of galaxies},
	volume = {365},
	issn = {0035-8711},
	shorttitle = {The many lives of active galactic nuclei},
	url = {https://ui.adsabs.harvard.edu/abs/2006MNRAS.365...11C/abstract},
	doi = {10.1111/j.1365-2966.2005.09675.x},
	language = {en},
	number = {1},
	urldate = {2026-02-08},
	journal = {Monthly Notices of the Royal Astronomical Society},
	author = {Croton, Darren J. and Springel, Volker and White, Simon D. M. and De Lucia, G. and Frenk, C. S. and Gao, L. and Jenkins, A. and Kauffmann, G. and Navarro, J. F. and Yoshida, N.},
	month = jan,
	year = {2006},
	pages = {11--28},
}

@article{birnboim_virial_2003,
	title = {Virial shocks in galactic haloes?},
	volume = {345},
	issn = {0035-8711},
	url = {https://ui.adsabs.harvard.edu/abs/2003MNRAS.345..349B/abstract},
	doi = {10.1046/j.1365-8711.2003.06955.x},
	language = {en},
	number = {1},
	urldate = {2026-02-08},
	journal = {Monthly Notices of the Royal Astronomical Society},
	author = {Birnboim, Yuval and Dekel, Avishai},
	month = oct,
	year = {2003},
	pages = {349--364},
}

@article{muzzin_gemini_2012,
	title = {The {Gemini} {Cluster} {Astrophysics} {Spectroscopic} {Survey} ({GCLASS}): {The} {Role} of {Environment} and {Self}-regulation in {Galaxy} {Evolution} at z {\textasciitilde} 1},
	volume = {746},
	issn = {0004-637X},
	shorttitle = {The {Gemini} {Cluster} {Astrophysics} {Spectroscopic} {Survey} ({GCLASS})},
	url = {https://ui.adsabs.harvard.edu/abs/2012ApJ...746..188M/abstract},
	doi = {10.1088/0004-637X/746/2/188},
	language = {en},
	number = {2},
	urldate = {2026-02-08},
	journal = {The Astrophysical Journal},
	author = {Muzzin, Adam and Wilson, Gillian and Yee, H. K. C. and Gilbank, David and Hoekstra, Henk and Demarco, Ricardo and Balogh, Michael and van Dokkum, Pieter and Franx, Marijn and Ellingson, Erica and Hicks, Amalia and Nantais, Julie and Noble, Allison and Lacy, Mark and Lidman, Chris and Rettura, Alessandro and Surace, Jason and Webb, Tracy},
	month = feb,
	year = {2012},
	pages = {188},
}

@article{balogh_evidence_2016,
	title = {Evidence for a change in the dominant satellite galaxy quenching mechanism at z = 1},
	volume = {456},
	issn = {0035-8711},
	url = {https://ui.adsabs.harvard.edu/abs/2016MNRAS.456.4364B/abstract},
	doi = {10.1093/mnras/stv2949},
	language = {en},
	number = {4},
	urldate = {2026-02-08},
	journal = {Monthly Notices of the Royal Astronomical Society},
	author = {Balogh, Michael L. and McGee, Sean L. and Mok, Angus and Muzzin, Adam and van der Burg, Remco F. J. and Bower, Richard G. and Finoguenov, Alexis and Hoekstra, Henk and Lidman, Chris and Mulchaey, John S. and Noble, Allison and Parker, Laura C. and Tanaka, Masayuki and Wilman, David J. and Webb, Tracy and Wilson, Gillian and Yee, Howard K. C.},
	month = mar,
	year = {2016},
	pages = {4364--4376},
}

@article{ji_evidence_2018,
	title = {Evidence of {Environmental} {Quenching} at {Redshift} z ≈ 2},
	volume = {862},
	issn = {0004-637X},
	url = {https://ui.adsabs.harvard.edu/abs/2018ApJ...862..135J/abstract},
	doi = {10.3847/1538-4357/aacc2c},
	language = {en},
	number = {2},
	urldate = {2026-02-08},
	journal = {The Astrophysical Journal},
	author = {Ji, Zhiyuan and Giavalisco, Mauro and Williams, Christina C. and Faber, Sandra M. and Ferguson, Henry C. and Guo, Yicheng and Liu, Teng and Lee, Bomee},
	month = aug,
	year = {2018},
	pages = {135},
}

@article{thomas_environment_2010,
	title = {Environment and self-regulation in galaxy formation},
	volume = {404},
	issn = {0035-8711},
	url = {https://ui.adsabs.harvard.edu/abs/2010MNRAS.404.1775T/abstract},
	doi = {10.1111/j.1365-2966.2010.16427.x},
	language = {en},
	number = {4},
	urldate = {2026-02-08},
	journal = {Monthly Notices of the Royal Astronomical Society},
	author = {Thomas, Daniel and Maraston, Claudia and Schawinski, Kevin and Sarzi, Marc and Silk, Joseph},
	month = jun,
	year = {2010},
	pages = {1775--1789},
}

@article{baker_exploring_2025,
	title = {Exploring over 700 massive quiescent galaxies at z = 2─7: {Demographics} and stellar mass functions},
	volume = {702},
	issn = {0004-6361},
	shorttitle = {Exploring over 700 massive quiescent galaxies at z = 2─7},
	url = {https://ui.adsabs.harvard.edu/abs/2025A&A...702A.270B/abstract},
	doi = {10.1051/0004-6361/202555829},
	language = {en},
	urldate = {2026-01-25},
	journal = {Astronomy and Astrophysics},
	author = {Baker, William M. and Valentino, Francesco and Lagos, Claudia del P. and Ito, Kei and Jespersen, Christian Kragh and Gottumukkala, Rashmi and Hjorth, Jens and Langeroodi, Danial and Sedgewick, Aidan},
	month = oct,
	year = {2025},
	pages = {A270},
}

@article{carreira_jwst_2026,
	title = {{JWST} {Advanced} {Deep} {Extragalactic} {Survey} ({JADES}) {Data} {Release} 5: {Catalogs} of inferred morphological properties of galaxies from {JWST}/{NIRCam} imaging in {GOODS}-{N} and {GOODS}-{S}},
	shorttitle = {{JWST} {Advanced} {Deep} {Extragalactic} {Survey} ({JADES}) {Data} {Release} 5},
	url = {https://ui.adsabs.harvard.edu/abs/2026arXiv260115957C/abstract},
	doi = {10.48550/arXiv.2601.15957},
	language = {en},
	urldate = {2026-01-25},
	journal = {arXiv e-prints},
	author = {Carreira, Courtney and Robertson, Brant E. and Danhaive, A. Lola and Ji, Zhiyuan and Rieke, Marcia and Tacchella, Sandro and Villanueva, Natalia C. and Willmer, Christopher N. A. and Wu, Zihao and Zhu, Yongda and Baker, William M. and Bunker, Andrew J. and Cameron, Alex J. and Chevallard, Jacopo and Curtis-Lake, Emma and Duan, Qiao and Eisenstein, Daniel J. and Hainline, Kevin and Hausen, Ryan and Johnson, Benjamin D. and Maiolino, Roberto and Mengistu, Petra and Puskás, Dávid and Rinaldi, Pierluigi and Sun, Yang and Trussler, James A. A. and Übler, Hannah and Uppal, Anavi and Williams, Christina C.},
	month = jan,
	year = {2026},
	pages = {arXiv:2601.15957},
}

@article{baker_abundance_2025,
	title = {The abundance and nature of high-redshift quiescent galaxies from {JADES} spectroscopy and the {FLAMINGO} simulations},
	volume = {539},
	issn = {0035-8711},
	url = {https://ui.adsabs.harvard.edu/abs/2025MNRAS.539..557B/abstract},
	doi = {10.1093/mnras/staf475},
	language = {en},
	number = {1},
	urldate = {2026-01-18},
	journal = {Monthly Notices of the Royal Astronomical Society},
	author = {Baker, William M. and Lim, Seunghwan and D'Eugenio, Francesco and Maiolino, Roberto and Ji, Zhiyuan and Arribas, Santiago and Bunker, Andrew J. and Carniani, Stefano and Charlot, Stephane and de Graaff, Anna and Hainline, Kevin and Looser, Tobias J. and Lyu, Jianwei and Rinaldi, Pierluigi and Robertson, Brant and Schaller, Matthieu and Schaye, Joop and Scholtz, Jan and Übler, Hannah and Williams, Christina C. and Willmer, Christopher N. A. and Willott, Chris and Zhu, Yongda},
	month = may,
	year = {2025},
	pages = {557--589},
}

@article{peng_mass_2012,
	title = {Mass and {Environment} as {Drivers} of {Galaxy} {Evolution}. {II}. {The} {Quenching} of {Satellite} {Galaxies} as the {Origin} of {Environmental} {Effects}},
	volume = {757},
	issn = {0004-637X},
	url = {https://ui.adsabs.harvard.edu/abs/2012ApJ...757....4P/abstract},
	doi = {10.1088/0004-637X/757/1/4},
	language = {en},
	number = {1},
	urldate = {2026-01-16},
	journal = {The Astrophysical Journal},
	author = {Peng, Ying-jie and Lilly, Simon J. and Renzini, Alvio and Carollo, Marcella},
	month = sep,
	year = {2012},
	pages = {4},
}

@article{peng_mass_2010,
	title = {Mass and {Environment} as {Drivers} of {Galaxy} {Evolution} in {SDSS} and {zCOSMOS} and the {Origin} of the {Schechter} {Function}},
	volume = {721},
	issn = {0004-637X},
	url = {https://ui.adsabs.harvard.edu/abs/2010ApJ...721..193P/abstract},
	doi = {10.1088/0004-637X/721/1/193},
	language = {en},
	number = {1},
	urldate = {2026-01-16},
	journal = {The Astrophysical Journal},
	author = {Peng, Ying-jie and Lilly, Simon J. and Kovač, Katarina and Bolzonella, Micol and Pozzetti, Lucia and Renzini, Alvio and Zamorani, Gianni and Ilbert, Olivier and Knobel, Christian and Iovino, Angela and Maier, Christian and Cucciati, Olga and Tasca, Lidia and Carollo, C. Marcella and Silverman, John and Kampczyk, Pawel and de Ravel, Loic and Sanders, David and Scoville, Nicholas and Contini, Thierry and Mainieri, Vincenzo and Scodeggio, Marco and Kneib, Jean-Paul and Le Fèvre, Olivier and Bardelli, Sandro and Bongiorno, Angela and Caputi, Karina and Coppa, Graziano and de la Torre, Sylvain and Franzetti, Paolo and Garilli, Bianca and Lamareille, Fabrice and Le Borgne, Jean-Francois and Le Brun, Vincent and Mignoli, Marco and Perez Montero, Enrique and Pello, Roser and Ricciardelli, Elena and Tanaka, Masayuki and Tresse, Laurence and Vergani, Daniela and Welikala, Niraj and Zucca, Elena and Oesch, Pascal and Abbas, Ummi and Barnes, Luke and Bordoloi, Rongmon and Bottini, Dario and Cappi, Alberto and Cassata, Paolo and Cimatti, Andrea and Fumana, Marco and Hasinger, Gunther and Koekemoer, Anton and Leauthaud, Alexei and Maccagni, Dario and Marinoni, Christian and McCracken, Henry and Memeo, Pierdomenico and Meneux, Baptiste and Nair, Preethi and Porciani, Cristiano and Presotto, Valentina and Scaramella, Roberto},
	month = sep,
	year = {2010},
	pages = {193--221},
}

@article{geha_stellar_2012,
	title = {A {Stellar} {Mass} {Threshold} for {Quenching} of {Field} {Galaxies}},
	volume = {757},
	issn = {0004-637X},
	url = {https://ui.adsabs.harvard.edu/abs/2012ApJ...757...85G/abstract},
	doi = {10.1088/0004-637X/757/1/85},
	language = {en},
	number = {1},
	urldate = {2026-01-16},
	journal = {The Astrophysical Journal},
	author = {Geha, M. and Blanton, M. R. and Yan, R. and Tinker, J. L.},
	month = sep,
	year = {2012},
	pages = {85},
}

@article{mcquinn_galactic_2019,
	title = {Galactic {Winds} in {Low}-mass {Galaxies}},
	volume = {886},
	issn = {0004-637X},
	url = {https://ui.adsabs.harvard.edu/abs/2019ApJ...886...74M/abstract},
	doi = {10.3847/1538-4357/ab4c37},
	language = {en},
	number = {1},
	urldate = {2026-01-16},
	journal = {The Astrophysical Journal},
	author = {McQuinn, Kristen B. W. and van Zee, Liese and Skillman, Evan D.},
	month = nov,
	year = {2019},
	pages = {74},
}

@article{boselli_origin_2008,
	title = {The {Origin} of {Dwarf} {Ellipticals} in the {Virgo} {Cluster}},
	volume = {674},
	issn = {0004-637X},
	url = {https://ui.adsabs.harvard.edu/abs/2008ApJ...674..742B/abstract},
	doi = {10.1086/525513},
	language = {en},
	number = {2},
	urldate = {2026-01-16},
	journal = {The Astrophysical Journal},
	author = {Boselli, A. and Boissier, S. and Cortese, L. and Gavazzi, G.},
	month = feb,
	year = {2008},
	pages = {742--767},
}

@article{mcnab_gogreen_2021,
	title = {The {GOGREEN} survey: transition galaxies and the evolution of environmental quenching},
	volume = {508},
	issn = {0035-8711},
	shorttitle = {The {GOGREEN} survey},
	url = {https://ui.adsabs.harvard.edu/abs/2021MNRAS.508..157M/abstract},
	doi = {10.1093/mnras/stab2558},
	language = {en},
	number = {1},
	urldate = {2026-01-16},
	journal = {Monthly Notices of the Royal Astronomical Society},
	author = {McNab, Karen and Balogh, Michael L. and van der Burg, Remco F. J. and Forestell, Anya and Webb, Kristi and Vulcani, Benedetta and Rudnick, Gregory and Muzzin, Adam and Cooper, M. C. and McGee, Sean and Biviano, Andrea and Cerulo, Pierluigi and Chan, Jeffrey C. C. and De Lucia, Gabriella and Demarco, Ricardo and Finoguenov, Alexis and Forrest, Ben and Golledge, Caelan and Jablonka, Pascale and Lidman, Chris and Nantais, Julie and Old, Lyndsay and Pintos-Castro, Irene and Poggianti, Bianca and Reeves, Andrew M. M. and Wilson, Gillian and Yee, Howard K. C. and Zaritsky, Dennis},
	month = nov,
	year = {2021},
	pages = {157--174},
}

@article{boselli_virgo_2020,
	title = {A {Virgo} {Environmental} {Survey} {Tracing} {Ionised} {Gas} {Emission} ({VESTIGE}). {VI}. {Environmental} quenching on {HII}-region scales},
	volume = {634},
	issn = {0004-6361},
	url = {https://ui.adsabs.harvard.edu/abs/2020A&A...634L...1B/abstract},
	doi = {10.1051/0004-6361/201937310},
	language = {en},
	urldate = {2026-01-16},
	journal = {Astronomy and Astrophysics},
	author = {Boselli, A. and Fossati, M. and Longobardi, A. and Boissier, S. and Boquien, M. and Braine, J. and Côté, P. and Cuillandre, J. C. and Epinat, B. and Ferrarese, L. and Gavazzi, G. and Gwyn, S. and Hensler, G. and Plana, H. and Roehlly, Y. and Schimd, C. and Sun, M. and Trinchieri, G.},
	month = feb,
	year = {2020},
	pages = {L1},
}

@article{sun_jwsts_2024,
	title = {{JWST}'s {First} {Glimpse} of a z \&gt; 2 {Forming} {Cluster} {Reveals} a {Top}-heavy {Stellar} {Mass} {Function}},
	volume = {967},
	issn = {0004-637X},
	url = {https://ui.adsabs.harvard.edu/abs/2024ApJ...967L..34S/abstract},
	doi = {10.3847/2041-8213/ad4986},
	language = {en},
	number = {2},
	urldate = {2026-01-16},
	journal = {The Astrophysical Journal},
	author = {Sun, Hanwen and Wang, Tao and Xu, Ke and Daddi, Emanuele and Gu, Qing and Kodama, Tadayuki and Zanella, Anita and Elbaz, David and Tanaka, Ichi and Gobat, Raphael and Guo, Qi and Han, Jiaxin and Lu, Shiying and Zhou, Luwenjia},
	month = jun,
	year = {2024},
	pages = {L34},
}

@article{borrow_there_2023,
	title = {There and back again: {Understanding} the critical properties of backsplash galaxies},
	volume = {520},
	issn = {0035-8711},
	shorttitle = {There and back again},
	url = {https://ui.adsabs.harvard.edu/abs/2023MNRAS.520..649B/abstract},
	doi = {10.1093/mnras/stad045},
	language = {en},
	number = {1},
	urldate = {2026-01-16},
	journal = {Monthly Notices of the Royal Astronomical Society},
	author = {Borrow, Josh and Vogelsberger, Mark and O'Neil, Stephanie and McDonald, Michael A. and Smith, Aaron},
	month = mar,
	year = {2023},
	pages = {649--667},
}

@article{harrold_role_2026,
	title = {The role of mergers and rejuvenation in the buildup of the quiescent population at cosmic noon},
	volume = {545},
	issn = {0035-8711},
	url = {https://ui.adsabs.harvard.edu/abs/2026MNRAS.545f2191H/abstract},
	doi = {10.1093/mnras/staf2191},
	language = {en},
	number = {3},
	urldate = {2026-01-03},
	journal = {Monthly Notices of the Royal Astronomical Society},
	author = {Harrold, Jimi E. and Almaini, Omar and Pearce, Frazer R. and Yates, Robert M. and Maltby, Dave and Rowlands, Kate and Wild, Vivienne and Skarbinski, Maya and de Lisle, Thomas},
	month = jan,
	year = {2026},
	pages = {staf2191},
}

@article{wylezalek_galaxy_2013,
	title = {Galaxy {Clusters} around {Radio}-loud {Active} {Galactic} {Nuclei} at 1.3 \&lt; z \&lt; 3.2 as {Seen} by {Spitzer}},
	volume = {769},
	issn = {0004-637X},
	url = {https://ui.adsabs.harvard.edu/abs/2013ApJ...769...79W/abstract},
	doi = {10.1088/0004-637X/769/1/79},
	language = {en},
	number = {1},
	urldate = {2025-12-22},
	journal = {The Astrophysical Journal},
	author = {Wylezalek, Dominika and Galametz, Audrey and Stern, Daniel and Vernet, Joël and De Breuck, Carlos and Seymour, Nick and Brodwin, Mark and Eisenhardt, Peter R. M. and Gonzalez, Anthony H. and Hatch, Nina and Jarvis, Matt and Rettura, Alessandro and Stanford, Spencer A. and Stevens, Jason A.},
	month = may,
	year = {2013},
	pages = {79},
}

@article{werner_satellite_2022,
	title = {Satellite quenching was not important for z ∼ 1 clusters: most quenching occurred during infall},
	volume = {510},
	issn = {0035-8711},
	shorttitle = {Satellite quenching was not important for z ∼ 1 clusters},
	url = {https://ui.adsabs.harvard.edu/abs/2022MNRAS.510..674W/abstract},
	doi = {10.1093/mnras/stab3484},
	language = {en},
	number = {1},
	urldate = {2025-12-22},
	journal = {Monthly Notices of the Royal Astronomical Society},
	author = {Werner, S. V. and Hatch, N. A. and Muzzin, A. and van der Burg, R. F. J. and Balogh, M. L. and Rudnick, G. and Wilson, G.},
	month = feb,
	year = {2022},
	pages = {674--686},
}

@article{hamadouche_jwst_2025,
	title = {{JWST} {PRIMER}: strong evidence for the environmental quenching of low-mass galaxies out to z≃ 2},
	volume = {541},
	issn = {0035-8711},
	shorttitle = {{JWST} {PRIMER}},
	url = {https://ui.adsabs.harvard.edu/abs/2025MNRAS.541..463H/abstract},
	doi = {10.1093/mnras/staf971},
	language = {en},
	number = {1},
	urldate = {2025-12-22},
	journal = {Monthly Notices of the Royal Astronomical Society},
	author = {Hamadouche, M. L. and McLure, R. J. and Carnall, A. C. and McLeod, D. J. and Dunlop, J. S. and Whitaker, K. E. and Donnan, C. T. and Begley, R. and Stanton, T. M. and Almaini, O. and Aird, J. and Cullen, F. and Cutler, S. and Grogin, N. A. and Koekemoer, A. M.},
	month = jul,
	year = {2025},
	pages = {463--475},
}

@article{ahad_environment-dependent_2024,
	title = {An environment-dependent halo mass function as a driver for the early quenching of z ≥ 1.5 cluster galaxies},
	volume = {528},
	issn = {0035-8711},
	url = {https://ui.adsabs.harvard.edu/abs/2024MNRAS.528.6329A/abstract},
	doi = {10.1093/mnras/stae341},
	language = {en},
	number = {4},
	urldate = {2025-12-21},
	journal = {Monthly Notices of the Royal Astronomical Society},
	author = {Ahad, Syeda Lammim and Muzzin, Adam and Bahé, Yannick M. and Hoekstra, Henk},
	month = mar,
	year = {2024},
	pages = {6329--6339},
}

@article{webb_gogreen_2020,
	title = {The {GOGREEN} survey: post-infall environmental quenching fails to predict the observed age difference between quiescent field and cluster galaxies at z \&gt; 1},
	volume = {498},
	issn = {0035-8711},
	shorttitle = {The {GOGREEN} survey},
	url = {https://ui.adsabs.harvard.edu/abs/2020MNRAS.498.5317W/abstract},
	doi = {10.1093/mnras/staa2752},
	language = {en},
	number = {4},
	urldate = {2025-12-21},
	journal = {Monthly Notices of the Royal Astronomical Society},
	author = {Webb, Kristi and Balogh, Michael L. and Leja, Joel and van der Burg, Remco F. J. and Rudnick, Gregory and Muzzin, Adam and Boak, Kevin and Cerulo, Pierluigi and Gilbank, David and Lidman, Chris and Old, Lyndsay J. and Pintos-Castro, Irene and McGee, Sean and Shipley, Heath and Biviano, Andrea and Chan, Jeffrey C. C. and Cooper, Michael and De Lucia, Gabriella and Demarco, Ricardo and Forrest, Ben and Jablonka, Pascale and Kukstas, Egidijus and McCarthy, Ian G. and McNab, Karen and Nantais, Julie and Noble, Allison and Poggianti, Bianca and Reeves, Andrew M. M. and Vulcani, Benedetta and Wilson, Gillian and Yee, Howard K. C. and Zaritsky, Dennis},
	month = nov,
	year = {2020},
	pages = {5317--5342},
}

@article{pintos-castro_evolution_2019,
	title = {The {Evolution} of the {Quenching} of {Star} {Formation} in {Cluster} {Galaxies} since z ∼ 1},
	volume = {876},
	issn = {0004-637X},
	url = {https://ui.adsabs.harvard.edu/abs/2019ApJ...876...40P/abstract},
	doi = {10.3847/1538-4357/ab14ee},
	language = {en},
	number = {1},
	urldate = {2025-12-21},
	journal = {The Astrophysical Journal},
	author = {Pintos-Castro, I. and Yee, H. K. C. and Muzzin, A. and Old, L. and Wilson, G.},
	month = may,
	year = {2019},
	pages = {40},
}

@article{kiyota_cluster_2025,
	title = {Cluster {Candidates} with {Massive} {Quiescent} {Galaxies} at z ∼ 2},
	volume = {980},
	issn = {0004-637X},
	url = {https://ui.adsabs.harvard.edu/abs/2025ApJ...980..104K/abstract},
	doi = {10.3847/1538-4357/ada5f4},
	language = {en},
	number = {1},
	urldate = {2025-12-21},
	journal = {The Astrophysical Journal},
	author = {Kiyota, Tomokazu and Ando, Makoto and Tanaka, Masayuki and Finoguenov, Alexis and Ali, Sadman Shariar and Coupon, Jean and Desprez, Guillaume and Gwyn, Stephen and Sawicki, Marcin and Shimakawa, Rhythm},
	month = feb,
	year = {2025},
	pages = {104},
}

@article{cooke_mature_2016,
	title = {A {Mature} {Galaxy} {Cluster} at z=1.58 around the {Radio} {Galaxy} {7C1753}+6311},
	volume = {816},
	issn = {0004-637X},
	url = {https://ui.adsabs.harvard.edu/abs/2016ApJ...816...83C/abstract},
	doi = {10.3847/0004-637X/816/2/83},
	language = {en},
	number = {2},
	urldate = {2025-12-21},
	journal = {The Astrophysical Journal},
	author = {Cooke, E. A. and Hatch, N. A. and Stern, D. and Rettura, A. and Brodwin, M. and Galametz, A. and Wylezalek, D. and Bridge, C. and Conselice, C. J. and De Breuck, C. and Gonzalez, A. H. and Jarvis, M.},
	month = jan,
	year = {2016},
	pages = {83},
}

@article{newman_spectroscopic_2014,
	title = {Spectroscopic {Confirmation} of the {Rich} z = 1.80 {Galaxy} {Cluster} {JKCS} 041 using the {WFC3} {Grism}: {Environmental} {Trends} in the {Ages} and {Structure} of {Quiescent} {Galaxies}},
	volume = {788},
	issn = {0004-637X},
	shorttitle = {Spectroscopic {Confirmation} of the {Rich} z = 1.80 {Galaxy} {Cluster} {JKCS} 041 using the {WFC3} {Grism}},
	url = {https://ui.adsabs.harvard.edu/abs/2014ApJ...788...51N/abstract},
	doi = {10.1088/0004-637X/788/1/51},
	language = {en},
	number = {1},
	urldate = {2025-12-21},
	journal = {The Astrophysical Journal},
	author = {Newman, Andrew B. and Ellis, Richard S. and Andreon, Stefano and Treu, Tommaso and Raichoor, Anand and Trinchieri, Ginevra},
	month = jun,
	year = {2014},
	pages = {51},
}

@article{van_der_burg_gogreen_2020,
	title = {The {GOGREEN} {Survey}: {A} deep stellar mass function of cluster galaxies at 1.0 \&lt; z \&lt; 1.4 and the complex nature of satellite quenching},
	volume = {638},
	issn = {0004-6361},
	shorttitle = {The {GOGREEN} {Survey}},
	url = {https://ui.adsabs.harvard.edu/abs/2020A&A...638A.112V/abstract},
	doi = {10.1051/0004-6361/202037754},
	language = {en},
	urldate = {2025-12-21},
	journal = {Astronomy and Astrophysics},
	author = {van der Burg, Remco F. J. and Rudnick, Gregory and Balogh, Michael L. and Muzzin, Adam and Lidman, Chris and Old, Lyndsay J. and Shipley, Heath and Gilbank, David and McGee, Sean and Biviano, Andrea and Cerulo, Pierluigi and Chan, Jeffrey C. C. and Cooper, Michael and De Lucia, Gabriella and Demarco, Ricardo and Forrest, Ben and Gwyn, Stephen and Jablonka, Pascale and Kukstas, Egidijus and Marchesini, Danilo and Nantais, Julie and Noble, Allison and Pintos-Castro, Irene and Poggianti, Bianca and Reeves, Andrew M. M. and Stefanon, Mauro and Vulcani, Benedetta and Webb, Kristi and Wilson, Gillian and Yee, Howard and Zaritsky, Dennis},
	month = jun,
	year = {2020},
	pages = {A112},
}

@article{zhang_halo_2024,
	title = {Halo {Mass}-observable {Proxy} {Scaling} {Relations} and {Their} {Dependencies} on {Galaxy} and {Group} {Properties}},
	volume = {960},
	issn = {0004-637X},
	url = {https://ui.adsabs.harvard.edu/abs/2024ApJ...960...71Z/abstract},
	doi = {10.3847/1538-4357/ad0892},
	language = {en},
	number = {1},
	urldate = {2025-12-21},
	journal = {The Astrophysical Journal},
	author = {Zhang, Ziwen and Wang, Huiyuan and Luo, Wentao and Mo, Houjun and Zhang, Jun and Yang, Xiaohu and Li, Hao and Li, Qinxun},
	month = jan,
	year = {2024},
	pages = {71},
}

@article{demaio_growth_2020,
	title = {The growth of brightest cluster galaxies and intracluster light over the past 10 billion years},
	volume = {491},
	issn = {0035-8711},
	url = {https://ui.adsabs.harvard.edu/abs/2020MNRAS.491.3751D/abstract},
	doi = {10.1093/mnras/stz3236},
	language = {en},
	number = {3},
	urldate = {2025-12-21},
	journal = {Monthly Notices of the Royal Astronomical Society},
	author = {DeMaio, Tahlia and Gonzalez, Anthony H. and Zabludoff, Ann and Zaritsky, Dennis and Aldering, Greg and Brodwin, Mark and Connor, Thomas and Donahue, Megan and Hayden, Brian and Mulchaey, John S. and Perlmutter, Saul and Stanford, S. A.},
	month = jan,
	year = {2020},
	pages = {3751--3759},
}

@article{shuntov_cosmos2020_2022,
	title = {{COSMOS2020}: {Cosmic} evolution of the stellar-to-halo mass relation for central and satellite galaxies up to z ∼ 5},
	volume = {664},
	issn = {0004-6361},
	shorttitle = {{COSMOS2020}},
	url = {https://ui.adsabs.harvard.edu/abs/2022A&A...664A..61S/abstract},
	doi = {10.1051/0004-6361/202243136},
	language = {en},
	urldate = {2025-12-21},
	journal = {Astronomy and Astrophysics},
	author = {Shuntov, M. and McCracken, H. J. and Gavazzi, R. and Laigle, C. and Weaver, J. R. and Davidzon, I. and Ilbert, O. and Kauffmann, O. B. and Faisst, A. and Dubois, Y. and Koekemoer, A. M. and Moneti, A. and Milvang-Jensen, B. and Mobasher, B. and Sanders, D. B. and Toft, S.},
	month = aug,
	year = {2022},
	pages = {A61},
}

@article{wang_discovery_2016,
	title = {Discovery of a {Galaxy} {Cluster} with a {Violently} {Starbursting} {Core} at z = 2.506},
	volume = {828},
	issn = {0004-637X},
	url = {https://ui.adsabs.harvard.edu/abs/2016ApJ...828...56W/abstract},
	doi = {10.3847/0004-637X/828/1/56},
	language = {en},
	number = {1},
	urldate = {2025-12-21},
	journal = {The Astrophysical Journal},
	author = {Wang, Tao and Elbaz, David and Daddi, Emanuele and Finoguenov, Alexis and Liu, Daizhong and Schreiber, Corentin and Martín, Sergio and Strazzullo, Veronica and Valentino, Francesco and van der Burg, Remco and Zanella, Anita and Ciesla, Laure and Gobat, Raphael and Le Brun, Amandine and Pannella, Maurilio and Sargent, Mark and Shu, Xinwen and Tan, Qinghua and Cappelluti, Nico and Li, Yanxia},
	month = sep,
	year = {2016},
	pages = {56},
}

@article{muldrew_what_2015,
	title = {What are protoclusters? - {Defining} high-redshift galaxy clusters and protoclusters},
	volume = {452},
	issn = {0035-8711},
	shorttitle = {What are protoclusters?},
	url = {https://ui.adsabs.harvard.edu/abs/2015MNRAS.452.2528M/abstract},
	doi = {10.1093/mnras/stv1449},
	language = {en},
	number = {3},
	urldate = {2025-12-21},
	journal = {Monthly Notices of the Royal Astronomical Society},
	author = {Muldrew, Stuart I. and Hatch, Nina A. and Cooke, Elizabeth A.},
	month = sep,
	year = {2015},
	pages = {2528--2539},
}

@article{overzier_realm_2016,
	title = {The realm of the galaxy protoclusters. {A} review},
	volume = {24},
	issn = {0935-4956},
	url = {https://ui.adsabs.harvard.edu/abs/2016A&ARv..24...14O/abstract},
	doi = {10.1007/s00159-016-0100-3},
	language = {en},
	number = {1},
	urldate = {2025-12-21},
	journal = {Astronomy and Astrophysics Review},
	author = {Overzier, Roderik A.},
	month = nov,
	year = {2016},
	pages = {14},
}

@article{de_lucia_hierarchical_2007,
	title = {The hierarchical formation of the brightest cluster galaxies},
	volume = {375},
	issn = {0035-8711},
	url = {https://ui.adsabs.harvard.edu/abs/2007MNRAS.375....2D/abstract},
	doi = {10.1111/j.1365-2966.2006.11287.x},
	language = {en},
	number = {1},
	urldate = {2025-12-21},
	journal = {Monthly Notices of the Royal Astronomical Society},
	author = {De Lucia, Gabriella and Blaizot, Jérémy},
	month = feb,
	year = {2007},
	pages = {2--14},
}

@article{thomas_epochs_2005,
	title = {The {Epochs} of {Early}-{Type} {Galaxy} {Formation} as a {Function} of {Environment}},
	volume = {621},
	issn = {0004-637X},
	url = {https://ui.adsabs.harvard.edu/abs/2005ApJ...621..673T/abstract},
	doi = {10.1086/426932},
	language = {en},
	number = {2},
	urldate = {2025-12-21},
	journal = {The Astrophysical Journal},
	author = {Thomas, Daniel and Maraston, Claudia and Bender, Ralf and Mendes de Oliveira, Claudia},
	month = mar,
	year = {2005},
	pages = {673--694},
}

@article{liu_morphological_2021,
	title = {Morphological {Transformation} and {Star} {Formation} {Quenching} of {Massive} {Galaxies} at 0.5 ≤ z ≤ 2.5 in {3D}-{HST}/{CANDELS}},
	volume = {923},
	issn = {0004-637X},
	url = {https://ui.adsabs.harvard.edu/abs/2021ApJ...923...46L/abstract},
	doi = {10.3847/1538-4357/ac2817},
	language = {en},
	number = {1},
	urldate = {2025-12-21},
	journal = {The Astrophysical Journal},
	author = {Liu, Shuang and Gu, Yizhou and Yuan, Qirong and Lu, Shiying and Bao, Min and Fang, Guanwen and Fan, Lulu},
	month = dec,
	year = {2021},
	pages = {46},
}

@article{martorano_sizemass_2024,
	title = {The {Size}–{Mass} {Relation} at {Rest}-frame 1.5 μm from {JWST}/{NIRCam} in the {COSMOS}-{WEB} and {PRIMER}-{COSMOS} {Fields}},
	volume = {972},
	issn = {0004-637X},
	url = {https://ui.adsabs.harvard.edu/abs/2024ApJ...972..134M/abstract},
	doi = {10.3847/1538-4357/ad5c6a},
	language = {en},
	number = {2},
	urldate = {2025-12-21},
	journal = {The Astrophysical Journal},
	author = {Martorano, Marco and van der Wel, Arjen and Baes, Maarten and Bell, Eric F. and Brammer, Gabriel and Franx, Marijn and Nersesian, Angelos},
	month = sep,
	year = {2024},
	pages = {134},
}

@article{ferreira_jwst_2023,
	title = {The {JWST} {Hubble} {Sequence}: {The} {Rest}-frame {Optical} {Evolution} of {Galaxy} {Structure} at 1.5 \&lt; z \&lt; 6.5},
	volume = {955},
	issn = {0004-637X},
	shorttitle = {The {JWST} {Hubble} {Sequence}},
	url = {https://ui.adsabs.harvard.edu/abs/2023ApJ...955...94F/abstract},
	doi = {10.3847/1538-4357/acec76},
	language = {en},
	number = {2},
	urldate = {2025-12-21},
	journal = {The Astrophysical Journal},
	author = {Ferreira, Leonardo and Conselice, Christopher J. and Sazonova, Elizaveta and Ferrari, Fabricio and Caruana, Joseph and Tohill, Clár-Bríd and Lucatelli, Geferson and Adams, Nathan and Irodotou, Dimitrios and Marshall, Madeline A. and Roper, Will J. and Lovell, Christopher C. and Verma, Aprajita and Austin, Duncan and Trussler, James and Wilkins, Stephen M.},
	month = oct,
	year = {2023},
	pages = {94},
}

@article{wagner_star_2015,
	title = {Star {Formation} in {High}-redshift {Cluster} {Ellipticals}},
	volume = {800},
	issn = {0004-637X},
	url = {https://ui.adsabs.harvard.edu/abs/2015ApJ...800..107W/abstract},
	doi = {10.1088/0004-637X/800/2/107},
	language = {en},
	number = {2},
	urldate = {2025-12-21},
	journal = {The Astrophysical Journal},
	author = {Wagner, Cory R. and Brodwin, Mark and Snyder, Gregory F. and Gonzalez, Anthony H. and Stanford, S. A. and Alberts, Stacey and Pope, Alexandra and Stern, Daniel and Zeimann, Gregory R. and Chary, Ranga-Ram and Dey, Arjun and Eisenhardt, Peter R. M. and Mancone, Conor L. and Moustakas, John},
	month = feb,
	year = {2015},
	pages = {107},
}

@article{noordeh_quiescent_2021,
	title = {Quiescent galaxies in a virialized cluster at redshift 2: evidence for accelerated size growth},
	volume = {507},
	issn = {0035-8711},
	shorttitle = {Quiescent galaxies in a virialized cluster at redshift 2},
	url = {https://ui.adsabs.harvard.edu/abs/2021MNRAS.507.5272N/abstract},
	doi = {10.1093/mnras/stab2459},
	language = {en},
	number = {4},
	urldate = {2025-12-21},
	journal = {Monthly Notices of the Royal Astronomical Society},
	author = {Noordeh, E. and Canning, R. E. A. and Willis, J. P. and Allen, S. W. and Mantz, A. and Stanford, S. A. and Brammer, G.},
	month = nov,
	year = {2021},
	pages = {5272--5280},
}

@article{baker_double_2025,
	title = {Double {Trouble}: {Two} spectroscopically confirmed low-mass quiescent galaxies at z\&gt;5 in overdensities},
	shorttitle = {Double {Trouble}},
	url = {https://ui.adsabs.harvard.edu/abs/2025arXiv250909761B/abstract},
	doi = {10.48550/arXiv.2509.09761},
	language = {en},
	urldate = {2025-12-21},
	journal = {arXiv e-prints},
	author = {Baker, William M. and Ito, Kei and Valentino, Francesco and Zhu, Pengpei and Scarpe, Gianluca and Gottumukkala, Rashmi and Hjorth, Jens and Barrufet, Laia and Langeroodi, Danial},
	month = sep,
	year = {2025},
	pages = {arXiv:2509.09761},
}

@article{cautun_subhalo_2014,
	title = {Subhalo statistics of galactic haloes: beyond the resolution limit},
	volume = {445},
	issn = {0035-8711},
	shorttitle = {Subhalo statistics of galactic haloes},
	url = {https://ui.adsabs.harvard.edu/abs/2014MNRAS.445.1820C/abstract},
	doi = {10.1093/mnras/stu1829},
	language = {en},
	number = {2},
	urldate = {2025-11-26},
	journal = {Monthly Notices of the Royal Astronomical Society},
	author = {Cautun, Marius and Hellwing, Wojciech A. and van de Weygaert, Rien and Frenk, Carlos S. and Jones, Bernard J. T. and Sawala, Till},
	month = dec,
	year = {2014},
	pages = {1820--1835},
}

@article{hahn_evolution_2007,
	title = {The evolution of dark matter halo properties in clusters, filaments, sheets and voids},
	volume = {381},
	issn = {0035-8711},
	url = {https://ui.adsabs.harvard.edu/abs/2007MNRAS.381...41H/abstract},
	doi = {10.1111/j.1365-2966.2007.12249.x},
	language = {en},
	number = {1},
	urldate = {2025-11-26},
	journal = {Monthly Notices of the Royal Astronomical Society},
	author = {Hahn, Oliver and Carollo, C. Marcella and Porciani, Cristiano and Dekel, Avishai},
	month = oct,
	year = {2007},
	pages = {41--51},
}

@article{kim_gradual_2023,
	title = {A {Gradual} {Decline} of {Star} {Formation} since {Cluster} {Infall}: {New} {Kinematic} {Insights} into {Environmental} {Quenching} at 0.3 \&lt; z \&lt; 1.1},
	volume = {955},
	issn = {0004-637X},
	shorttitle = {A {Gradual} {Decline} of {Star} {Formation} since {Cluster} {Infall}},
	url = {https://ui.adsabs.harvard.edu/abs/2023ApJ...955...32K/abstract},
	doi = {10.3847/1538-4357/acecff},
	language = {en},
	number = {1},
	urldate = {2025-11-25},
	journal = {The Astrophysical Journal},
	author = {Kim, Keunho J. and Bayliss, Matthew B. and Noble, Allison G. and Khullar, Gourav and Cronk, Ethan and Roberson, Joshua and Ansarinejad, Behzad and Bleem, Lindsey E. and Floyd, Benjamin and Grandis, Sebastian and Mahler, Guillaume and McDonald, Michael A. and Reichardt, Christian L. and Saro, Alexandro and Sharon, Keren and Somboonpanyakul, Taweewat and Strazzullo, Veronica},
	month = sep,
	year = {2023},
	pages = {32},
}

@article{dou_estimating_2025,
	title = {Estimating infall times of galaxies around clusters: {Working} to achieve accurate results based on observational data},
	volume = {699},
	issn = {0004-6361},
	shorttitle = {Estimating infall times of galaxies around clusters},
	url = {https://ui.adsabs.harvard.edu/abs/2025A&A...699A..95D/abstract},
	doi = {10.1051/0004-6361/202555240},
	language = {en},
	urldate = {2025-11-25},
	journal = {Astronomy and Astrophysics},
	author = {Dou, Haoran and Yu, Heng},
	month = jul,
	year = {2025},
	pages = {A95},
}

@article{noble_phase_2016,
	title = {The {Phase} {Space} of z∼1.2 {SpARCS} {Clusters}: {Using} {Herschel} to {Probe} {Dust} {Temperature} as a {Function} of {Environment} and {Accretion} {History}},
	volume = {816},
	issn = {0004-637X},
	shorttitle = {The {Phase} {Space} of z∼1.2 {SpARCS} {Clusters}},
	url = {https://ui.adsabs.harvard.edu/abs/2016ApJ...816...48N/abstract},
	doi = {10.3847/0004-637X/816/2/48},
	language = {en},
	number = {2},
	urldate = {2025-11-25},
	journal = {The Astrophysical Journal},
	author = {Noble, A. G. and Webb, T. M. A. and Yee, H. K. C. and Muzzin, A. and Wilson, G. and van der Burg, R. F. J. and Balogh, M. L. and Shupe, D. L.},
	month = jan,
	year = {2016},
	pages = {48},
}

@article{noble_kinematic_2013,
	title = {A {Kinematic} {Approach} to {Assessing} {Environmental} {Effects}: {Star}-forming {Galaxies} in a z {\textasciitilde} 0.9 {SpARCS} {Cluster} {Using} {Spitzer} 24 μm {Observations}},
	volume = {768},
	issn = {0004-637X},
	shorttitle = {A {Kinematic} {Approach} to {Assessing} {Environmental} {Effects}},
	url = {https://ui.adsabs.harvard.edu/abs/2013ApJ...768..118N/abstract},
	doi = {10.1088/0004-637X/768/2/118},
	language = {en},
	number = {2},
	urldate = {2025-11-25},
	journal = {The Astrophysical Journal},
	author = {Noble, A. G. and Webb, T. M. A. and Muzzin, A. and Wilson, G. and Yee, H. K. C. and van der Burg, R. F. J.},
	month = may,
	year = {2013},
	pages = {118},
}

@article{jaffe_budhies_2015,
	title = {{BUDHIES} {II}: a phase-space view of {H} {I} gas stripping and star formation quenching in cluster galaxies},
	volume = {448},
	issn = {0035-8711},
	shorttitle = {{BUDHIES} {II}},
	url = {https://ui.adsabs.harvard.edu/abs/2015MNRAS.448.1715J/abstract},
	doi = {10.1093/mnras/stv100},
	language = {en},
	number = {2},
	urldate = {2025-11-25},
	journal = {Monthly Notices of the Royal Astronomical Society},
	author = {Jaffé, Yara L. and Smith, Rory and Candlish, Graeme N. and Poggianti, Bianca M. and Sheen, Yun-Kyeong and Verheijen, Marc A. W.},
	month = apr,
	year = {2015},
	pages = {1715--1728},
}

@article{rhee_phase-space_2017,
	title = {Phase-space {Analysis} in the {Group} and {Cluster} {Environment}: {Time} {Since} {Infall} and {Tidal} {Mass} {Loss}},
	volume = {843},
	issn = {0004-637X},
	shorttitle = {Phase-space {Analysis} in the {Group} and {Cluster} {Environment}},
	url = {https://ui.adsabs.harvard.edu/abs/2017ApJ...843..128R/abstract},
	doi = {10.3847/1538-4357/aa6d6c},
	language = {en},
	number = {2},
	urldate = {2025-11-25},
	journal = {The Astrophysical Journal},
	author = {Rhee, Jinsu and Smith, Rory and Choi, Hoseung and Yi, Sukyoung K. and Jaffé, Yara and Candlish, Graeme and Sánchez-Jánssen, Ruben},
	month = jul,
	year = {2017},
	pages = {128},
}

@article{belli_mosfire_2019,
	title = {{MOSFIRE} {Spectroscopy} of {Quiescent} {Galaxies} at 1.5 \&lt; z \&lt; 2.5. {II}. {Star} {Formation} {Histories} and {Galaxy} {Quenching}},
	volume = {874},
	issn = {0004-637X},
	url = {https://ui.adsabs.harvard.edu/abs/2019ApJ...874...17B/abstract},
	doi = {10.3847/1538-4357/ab07af},
	language = {en},
	number = {1},
	urldate = {2025-11-25},
	journal = {The Astrophysical Journal},
	author = {Belli, Sirio and Newman, Andrew B. and Ellis, Richard S.},
	month = mar,
	year = {2019},
	pages = {17},
}

@article{whitaker_newfirm_2011,
	title = {The {NEWFIRM} {Medium}-band {Survey}: {Photometric} {Catalogs}, {Redshifts}, and the {Bimodal} {Color} {Distribution} of {Galaxies} out to z {\textasciitilde} 3},
	volume = {735},
	issn = {0004-637X},
	shorttitle = {The {NEWFIRM} {Medium}-band {Survey}},
	url = {https://ui.adsabs.harvard.edu/abs/2011ApJ...735...86W/abstract},
	doi = {10.1088/0004-637X/735/2/86},
	language = {en},
	number = {2},
	urldate = {2025-11-25},
	journal = {The Astrophysical Journal},
	author = {Whitaker, Katherine E. and Labbé, Ivo and van Dokkum, Pieter G. and Brammer, Gabriel and Kriek, Mariska and Marchesini, Danilo and Quadri, Ryan F. and Franx, Marijn and Muzzin, Adam and Williams, Rik J. and Bezanson, Rachel and Illingworth, Garth D. and Lee, Kyoung-Soo and Lundgren, Britt and Nelson, Erica J. and Rudnick, Gregory and Tal, Tomer and Wake, David A.},
	month = jul,
	year = {2011},
	pages = {86},
}

@article{guo_candels_2013,
	title = {{CANDELS} {Multi}-wavelength {Catalogs}: {Source} {Detection} and {Photometry} in the {GOODS}-{South} {Field}},
	volume = {207},
	issn = {0067-0049},
	shorttitle = {{CANDELS} {Multi}-wavelength {Catalogs}},
	url = {https://ui.adsabs.harvard.edu/abs/2013ApJS..207...24G/abstract},
	doi = {10.1088/0067-0049/207/2/24},
	language = {en},
	number = {2},
	urldate = {2025-11-25},
	journal = {The Astrophysical Journal Supplement Series},
	author = {Guo, Yicheng and Ferguson, Henry C. and Giavalisco, Mauro and Barro, Guillermo and Willner, S. P. and Ashby, Matthew L. N. and Dahlen, Tomas and Donley, Jennifer L. and Faber, Sandra M. and Fontana, Adriano and Galametz, Audrey and Grazian, Andrea and Huang, Kuang-Han and Kocevski, Dale D. and Koekemoer, Anton M. and Koo, David C. and McGrath, Elizabeth J. and Peth, Michael and Salvato, Mara and Wuyts, Stijn and Castellano, Marco and Cooray, Asantha R. and Dickinson, Mark E. and Dunlop, James S. and Fazio, G. G. and Gardner, Jonathan P. and Gawiser, Eric and Grogin, Norman A. and Hathi, Nimish P. and Hsu, Li-Ting and Lee, Kyoung-Soo and Lucas, Ray A. and Mobasher, Bahram and Nandra, Kirpal and Newman, Jeffery A. and van der Wel, Arjen},
	month = aug,
	year = {2013},
	pages = {24},
}

@article{chiang_galaxy_2017,
	title = {Galaxy {Protoclusters} as {Drivers} of {Cosmic} {Star} {Formation} {History} in the {First} 2 {Gyr}},
	volume = {844},
	issn = {0004-637X},
	url = {https://ui.adsabs.harvard.edu/abs/2017ApJ...844L..23C/abstract},
	doi = {10.3847/2041-8213/aa7e7b},
	language = {en},
	number = {2},
	urldate = {2025-11-25},
	journal = {The Astrophysical Journal},
	author = {Chiang, Yi-Kuan and Overzier, Roderik A. and Gebhardt, Karl and Henriques, Bruno},
	month = aug,
	year = {2017},
	pages = {L23},
}

@article{simmonds_bursting_2025,
	title = {Bursting at the seams: the star-forming main sequence and its scatter at z=3-9 using {NIRCam} photometry from {JADES}},
	issn = {0035-8711},
	shorttitle = {Bursting at the seams},
	url = {https://ui.adsabs.harvard.edu/abs/2025MNRAS.tmp.1832S/abstract},
	doi = {10.1093/mnras/staf1950},
	language = {en},
	urldate = {2025-11-25},
	journal = {Monthly Notices of the Royal Astronomical Society},
	author = {Simmonds, C. and Tacchella, S. and Curtis-Lake, W. McClymont E. and D'Eugenio, F. and Hainline, K. and Johnson, B. D. and Kravtsov, A. and PuskÃ¡s, D. and Robertson, B. and Stoffers, A. and Willott, C. and Baker, W. M. and Belokurov, V. A. and Bhatawdekar, R. and Bunker, A. J. and Carniani, S. and Chevallard, J. and Curti, M. and Duan, Q. and Helton, J. M. and Ji, Z. and Looser, T. J. and Maiolino, R. and Maseda, M. V. and Shivaei, I. and Williams, C. C.},
	month = nov,
	year = {2025},
}

@article{pasha_pysersic_2023,
	title = {pysersic: {A} {Python} package for determining galaxy structural properties via {Bayesian} inference, accelerated with jax},
	volume = {8},
	shorttitle = {pysersic},
	url = {https://ui.adsabs.harvard.edu/abs/2023JOSS....8.5703P/abstract},
	doi = {10.21105/joss.05703},
	language = {en},
	number = {89},
	urldate = {2025-11-25},
	journal = {The Journal of Open Source Software},
	author = {Pasha, Imad and Miller, Tim B.},
	month = sep,
	year = {2023},
	pages = {5703},
}

@article{meidt_reconstructing_2014,
	title = {Reconstructing the {Stellar} {Mass} {Distributions} of {Galaxies} {Using} {S}{\textless}{SUP}{\textgreater}4{\textless}/{SUP}{\textgreater}{G} {IRAC} 3.6 and 4.5 μm {Images}. {II}. {The} {Conversion} from {Light} to {Mass}},
	volume = {788},
	issn = {0004-637X},
	url = {https://ui.adsabs.harvard.edu/abs/2014ApJ...788..144M/abstract},
	doi = {10.1088/0004-637X/788/2/144},
	language = {en},
	number = {2},
	urldate = {2025-11-25},
	journal = {The Astrophysical Journal},
	author = {Meidt, Sharon E. and Schinnerer, Eva and van de Ven, Glenn and Zaritsky, Dennis and Peletier, Reynier and Knapen, Johan H. and Sheth, Kartik and Regan, Michael and Querejeta, Miguel and Muñoz-Mateos, Juan-Carlos and Kim, Taehyun and Hinz, Joannah L. and Gil de Paz, Armando and Athanassoula, E. and Bosma, Albert and Buta, Ronald J. and Cisternas, Mauricio and Ho, Luis C. and Holwerda, Benne and Skibba, Ramin and Laurikainen, E. and Salo, H. and Gadotti, D. A. and Laine, Jarkko and Erroz-Ferrer, S. and Comerón, Sébastien and Menéndez-Delmestre, K. and Seibert, M. and Mizusawa, T.},
	month = jun,
	year = {2014},
	pages = {144},
}

@article{carlsten_structures_2021,
	title = {Structures of {Dwarf} {Satellites} of {Milky} {Way}-like {Galaxies}: {Morphology}, {Scaling} {Relations}, and {Intrinsic} {Shapes}},
	volume = {922},
	issn = {0004-637X},
	shorttitle = {Structures of {Dwarf} {Satellites} of {Milky} {Way}-like {Galaxies}},
	url = {https://ui.adsabs.harvard.edu/abs/2021ApJ...922..267C/abstract},
	doi = {10.3847/1538-4357/ac2581},
	language = {en},
	number = {2},
	urldate = {2025-11-25},
	journal = {The Astrophysical Journal},
	author = {Carlsten, Scott G. and Greene, Jenny E. and Greco, Johnny P. and Beaton, Rachael L. and Kado-Fong, Erin},
	month = dec,
	year = {2021},
	pages = {267},
}

@article{kuchner_effects_2017,
	title = {The effects of the cluster environment on the galaxy mass-size relation in {MACS} {J1206}.2-0847},
	volume = {604},
	issn = {0004-6361},
	url = {https://ui.adsabs.harvard.edu/abs/2017A&A...604A..54K/abstract},
	doi = {10.1051/0004-6361/201630252},
	language = {en},
	urldate = {2025-11-25},
	journal = {Astronomy and Astrophysics},
	author = {Kuchner, U. and Ziegler, B. and Verdugo, M. and Bamford, S. and Häußler, B.},
	month = aug,
	year = {2017},
	pages = {A54},
}

@article{dimauro_coincidence_2022,
	title = {Coincidence between morphology and star formation activity through cosmic time: the impact of the bulge growth},
	volume = {513},
	issn = {0035-8711},
	shorttitle = {Coincidence between morphology and star formation activity through cosmic time},
	url = {https://ui.adsabs.harvard.edu/abs/2022MNRAS.513..256D/abstract},
	doi = {10.1093/mnras/stac884},
	language = {en},
	number = {1},
	urldate = {2025-11-24},
	journal = {Monthly Notices of the Royal Astronomical Society},
	author = {Dimauro, Paola and Daddi, Emanuele and Shankar, Francesco and Cattaneo, Andrea and Huertas-Company, Marc and Bernardi, Mariangela and Caro, Fernando and Dupke, Renato and Häußler, Boris and Johnston, Evelyn and Cortesi, Arianna and Mei, Simona and Peletier, Reynier},
	month = jun,
	year = {2022},
	pages = {256--281},
}

@article{george_deep_2023,
	title = {Deep optical imaging of star-forming blue early-type galaxies. {Color} map structures and faint features indicative of recent mergers},
	volume = {678},
	issn = {0004-6361},
	url = {https://ui.adsabs.harvard.edu/abs/2023A&A...678A..10G/abstract},
	doi = {10.1051/0004-6361/202245621},
	language = {en},
	urldate = {2025-11-24},
	journal = {Astronomy and Astrophysics},
	author = {George, Koshy},
	month = oct,
	year = {2023},
	pages = {A10},
}

@article{george_structural_2017,
	title = {Structural analysis of star-forming blue early-type galaxies. {Merger}-driven star formation in elliptical galaxies},
	volume = {598},
	issn = {0004-6361},
	url = {https://ui.adsabs.harvard.edu/abs/2017A&A...598A..45G/abstract},
	doi = {10.1051/0004-6361/201629667},
	language = {en},
	urldate = {2025-11-24},
	journal = {Astronomy and Astrophysics},
	author = {George, Koshy},
	month = feb,
	year = {2017},
	pages = {A45},
}

@article{nolan_young_2007,
	title = {Young stellar populations in early-type galaxies in the {Sloan} {Digital} {Sky} {Survey}},
	volume = {375},
	issn = {0035-8711},
	url = {https://ui.adsabs.harvard.edu/abs/2007MNRAS.375..381N/abstract},
	doi = {10.1111/j.1365-2966.2006.11326.x},
	language = {en},
	number = {1},
	urldate = {2025-11-24},
	journal = {Monthly Notices of the Royal Astronomical Society},
	author = {Nolan, Louisa A. and Raychaudhury, Somak and Kabán, Ata},
	month = feb,
	year = {2007},
	pages = {381--387},
}

@article{kollmeier_sloan_2025,
	title = {Sloan {Digital} {Sky} {Survey}-{V}: {Pioneering} {Panoptic} {Spectroscopy}},
	shorttitle = {Sloan {Digital} {Sky} {Survey}-{V}},
	url = {https://ui.adsabs.harvard.edu/abs/2025arXiv250706989K/abstract},
	doi = {10.48550/arXiv.2507.06989},
	language = {en},
	urldate = {2025-11-24},
	journal = {arXiv e-prints},
	author = {Kollmeier, Juna A. and Rix, Hans-Walter and Aerts, Conny and Aird, James and Alfaro, Pablo Vera and Almeida, Andrés and Anderson, Scott F. and Jiménez Arranz, Óscar and Arseneau, Stefan M. and Assef, Roberto and Aviram, Shir and Aydar, Catarina and Badenes, Carles and Bandyopadhyay, Avrajit and Barger, Kat and Barkhouser, Robert H. and Bauer, Franz E. and Bender, Chad and Besser, Felipe and Bhattarai, Binod and Bilgi, Pavaman and Bird, Jonathan and Bizyaev, Dmitry and Blanc, Guillermo A. and Blanton, Michael R. and Bochanski, John and Bovy, Jo and Brandon, Christopher and Brandt, William Nielsen and Brownstein, Joel R. and Buchner, Johannes and Burchett, Joseph N. and Carlberg, Joleen and Casey, Andrew R. and Castaneda-Carlos, Lesly and Chakraborty, Priyanka and Chanamé, Julio and Chandra, Vedant and Cherinka, Brian and Chilingarian, Igor and Comparat, Johan and Cosens, Maren and Covey, Kevin and Crane, Jeffrey D. and Crumpler, Nicole R. and Cunha, Katia and Cunningham, Tim and Dai, Xinyu and Darling, Jeremy and Davidson, James W. and Davis, Megan C. and De Lee, Nathan and Deacon, Niall and Méndez Delgado, José Eduardo and Demasi, Sebastian and Demianenko, Mariia and Derwent, Mark and D'Onghia, Elena and Di Mille, Francesco and Dias, Bruno and Donor, John and Drory, Niv and Dwelly, Tom and Egorov, Oleg and Egorova, Evgeniya and El-Badry, Kareem and Engelman, Mike and Eracleous, Mike and Fan, Xiaohui and Farr, Emily and Fries, Logan and Frinchaboy, Peter and Froning, Cynthia S. and Gänsicke, Boris T. and García, Pablo and Gelfand, Joseph and Gentile Fusillo, Nicola Pietro and Glover, Simon and Grabowski, Katie and Grebel, Eva K. and Green, Paul J. and Grier, Catherine and Gupta, Pramod and Gray, Aidan C. and Häberle, Maximilian and Hall, Patrick B. and Hammond, Randolph P. and Hawkins, Keith and Harding, Albert C. and Hegedűs, Viola and Herbst, Tom and Hermes, J. J. and Rodríguez Hidalgo, Paola and Hilder, Thomas and Hogg, David W. and Holtzman, Jon A. and Horta, Danny and Huang, Yang and Hwang, Hsiang-Chih and Ibarra-Medel, Hector Javier and Imig, Julie and Inight, Keith and Jana, Arghajit and Ji, Alexander P. and Jofre, Paula and Johns, Matt and Johnson, Jennifer and Johnson, James W. and Johnston, Evelyn J. and Jones, Amy M. and Katkov, Ivan and Koekemoer, Anton M. and Kounkel, Marina and Kreckel, Kathryn and Krishnarao, Dhanesh and Krumpe, Mirko and Kumari, Nimisha and Kupfer, Thomas and Lacerna, Ivan and Laporte, Chervin and Lepine, Sebastien and Li, Jing and Liu, Xin and Loebman, Sarah and Long, Knox and Roman-Lopes, Alexandre and Lu, Yuxi and Majewski, Steven Raymond and Maoz, Dan and McKinnon, Kevin A. and Medan, Ilija and Merloni, Andrea and Minniti, Dante and Morrison, Sean and Myers, Natalie and Mészáros, Szabolcs and Nandra, Kirpal and Nayak, Prasanta K. and Ness, Melissa K. and Nidever, David L. and O'Brien, Thomas and Oeur, Micah and Oravetz, Audrey and Oravetz, Daniel and Otto, Jonah and Adamane Pallathadka, Gautham and Palunas, Povilas and Pan, Kaike and Pappalardo, Daniel and Pandey, Rakesh and Negrete Peñaloza, Castalia Alenka and Pinsonneault, Marc H. and Pogge, Richard W. and Taghizadeh Popp, Manuchehr and Price-Whelan, Adrian M. and Pulatova, Nadiia and Qiu, Dan and Ramirez, Solange and Rankine, Amy and Ricci, Claudio and Runnoe, Jessie C. and Sanchez, Sebastian and Salvato, Mara and Sattler, Natascha and Saydjari, Andrew K. and Sayres, Conor and Schlaufman, Kevin C. and Schneider, Donald P. and Schreiber, Matthias R. and Schwope, Axel and Serna, Javier and Shen, Yue and Sifón, Cristóbal and Singh, Amrita and Sinha, Amaya and Smee, Stephen and Song, Ying-Yi and Souto, Diogo and Stassun, Keivan G. and Steinmetz, Matthias and Stone-Martinez, Alexander and Stringfellow, Guy and Stutz, Amelia and José and Sá and nchez-Gallego and Tan, Jonathan C. and Tayar, Jamie and Thai, Riley and Thakar, Ani and Ting, Yuan-Sen and Tkachenko, Andrew and Tovmasian, Gagik and Trakhtenbrot, Benny and Fernández-Trincado, José G. and Troup, Nicholas and Trump, Jonathan and Tuttle, Sarah and van der Marel, Roeland P. and Villanova, Sandro and Villaseñor, Jaime and Wachter, Stefanie and Way, Zachary and Weijmans, Anne-Marie and Weinberg, David and Wheeler, Adam and Wilson, John and Wiggins, Alessa I. and Wong, Tony and Wu, Qiaoya and Wylezalek, Dominika and Xue, Xiang-Xiang and Yang, Qian and Zakamska, Nadia and Zari, Eleonora and Zasowski, Gail and Zeltyn, Grisha and Zucker, Catherine and Román Zúñiga, Carlos G. and Zermeño, Rodolfo de J.},
	month = jul,
	year = {2025},
	pages = {arXiv:2507.06989},
}

@article{george_revealing_2015,
	title = {Revealing the nature of star forming blue early-type galaxies at low redshift},
	volume = {583},
	issn = {0004-6361},
	url = {https://ui.adsabs.harvard.edu/abs/2015A&A...583A.103G/abstract},
	doi = {10.1051/0004-6361/201424826},
	language = {en},
	urldate = {2025-11-24},
	journal = {Astronomy and Astrophysics},
	author = {George, Koshy and Zingade, Kshama},
	month = nov,
	year = {2015},
	pages = {A103},
}

@article{mei_morphology-density_2023,
	title = {Morphology-density relation, quenching, and mergers in {CARLA} clusters and protoclusters at 1.4 \&lt; z \&lt; 2.8},
	volume = {670},
	issn = {0004-6361},
	url = {https://ui.adsabs.harvard.edu/abs/2023A&A...670A..58M/abstract},
	doi = {10.1051/0004-6361/202243551},
	language = {en},
	urldate = {2025-11-24},
	journal = {Astronomy and Astrophysics},
	author = {Mei, Simona and Hatch, Nina A. and Amodeo, Stefania and Afanasiev, Anton V. and De Breuck, Carlos and Stern, Daniel and Cooke, Elizabeth A. and Gonzalez, Anthony H. and Noirot, Gaël and Rettura, Alessandro and Seymour, Nick and Stanford, Spencer A. and Vernet, Joël and Wylezalek, Dominika},
	month = feb,
	year = {2023},
	pages = {A58},
}

@article{behroozi_average_2013,
	title = {The {Average} {Star} {Formation} {Histories} of {Galaxies} in {Dark} {Matter} {Halos} from z = 0-8},
	volume = {770},
	issn = {0004-637X},
	url = {https://ui.adsabs.harvard.edu/abs/2013ApJ...770...57B/abstract},
	doi = {10.1088/0004-637X/770/1/57},
	language = {en},
	number = {1},
	urldate = {2025-11-24},
	journal = {The Astrophysical Journal},
	author = {Behroozi, Peter S. and Wechsler, Risa H. and Conroy, Charlie},
	month = jun,
	year = {2013},
	pages = {57},
}

@article{smail_obscured_2024,
	title = {Obscured star formation in clusters at z = 1.6-2.0: massive galaxy formation and the reversal of the star formation-density relation},
	volume = {529},
	issn = {0035-8711},
	shorttitle = {Obscured star formation in clusters at z = 1.6-2.0},
	url = {https://ui.adsabs.harvard.edu/abs/2024MNRAS.529.2290S/abstract},
	doi = {10.1093/mnras/stae692},
	language = {en},
	number = {3},
	urldate = {2025-11-24},
	journal = {Monthly Notices of the Royal Astronomical Society},
	author = {Smail, Ian},
	month = apr,
	year = {2024},
	pages = {2290--2308},
}

@article{cooke_submillimetre_2019,
	title = {The submillimetre view of massive clusters at z ∼ 0.8-1.6},
	volume = {486},
	issn = {0035-8711},
	url = {https://ui.adsabs.harvard.edu/abs/2019MNRAS.486.3047C/abstract},
	doi = {10.1093/mnras/stz955},
	language = {en},
	number = {3},
	urldate = {2025-11-24},
	journal = {Monthly Notices of the Royal Astronomical Society},
	author = {Cooke, E. A. and Smail, Ian and Stach, S. M. and Swinbank, A. M. and Bower, R. G. and Chen, Chian-Chou and Koyama, Y. and Thomson, A. P.},
	month = jul,
	year = {2019},
	pages = {3047--3058},
}

@article{popesso_evolution_2012,
	title = {The evolution of the star formation activity per halo mass up to redshift {\textasciitilde}1.6 as seen by {Herschel}},
	volume = {537},
	issn = {0004-6361},
	url = {https://ui.adsabs.harvard.edu/abs/2012A&A...537A..58P/abstract},
	doi = {10.1051/0004-6361/201117973},
	language = {en},
	urldate = {2025-11-24},
	journal = {Astronomy and Astrophysics},
	author = {Popesso, P. and Biviano, A. and Rodighiero, G. and Baronchelli, I. and Salvato, M. and Saintonge, A. and Finoguenov, A. and Magnelli, B. and Gruppioni, C. and Pozzi, F. and Lutz, D. and Elbaz, D. and Altieri, B. and Andreani, P. and Aussel, H. and Berta, S. and Capak, P. and Cava, A. and Cimatti, A. and Coia, D. and Daddi, E. and Dannerbauer, H. and Dickinson, M. and Dasyra, K. and Fadda, D. and Förster Schreiber, N. and Genzel, R. and Hwang, H. S. and Kartaltepe, J. and Ilbert, O. and Le Floch, E. and Leiton, R. and Magdis, G. and Nordon, R. and Patel, S. and Poglitsch, A. and Riguccini, L. and Sanchez Portal, M. and Shao, L. and Tacconi, L. and Tomczak, A. and Tran, K. and Valtchanov, I.},
	month = jan,
	year = {2012},
	pages = {A58},
}

@article{andrews_galaxy_2025,
	title = {Galaxy populations in protoclusters at cosmic noon},
	volume = {698},
	issn = {0004-6361},
	url = {https://ui.adsabs.harvard.edu/abs/2025A&A...698A.280A/abstract},
	doi = {10.1051/0004-6361/202452628},
	language = {en},
	urldate = {2025-11-24},
	journal = {Astronomy and Astrophysics},
	author = {Andrews, Moira and Artale, M. Celeste and Kumar, Ankit and Lee, Kyoung-Soo and Florek, Tess and Anand, Kaustub and Cerdosino, Candela and Ciardullo, Robin and Firestone, Nicole and Gawiser, Eric and Gronwall, Caryl and Guaita, Lucia and Hong, Sungryong and Hwang, Ho Seong and Lee, Jaehyun and Lee, Seong-Kook and Padilla, Nelson and Park, Jaehong and Popescu, Roxana and Ramakrishnan, Vandana and Song, Hyunmi and Vivanco Cádiz, F. and Vogelsberger, Mark},
	month = jun,
	year = {2025},
	pages = {A280},
}

@article{alberts_star_2016,
	title = {Star {Formation} and {AGN} {Activity} in {Galaxy} {Clusters} from z=1-2: a {Multi}-{Wavelength} {Analysis} {Featuring} {Herschel}/{PACS}},
	volume = {825},
	issn = {0004-637X},
	shorttitle = {Star {Formation} and {AGN} {Activity} in {Galaxy} {Clusters} from z=1-2},
	url = {https://ui.adsabs.harvard.edu/abs/2016ApJ...825...72A/abstract},
	doi = {10.3847/0004-637X/825/1/72},
	language = {en},
	number = {1},
	urldate = {2025-11-24},
	journal = {The Astrophysical Journal},
	author = {Alberts, Stacey and Pope, Alexandra and Brodwin, Mark and Chung, Sun Mi and Cybulski, Ryan and Dey, Arjun and Eisenhardt, Peter R. M. and Galametz, Audrey and Gonzalez, Anthony H. and Jannuzi, Buell T. and Stanford, S. Adam and Snyder, Gregory F. and Stern, Daniel and Zeimann, Gregory R.},
	month = jul,
	year = {2016},
	pages = {72},
}

@article{whitaker_star_2012,
	title = {The {Star} {Formation} {Mass} {Sequence} {Out} to z = 2.5},
	volume = {754},
	issn = {0004-637X},
	url = {https://ui.adsabs.harvard.edu/abs/2012ApJ...754L..29W/abstract},
	doi = {10.1088/2041-8205/754/2/L29},
	language = {en},
	number = {2},
	urldate = {2025-11-18},
	journal = {The Astrophysical Journal},
	author = {Whitaker, Katherine E. and van Dokkum, Pieter G. and Brammer, Gabriel and Franx, Marijn},
	month = aug,
	year = {2012},
	pages = {L29},
}

@article{speagle_highly_2014,
	title = {A {Highly} {Consistent} {Framework} for the {Evolution} of the {Star}-{Forming} "{Main} {Sequence}" from z {\textasciitilde} 0-6},
	volume = {214},
	issn = {0067-0049},
	url = {https://ui.adsabs.harvard.edu/abs/2014ApJS..214...15S/abstract},
	doi = {10.1088/0067-0049/214/2/15},
	language = {en},
	number = {2},
	urldate = {2025-11-18},
	journal = {The Astrophysical Journal Supplement Series},
	author = {Speagle, J. S. and Steinhardt, C. L. and Capak, P. L. and Silverman, J. D.},
	month = oct,
	year = {2014},
	pages = {15},
}

@article{brinchmann_physical_2004,
	title = {The physical properties of star-forming galaxies in the low-redshift {Universe}},
	volume = {351},
	issn = {0035-8711},
	url = {https://ui.adsabs.harvard.edu/abs/2004MNRAS.351.1151B/abstract},
	doi = {10.1111/j.1365-2966.2004.07881.x},
	language = {en},
	number = {4},
	urldate = {2025-11-18},
	journal = {Monthly Notices of the Royal Astronomical Society},
	author = {Brinchmann, J. and Charlot, S. and White, S. D. M. and Tremonti, C. and Kauffmann, G. and Heckman, T. and Brinkmann, J.},
	month = jul,
	year = {2004},
	pages = {1151--1179},
}

@article{leja_new_2022,
	title = {A {New} {Census} of the 0.2 \&lt; z \&lt; 3.0 {Universe}. {II}. {The} {Star}-forming {Sequence}},
	volume = {936},
	issn = {0004-637X},
	url = {https://ui.adsabs.harvard.edu/abs/2022ApJ...936..165L/abstract},
	doi = {10.3847/1538-4357/ac887d},
	language = {en},
	number = {2},
	urldate = {2025-11-18},
	journal = {The Astrophysical Journal},
	author = {Leja, Joel and Speagle, Joshua S. and Ting, Yuan-Sen and Johnson, Benjamin D. and Conroy, Charlie and Whitaker, Katherine E. and Nelson, Erica J. and van Dokkum, Pieter and Franx, Marijn},
	month = sep,
	year = {2022},
	pages = {165},
}

@article{papovich_effects_2018,
	title = {The {Effects} of {Environment} on the {Evolution} of the {Galaxy} {Stellar} {Mass} {Function}},
	volume = {854},
	issn = {0004-637X},
	url = {https://ui.adsabs.harvard.edu/abs/2018ApJ...854...30P/abstract},
	doi = {10.3847/1538-4357/aaa766},
	language = {en},
	number = {1},
	urldate = {2025-11-18},
	journal = {The Astrophysical Journal},
	author = {Papovich, Casey and Kawinwanichakij, Lalitwadee and Quadri, Ryan F. and Glazebrook, Karl and Labbé, Ivo and Tran, Kim-Vy H. and Forrest, Ben and Kacprzak, Glenn G. and Spitler, Lee R. and Straatman, Caroline M. S. and Tomczak, Adam R.},
	month = feb,
	year = {2018},
	pages = {30},
}

@article{tomczak_glimpsing_2017,
	title = {Glimpsing the imprint of local environment on the galaxy stellar mass function},
	volume = {472},
	issn = {0035-8711},
	url = {https://ui.adsabs.harvard.edu/abs/2017MNRAS.472.3512T/abstract},
	doi = {10.1093/mnras/stx2245},
	language = {en},
	number = {3},
	urldate = {2025-11-18},
	journal = {Monthly Notices of the Royal Astronomical Society},
	author = {Tomczak, Adam R. and Lemaux, Brian C. and Lubin, Lori M. and Gal, Roy R. and Wu, Po-Feng and Holden, Bradford and Kocevski, Dale D. and Mei, Simona and Pelliccia, Debora and Rumbaugh, Nicholas and Shen, Lu},
	month = dec,
	year = {2017},
	pages = {3512--3531},
}

@article{kawinwanichakij_effect_2017,
	title = {Effect of {Local} {Environment} and {Stellar} {Mass} on {Galaxy} {Quenching} and {Morphology} at 0.5 \&lt; z \&lt; 2.0},
	volume = {847},
	issn = {0004-637X},
	url = {https://ui.adsabs.harvard.edu/abs/2017ApJ...847..134K/abstract},
	doi = {10.3847/1538-4357/aa8b75},
	language = {en},
	number = {2},
	urldate = {2025-11-18},
	journal = {The Astrophysical Journal},
	author = {Kawinwanichakij, Lalitwadee and Papovich, Casey and Quadri, Ryan F. and Glazebrook, Karl and Kacprzak, Glenn G. and Allen, Rebecca J. and Bell, Eric F. and Croton, Darren J. and Dekel, Avishai and Ferguson, Henry C. and Forrest, Ben and Grogin, Norman A. and Guo, Yicheng and Kocevski, Dale D. and Koekemoer, Anton M. and Labbé, Ivo and Lucas, Ray A. and Nanayakkara, Themiya and Spitler, Lee R. and Straatman, Caroline M. S. and Tran, Kim-Vy H. and Tomczak, Adam and van Dokkum, Pieter},
	month = oct,
	year = {2017},
	pages = {134},
}

@article{baldry_galaxy_2006,
	title = {Galaxy bimodality versus stellar mass and environment},
	volume = {373},
	issn = {0035-8711},
	url = {https://ui.adsabs.harvard.edu/abs/2006MNRAS.373..469B/abstract},
	doi = {10.1111/j.1365-2966.2006.11081.x},
	language = {en},
	number = {2},
	urldate = {2025-11-18},
	journal = {Monthly Notices of the Royal Astronomical Society},
	author = {Baldry, I. K. and Balogh, M. L. and Bower, R. G. and Glazebrook, K. and Nichol, R. C. and Bamford, S. P. and Budavari, T.},
	month = dec,
	year = {2006},
	pages = {469--483},
}

@article{kauffmann_environmental_2004,
	title = {The environmental dependence of the relations between stellar mass, structure, star formation and nuclear activity in galaxies},
	volume = {353},
	issn = {0035-8711},
	url = {https://ui.adsabs.harvard.edu/abs/2004MNRAS.353..713K/abstract},
	doi = {10.1111/j.1365-2966.2004.08117.x},
	language = {en},
	number = {3},
	urldate = {2025-11-18},
	journal = {Monthly Notices of the Royal Astronomical Society},
	author = {Kauffmann, Guinevere and White, Simon D. M. and Heckman, Timothy M. and Ménard, Brice and Brinchmann, Jarle and Charlot, Stéphane and Tremonti, Christy and Brinkmann, Jon},
	month = sep,
	year = {2004},
	pages = {713--731},
}

@article{balogh_bimodal_2004,
	title = {The {Bimodal} {Galaxy} {Color} {Distribution}: {Dependence} on {Luminosity} and {Environment}},
	volume = {615},
	issn = {0004-637X},
	shorttitle = {The {Bimodal} {Galaxy} {Color} {Distribution}},
	url = {https://ui.adsabs.harvard.edu/abs/2004ApJ...615L.101B/abstract},
	doi = {10.1086/426079},
	language = {en},
	number = {2},
	urldate = {2025-11-18},
	journal = {The Astrophysical Journal},
	author = {Balogh, Michael L. and Baldry, Ivan K. and Nichol, Robert and Miller, Chris and Bower, Richard and Glazebrook, Karl},
	month = nov,
	year = {2004},
	pages = {L101--L104},
}

@article{shen_ngdeep_2024,
	title = {{NGDEEP} {Epoch} 1: {Spatially} {Resolved} {Hα} {Observations} of {Disk} and {Bulge} {Growth} in {Star}-forming {Galaxies} at z ∼ 0.6–2.2 from {JWST} {NIRISS} {Slitless} {Spectroscopy}},
	volume = {963},
	issn = {0004-637X},
	shorttitle = {{NGDEEP} {Epoch} 1},
	url = {https://ui.adsabs.harvard.edu/abs/2024ApJ...963L..49S/abstract},
	doi = {10.3847/2041-8213/ad28bd},
	language = {en},
	number = {2},
	urldate = {2025-11-18},
	journal = {The Astrophysical Journal},
	author = {Shen, Lu and Papovich, Casey and Matharu, Jasleen and Pirzkal, Nor and Hu, Weida and Backhaus, Bren E. and Bagley, Micaela B. and Cheng, Yingjie and Cleri, Nikko J. and Finkelstein, Steven L. and Huertas-Company, Marc and Giavalisco, Mauro and Grogin, Norman A. and Jung, Intae and Kartaltepe, Jeyhan S. and Koekemoer, Anton M. and Lotz, Jennifer M. and Maseda, Michael V. and Pérez-González, Pablo G. and Rothberg, Barry and Simons, Raymond C. and Tacchella, Sandro and Williams, Christina C. and Yung, L. Y. Aaron},
	month = mar,
	year = {2024},
	pages = {L49},
}

@article{shen_ngdeep_2025,
	title = {{NGDEEP}: {The} {Star} {Formation} and {Ionization} {Properties} of {Galaxies} at 1.7 \&lt; z \&lt; 3.4},
	volume = {980},
	issn = {0004-637X},
	shorttitle = {{NGDEEP}},
	url = {https://ui.adsabs.harvard.edu/abs/2025ApJ...980L..45S/abstract},
	doi = {10.3847/2041-8213/adb28d},
	language = {en},
	number = {2},
	urldate = {2025-11-18},
	journal = {The Astrophysical Journal},
	author = {Shen, Lu and Papovich, Casey and Matharu, Jasleen and Pirzkal, Nor and Hu, Weida and Berg, Danielle A. and Bagley, Micaela B. and Backhaus, Bren E. and Cleri, Nikko J. and Dickinson, Mark and Finkelstein, Steven L. and Hathi, Nimish P. and Huertas-Company, Marc and Hutchison, Taylor A. and Giavalisco, Mauro and Grogin, Norman A. and Jaskot, Anne E. and Jung, Intae and Kartaltepe, Jeyhan S. and Koekemoer, Anton M. and Lotz, Jennifer M. and Pérez-González, Pablo G. and Rothberg, Barry and Simons, Raymond C. and Vanderhoof, Brittany N. and Yung, L. Y. Aaron},
	month = feb,
	year = {2025},
	pages = {L45},
}

@article{pan_uncovermegascience_2025,
	title = {{UNCOVER}/{MegaScience}: {No} {Evidence} of {Environmental} {Quenching} in a z ∼ 2.6 {Protocluster}},
	volume = {990},
	issn = {0004-637X},
	shorttitle = {{UNCOVER}/{MegaScience}},
	url = {https://ui.adsabs.harvard.edu/abs/2025ApJ...990L..24P/abstract},
	doi = {10.3847/2041-8213/adf7ab},
	language = {en},
	number = {1},
	urldate = {2025-11-18},
	journal = {The Astrophysical Journal},
	author = {Pan, Richard and Suess, Katherine A. and Marchesini, Danilo and Wang, Bingjie and Leja, Joel and Cutler, Sam E. and Whitaker, Katherine E. and Bezanson, Rachel and Price, Sedona H. and Furtak, Lukas J. and Weaver, John R. and Labbé, Ivo and Brammer, Gabriel and Zhang, Yunchong and Dayal, Pratika and Feldmann, Robert and Greene, Jenny E. and Glazebrook, Karl and Miller, Tim B. and Mitsuhashi, Ikki and Muzzin, Adam and Nanayakkara, Themiya and Nelson, Erica J. and Setton, David J. and Zitrin, Adi},
	month = sep,
	year = {2025},
	pages = {L24},
}

@article{rinaldi_galaxy_2022,
	title = {The {Galaxy} {Starburst}/{Main}-sequence {Bimodality} over {Five} {Decades} in {Stellar} {Mass} at z ≈ 3-6.5},
	volume = {930},
	issn = {0004-637X},
	url = {https://ui.adsabs.harvard.edu/abs/2022ApJ...930..128R/abstract},
	doi = {10.3847/1538-4357/ac5d39},
	language = {en},
	number = {2},
	urldate = {2025-11-18},
	journal = {The Astrophysical Journal},
	author = {Rinaldi, Pierluigi and Caputi, Karina I. and van Mierlo, Sophie E. and Ashby, Matthew L. N. and Caminha, Gabriel B. and Iani, Edoardo},
	month = may,
	year = {2022},
	pages = {128},
}

@article{hamadouche_combined_2022,
	title = {A combined {VANDELS} and {LEGA}-{C} study: the evolution of quiescent galaxy size, stellar mass, and age from z = 0.6 to z = 1.3},
	volume = {512},
	issn = {0035-8711},
	shorttitle = {A combined {VANDELS} and {LEGA}-{C} study},
	url = {https://ui.adsabs.harvard.edu/abs/2022MNRAS.512.1262H/abstract},
	doi = {10.1093/mnras/stac535},
	language = {en},
	number = {1},
	urldate = {2025-11-18},
	journal = {Monthly Notices of the Royal Astronomical Society},
	author = {Hamadouche, M. L. and Carnall, A. C. and McLure, R. J. and Dunlop, J. S. and McLeod, D. J. and Cullen, F. and Begley, R. and Bolzonella, M. and Buitrago, F. and Castellano, M. and Cucciati, O. and Fontana, A. and Gargiulo, A. and Moresco, M. and Pozzetti, L. and Zamorani, G.},
	month = may,
	year = {2022},
	pages = {1262--1274},
}

@article{binggeli_balmer_2019,
	title = {Balmer breaks in simulated galaxies at z \&gt; 6},
	volume = {489},
	issn = {0035-8711},
	url = {https://ui.adsabs.harvard.edu/abs/2019MNRAS.489.3827B/abstract},
	doi = {10.1093/mnras/stz2387},
	language = {en},
	number = {3},
	urldate = {2025-11-18},
	journal = {Monthly Notices of the Royal Astronomical Society},
	author = {Binggeli, Christian and Zackrisson, Erik and Ma, Xiangcheng and Inoue, Akio K. and Vikaeus, Anton and Hashimoto, Takuya and Mawatari, Ken and Shimizu, Ikkoh and Ceverino, Daniel},
	month = nov,
	year = {2019},
	pages = {3827--3835},
}

@article{wilkins_first_2024,
	title = {First {Light} and {Reionization} {Epoch} {Simulations} ({FLARES}) - {XIV}. {The} {Balmer}/4000 Å breaks of distant galaxies},
	volume = {527},
	issn = {0035-8711},
	url = {https://ui.adsabs.harvard.edu/abs/2024MNRAS.527.7965W/abstract},
	doi = {10.1093/mnras/stad3558},
	language = {en},
	number = {3},
	urldate = {2025-11-18},
	journal = {Monthly Notices of the Royal Astronomical Society},
	author = {Wilkins, Stephen M. and Lovell, Christopher C. and Irodotou, Dimitrios and Vijayan, Aswin P. and Vikaeus, Anton and Zackrisson, Erik and Caruana, Joseph and Stanway, Elizabeth R. and Conselice, Christopher J. and Seeyave, Louise T. C. and Roper, William J. and Chworowsky, Katherine and Finkelstein, Steven L.},
	month = jan,
	year = {2024},
	pages = {7965--7973},
}

@article{bruzual_a_spectral_1983,
	title = {Spectral evolution of galaxies. {I}. {Early}-type systems.},
	volume = {273},
	issn = {0004-637X},
	url = {https://ui.adsabs.harvard.edu/abs/1983ApJ...273..105B/abstract},
	doi = {10.1086/161352},
	language = {en},
	urldate = {2025-11-18},
	journal = {The Astrophysical Journal},
	author = {Bruzual A., G.},
	month = oct,
	year = {1983},
	pages = {105--127},
}

@article{hainline_mid-infrared_2016,
	title = {Mid-infrared {Colors} of {Dwarf} {Galaxies}: {Young} {Starbursts} {Mimicking} {Active} {Galactic} {Nuclei}},
	volume = {832},
	issn = {0004-637X},
	shorttitle = {Mid-infrared {Colors} of {Dwarf} {Galaxies}},
	url = {https://ui.adsabs.harvard.edu/abs/2016ApJ...832..119H/abstract},
	doi = {10.3847/0004-637X/832/2/119},
	language = {en},
	number = {2},
	urldate = {2025-11-18},
	journal = {The Astrophysical Journal},
	author = {Hainline, Kevin N. and Reines, Amy E. and Greene, Jenny E. and Stern, Daniel},
	month = dec,
	year = {2016},
	pages = {119},
}

@article{schaye_flamingo_2023,
	title = {The {FLAMINGO} project: cosmological hydrodynamical simulations for large-scale structure and galaxy cluster surveys},
	volume = {526},
	issn = {0035-8711},
	shorttitle = {The {FLAMINGO} project},
	url = {https://ui.adsabs.harvard.edu/abs/2023MNRAS.526.4978S/abstract},
	doi = {10.1093/mnras/stad2419},
	language = {en},
	number = {4},
	urldate = {2025-11-18},
	journal = {Monthly Notices of the Royal Astronomical Society},
	author = {Schaye, Joop and Kugel, Roi and Schaller, Matthieu and Helly, John C. and Braspenning, Joey and Elbers, Willem and McCarthy, Ian G. and van Daalen, Marcel P. and Vandenbroucke, Bert and Frenk, Carlos S. and Kwan, Juliana and Salcido, Jaime and Bahé, Yannick M. and Borrow, Josh and Chaikin, Evgenii and Hahn, Oliver and Huško, Filip and Jenkins, Adrian and Lacey, Cedric G. and Nobels, Folkert S. J.},
	month = dec,
	year = {2023},
	pages = {4978--5020},
}

@article{nelson_illustristng_2019,
	title = {The {IllustrisTNG} simulations: public data release},
	volume = {6},
	shorttitle = {The {IllustrisTNG} simulations},
	url = {https://ui.adsabs.harvard.edu/abs/2019ComAC...6....2N/abstract},
	doi = {10.1186/s40668-019-0028-x},
	language = {en},
	number = {1},
	urldate = {2025-11-18},
	journal = {Computational Astrophysics and Cosmology},
	author = {Nelson, Dylan and Springel, Volker and Pillepich, Annalisa and Rodriguez-Gomez, Vicente and Torrey, Paul and Genel, Shy and Vogelsberger, Mark and Pakmor, Ruediger and Marinacci, Federico and Weinberger, Rainer and Kelley, Luke and Lovell, Mark and Diemer, Benedikt and Hernquist, Lars},
	month = may,
	year = {2019},
	pages = {2},
}

@article{dekel_galaxy_2006,
	title = {Galaxy bimodality due to cold flows and shock heating},
	volume = {368},
	issn = {0035-8711},
	url = {https://ui.adsabs.harvard.edu/abs/2006MNRAS.368....2D/abstract},
	doi = {10.1111/j.1365-2966.2006.10145.x},
	language = {en},
	number = {1},
	urldate = {2025-11-18},
	journal = {Monthly Notices of the Royal Astronomical Society},
	author = {Dekel, Avishai and Birnboim, Yuval},
	month = may,
	year = {2006},
	pages = {2--20},
}

@article{casey_massive_2015,
	title = {A {Massive}, {Distant} {Proto}-cluster at z = 2.47 {Caught} in a {Phase} of {Rapid} {Formation}?},
	volume = {808},
	issn = {0004-637X},
	url = {https://ui.adsabs.harvard.edu/abs/2015ApJ...808L..33C/abstract},
	doi = {10.1088/2041-8205/808/2/L33},
	language = {en},
	number = {2},
	urldate = {2025-11-17},
	journal = {The Astrophysical Journal},
	author = {Casey, C. M. and Cooray, A. and Capak, P. and Fu, H. and Kovac, K. and Lilly, S. and Sanders, D. B. and Scoville, N. Z. and Treister, E.},
	month = aug,
	year = {2015},
	pages = {L33},
}

@article{darvish_spectroscopic_2020,
	title = {Spectroscopic {Confirmation} of a {Coma} {Cluster} {Progenitor} at z ∼ 2.2},
	volume = {892},
	issn = {0004-637X},
	url = {https://ui.adsabs.harvard.edu/abs/2020ApJ...892....8D/abstract},
	doi = {10.3847/1538-4357/ab75c3},
	language = {en},
	number = {1},
	urldate = {2025-11-17},
	journal = {The Astrophysical Journal},
	author = {Darvish, Behnam and Scoville, Nick Z. and Martin, Christopher and Sobral, David and Mobasher, Bahram and Rettura, Alessandro and Matthee, Jorryt and Capak, Peter and Chartab, Nima and Hemmati, Shoubaneh and Masters, Daniel and Nayyeri, Hooshang and O'Sullivan, Donal and Paulino-Afonso, Ana and Sattari, Zahra and Shahidi, Abtin and Salvato, Mara and Lemaux, Brian C. and Fèvre, Olivier Le and Cucciati, Olga},
	month = mar,
	year = {2020},
	pages = {8},
}

@article{polletta_spectroscopic_2021,
	title = {Spectroscopic observations of {PHz} {G237}.01+42.50: {A} galaxy protocluster at z = 2.16 in the {Cosmos} field},
	volume = {654},
	issn = {0004-6361},
	shorttitle = {Spectroscopic observations of {PHz} {G237}.01+42.50},
	url = {https://ui.adsabs.harvard.edu/abs/2021A&A...654A.121P/abstract},
	doi = {10.1051/0004-6361/202140612},
	language = {en},
	urldate = {2025-11-17},
	journal = {Astronomy and Astrophysics},
	author = {Polletta, M. and Soucail, G. and Dole, H. and Lehnert, M. D. and Pointecouteau, E. and Vietri, G. and Scodeggio, M. and Montier, L. and Koyama, Y. and Lagache, G. and Frye, B. L. and Cusano, F. and Fumana, M.},
	month = oct,
	year = {2021},
	pages = {A121},
}

@article{miller_massive_2018,
	title = {A massive core for a cluster of galaxies at a redshift of 4.3},
	volume = {556},
	issn = {0028-0836},
	url = {https://ui.adsabs.harvard.edu/abs/2018Natur.556..469M/abstract},
	doi = {10.1038/s41586-018-0025-2},
	language = {en},
	number = {7702},
	urldate = {2025-11-17},
	journal = {Nature},
	author = {Miller, T. B. and Chapman, S. C. and Aravena, M. and Ashby, M. L. N. and Hayward, C. C. and Vieira, J. D. and Weiß, A. and Babul, A. and Béthermin, M. and Bradford, C. M. and Brodwin, M. and Carlstrom, J. E. and Chen, Chian-Chou and Cunningham, D. J. M. and De Breuck, C. and Gonzalez, A. H. and Greve, T. R. and Harnett, J. and Hezaveh, Y. and Lacaille, K. and Litke, K. C. and Ma, J. and Malkan, M. and Marrone, D. P. and Morningstar, W. and Murphy, E. J. and Narayanan, D. and Pass, E. and Perry, R. and Phadke, K. A. and Rennehan, D. and Rotermund, K. M. and Simpson, J. and Spilker, J. S. and Sreevani, J. and Stark, A. A. and Strandet, M. L. and Strom, A. L.},
	month = apr,
	year = {2018},
	pages = {469--472},
}

@article{menanteau_atacama_2012,
	title = {The {Atacama} {Cosmology} {Telescope}: {ACT}-{CL} {J0102}-4915 "{El} {Gordo}," a {Massive} {Merging} {Cluster} at {Redshift} 0.87},
	volume = {748},
	issn = {0004-637X},
	shorttitle = {The {Atacama} {Cosmology} {Telescope}},
	url = {https://ui.adsabs.harvard.edu/abs/2012ApJ...748....7M/abstract},
	doi = {10.1088/0004-637X/748/1/7},
	language = {en},
	number = {1},
	urldate = {2025-11-17},
	journal = {The Astrophysical Journal},
	author = {Menanteau, Felipe and Hughes, John P. and Sifón, Cristóbal and Hilton, Matt and González, Jorge and Infante, Leopoldo and Barrientos, L. Felipe and Baker, Andrew J. and Bond, John R. and Das, Sudeep and Devlin, Mark J. and Dunkley, Joanna and Hajian, Amir and Hincks, Adam D. and Kosowsky, Arthur and Marsden, Danica and Marriage, Tobias A. and Moodley, Kavilan and Niemack, Michael D. and Nolta, Michael R. and Page, Lyman A. and Reese, Erik D. and Sehgal, Neelima and Sievers, Jon and Spergel, David N. and Staggs, Suzanne T. and Wollack, Edward},
	month = mar,
	year = {2012},
	pages = {7},
}

@article{williamson_sunyaev-zeldovich-selected_2011,
	title = {A {Sunyaev}-{Zel}'dovich-selected {Sample} of the {Most} {Massive} {Galaxy} {Clusters} in the 2500 deg{\textless}{SUP}{\textgreater}2{\textless}/{SUP}{\textgreater} {South} {Pole} {Telescope} {Survey}},
	volume = {738},
	issn = {0004-637X},
	url = {https://ui.adsabs.harvard.edu/abs/2011ApJ...738..139W/abstract},
	doi = {10.1088/0004-637X/738/2/139},
	language = {en},
	number = {2},
	urldate = {2025-11-17},
	journal = {The Astrophysical Journal},
	author = {Williamson, R. and Benson, B. A. and High, F. W. and Vanderlinde, K. and Ade, P. a. R. and Aird, K. A. and Andersson, K. and Armstrong, R. and Ashby, M. L. N. and Bautz, M. and Bazin, G. and Bertin, E. and Bleem, L. E. and Bonamente, M. and Brodwin, M. and Carlstrom, J. E. and Chang, C. L. and Chapman, S. C. and Clocchiatti, A. and Crawford, T. M. and Crites, A. T. and de Haan, T. and Desai, S. and Dobbs, M. A. and Dudley, J. P. and Fazio, G. G. and Foley, R. J. and Forman, W. R. and Garmire, G. and George, E. M. and Gladders, M. D. and Gonzalez, A. H. and Halverson, N. W. and Holder, G. P. and Holzapfel, W. L. and Hoover, S. and Hrubes, J. D. and Jones, C. and Joy, M. and Keisler, R. and Knox, L. and Lee, A. T. and Leitch, E. M. and Lueker, M. and Luong-Van, D. and Marrone, D. P. and McMahon, J. J. and Mehl, J. and Meyer, S. S. and Mohr, J. J. and Montroy, T. E. and Murray, S. S. and Padin, S. and Plagge, T. and Pryke, C. and Reichardt, C. L. and Rest, A. and Ruel, J. and Ruhl, J. E. and Saliwanchik, B. R. and Saro, A. and Schaffer, K. K. and Shaw, L. and Shirokoff, E. and Song, J. and Spieler, H. G. and Stalder, B. and Stanford, S. A. and Staniszewski, Z. and Stark, A. A. and Story, K. and Stubbs, C. W. and Vieira, J. D. and Vikhlinin, A. and Zenteno, A.},
	month = sep,
	year = {2011},
	pages = {139},
}

@article{jee_scaling_2011,
	title = {Scaling {Relations} and {Overabundance} of {Massive} {Clusters} at z \&gt;{\textasciitilde} 1 from {Weak}-lensing {Studies} with the {Hubble} {Space} {Telescope}},
	volume = {737},
	issn = {0004-637X},
	url = {https://ui.adsabs.harvard.edu/abs/2011ApJ...737...59J/abstract},
	doi = {10.1088/0004-637X/737/2/59},
	language = {en},
	number = {2},
	urldate = {2025-11-17},
	journal = {The Astrophysical Journal},
	author = {Jee, M. J. and Dawson, K. S. and Hoekstra, H. and Perlmutter, S. and Rosati, P. and Brodwin, M. and Suzuki, N. and Koester, B. and Postman, M. and Lubin, L. and Meyers, J. and Stanford, S. A. and Barbary, K. and Barrientos, F. and Eisenhardt, P. and Ford, H. C. and Gilbank, D. G. and Gladders, M. D. and Gonzalez, A. and Harris, D. W. and Huang, X. and Lidman, C. and Rykoff, E. S. and Rubin, D. and Spadafora, A. L.},
	month = aug,
	year = {2011},
	pages = {59},
}

@article{brodwin_x-ray_2011,
	title = {X-ray {Emission} from {Two} {Infrared}-selected {Galaxy} {Clusters} at z \&gt; 1.4 in the {IRAC} {Shallow} {Cluster} {Survey}},
	volume = {732},
	issn = {0004-637X},
	url = {https://ui.adsabs.harvard.edu/abs/2011ApJ...732...33B/abstract},
	doi = {10.1088/0004-637X/732/1/33},
	language = {en},
	number = {1},
	urldate = {2025-11-17},
	journal = {The Astrophysical Journal},
	author = {Brodwin, M. and Stern, D. and Vikhlinin, A. and Stanford, S. A. and Gonzalez, A. H. and Eisenhardt, P. R. and Ashby, M. L. N. and Bautz, M. and Dey, A. and Forman, W. R. and Gettings, D. and Hickox, R. C. and Jannuzi, B. T. and Jones, C. and Mancone, C. and Miller, E. D. and Moustakas, L. A. and Ruel, J. and Snyder, G. and Zeimann, G.},
	month = may,
	year = {2011},
	pages = {33},
}

@article{helton_identification_2024,
	title = {Identification of {High}-redshift {Galaxy} {Overdensities} in {GOODS}-{N} and {GOODS}-{S}},
	volume = {974},
	issn = {0004-637X},
	url = {https://ui.adsabs.harvard.edu/abs/2024ApJ...974...41H/abstract},
	doi = {10.3847/1538-4357/ad6867},
	language = {en},
	number = {1},
	urldate = {2025-11-17},
	journal = {The Astrophysical Journal},
	author = {Helton, Jakob M. and Sun, Fengwu and Woodrum, Charity and Hainline, Kevin N. and Willmer, Christopher N. A. and Rieke, Marcia J. and Rieke, George H. and Alberts, Stacey and Eisenstein, Daniel J. and Tacchella, Sandro and Robertson, Brant and Johnson, Benjamin D. and Baker, William M. and Bhatawdekar, Rachana and Bunker, Andrew J. and Chen, Zuyi and Egami, Eiichi and Ji, Zhiyuan and Maiolino, Roberto and Willott, Chris and Witstok, Joris},
	month = oct,
	year = {2024},
	pages = {41},
}

@article{wang_revealing_2018,
	title = {Revealing the {Environmental} {Dependence} of {Molecular} {Gas} {Content} in a {Distant} {X}-{Ray} {Cluster} at z = 2.51},
	volume = {867},
	issn = {0004-637X},
	url = {https://ui.adsabs.harvard.edu/abs/2018ApJ...867L..29W/abstract},
	doi = {10.3847/2041-8213/aaeb2c},
	language = {en},
	number = {2},
	urldate = {2025-11-17},
	journal = {The Astrophysical Journal},
	author = {Wang, Tao and Elbaz, David and Daddi, Emanuele and Liu, Daizhong and Kodama, Tadayuki and Tanaka, Ichi and Schreiber, Corentin and Zanella, Anita and Valentino, Francesco and Sargent, Mark and Kohno, Kotaro and Xiao, Mengyuan and Pannella, Maurilio and Ciesla, Laure and Gobat, Raphael and Koyama, Yusei},
	month = nov,
	year = {2018},
	pages = {L29},
}

@article{gobat_mature_2011,
	title = {A mature cluster with {X}-ray emission at z = 2.07},
	volume = {526},
	issn = {0004-6361},
	url = {https://ui.adsabs.harvard.edu/abs/2011A&A...526A.133G/abstract},
	doi = {10.1051/0004-6361/201016084},
	language = {en},
	urldate = {2025-11-17},
	journal = {Astronomy and Astrophysics},
	author = {Gobat, R. and Daddi, E. and Onodera, M. and Finoguenov, A. and Renzini, A. and Arimoto, N. and Bouwens, R. and Brusa, M. and Chary, R.-R. and Cimatti, A. and Dickinson, M. and Kong, X. and Mignoli, M.},
	month = feb,
	year = {2011},
	pages = {A133},
}

@article{dannerbauer_excess_2014,
	title = {An excess of dusty starbursts related to the {Spiderweb} galaxy},
	volume = {570},
	issn = {0004-6361},
	url = {https://ui.adsabs.harvard.edu/abs/2014A&A...570A..55D/abstract},
	doi = {10.1051/0004-6361/201423771},
	language = {en},
	urldate = {2025-11-17},
	journal = {Astronomy and Astrophysics},
	author = {Dannerbauer, H. and Kurk, J. D. and De Breuck, C. and Wylezalek, D. and Santos, J. S. and Koyama, Y. and Seymour, N. and Tanaka, M. and Hatch, N. and Altieri, B. and Coia, D. and Galametz, A. and Kodama, T. and Miley, G. and Röttgering, H. and Sanchez-Portal, M. and Valtchanov, I. and Venemans, B. and Ziegler, B.},
	month = oct,
	year = {2014},
	pages = {A55},
}

@article{umehata_alma_2015,
	title = {{ALMA} {Deep} {Field} in {SSA22}: {A} {Concentration} of {Dusty} {Starbursts} in a z = 3.09 {Protocluster} {Core}},
	volume = {815},
	issn = {0004-637X},
	shorttitle = {{ALMA} {Deep} {Field} in {SSA22}},
	url = {https://ui.adsabs.harvard.edu/abs/2015ApJ...815L...8U/abstract},
	doi = {10.1088/2041-8205/815/1/L8},
	language = {en},
	number = {1},
	urldate = {2025-11-17},
	journal = {The Astrophysical Journal},
	author = {Umehata, H. and Tamura, Y. and Kohno, K. and Ivison, R. J. and Alexander, D. M. and Geach, J. E. and Hatsukade, B. and Hughes, D. H. and Ikarashi, S. and Kato, Y. and Izumi, T. and Kawabe, R. and Kubo, M. and Lee, M. and Lehmer, B. and Makiya, R. and Matsuda, Y. and Nakanishi, K. and Saito, T. and Smail, I. and Yamada, T. and Yamaguchi, Y. and Yun, M.},
	month = dec,
	year = {2015},
	pages = {L8},
}

@article{chapman_submillimeter_2009,
	title = {Do {Submillimeter} {Galaxies} {Really} {Trace} the {Most} {Massive} {Dark}-{Matter} {Halos}? {Discovery} of a {High}-z {Cluster} in a {Highly} {Active} {Phase} of {Evolution}},
	volume = {691},
	issn = {0004-637X},
	shorttitle = {Do {Submillimeter} {Galaxies} {Really} {Trace} the {Most} {Massive} {Dark}-{Matter} {Halos}?},
	url = {https://ui.adsabs.harvard.edu/abs/2009ApJ...691..560C/abstract},
	doi = {10.1088/0004-637X/691/1/560},
	language = {en},
	number = {1},
	urldate = {2025-11-17},
	journal = {The Astrophysical Journal},
	author = {Chapman, S. C. and Blain, A. and Ibata, R. and Ivison, R. J. and Smail, I. and Morrison, G.},
	month = jan,
	year = {2009},
	pages = {560--568},
}

@article{lim_is_2021,
	title = {Is there enough star formation in simulated protoclusters?},
	volume = {501},
	issn = {0035-8711},
	url = {https://ui.adsabs.harvard.edu/abs/2021MNRAS.501.1803L/abstract},
	doi = {10.1093/mnras/staa3693},
	language = {en},
	number = {2},
	urldate = {2025-11-17},
	journal = {Monthly Notices of the Royal Astronomical Society},
	author = {Lim, Seunghwan and Scott, Douglas and Babul, Arif and Barnes, David J. and Kay, Scott T. and McCarthy, Ian G. and Rennehan, Douglas and Vogelsberger, Mark},
	month = feb,
	year = {2021},
	pages = {1803--1822},
}

@article{nelson_introducing_2024,
	title = {Introducing the {TNG}-{Cluster} simulation: {Overview} and the physical properties of the gaseous intracluster medium},
	volume = {686},
	issn = {0004-6361},
	shorttitle = {Introducing the {TNG}-{Cluster} simulation},
	url = {https://ui.adsabs.harvard.edu/abs/2024A&A...686A.157N/abstract},
	doi = {10.1051/0004-6361/202348608},
	language = {en},
	urldate = {2025-11-17},
	journal = {Astronomy and Astrophysics},
	author = {Nelson, Dylan and Pillepich, Annalisa and Ayromlou, Mohammadreza and Lee, Wonki and Lehle, Katrin and Rohr, Eric and Truong, Nhut},
	month = jun,
	year = {2024},
	pages = {A157},
}

@article{zheng_measuring_2023,
	title = {Measuring the {X}-ray luminosities of {DESI} groups from {eROSITA} {Final} {Equatorial}-{Depth} {Survey} - {I}. {X}-ray luminosity-halo mass scaling relation},
	volume = {523},
	issn = {0035-8711},
	url = {https://ui.adsabs.harvard.edu/abs/2023MNRAS.523.4909Z/abstract},
	doi = {10.1093/mnras/stad1684},
	language = {en},
	number = {4},
	urldate = {2025-11-17},
	journal = {Monthly Notices of the Royal Astronomical Society},
	author = {Zheng, Yun-Liang and Yang, Xiaohu and He, Min and Shen, Shi-Yin and Li, Qingyang and Li, Xuejie},
	month = aug,
	year = {2023},
	pages = {4909--4922},
}

@article{xue_chandra_2011,
	title = {The {Chandra} {Deep} {Field}-{South} {Survey}: 4 {Ms} {Source} {Catalogs}},
	volume = {195},
	issn = {0067-0049},
	shorttitle = {The {Chandra} {Deep} {Field}-{South} {Survey}},
	url = {https://ui.adsabs.harvard.edu/abs/2011ApJS..195...10X/abstract},
	doi = {10.1088/0067-0049/195/1/10},
	language = {en},
	number = {1},
	urldate = {2025-11-17},
	journal = {The Astrophysical Journal Supplement Series},
	author = {Xue, Y. Q. and Luo, B. and Brandt, W. N. and Bauer, F. E. and Lehmer, B. D. and Broos, P. S. and Schneider, D. P. and Alexander, D. M. and Brusa, M. and Comastri, A. and Fabian, A. C. and Gilli, R. and Hasinger, G. and Hornschemeier, A. E. and Koekemoer, A. and Liu, T. and Mainieri, V. and Paolillo, M. and Rafferty, D. A. and Rosati, P. and Shemmer, O. and Silverman, J. D. and Smail, I. and Tozzi, P. and Vignali, C.},
	month = jul,
	year = {2011},
	pages = {10},
}

@article{carlberg_average_1997,
	title = {The {Average} {Mass} and {Light} {Profiles} of {Galaxy} {Clusters}},
	volume = {478},
	issn = {0004-637X},
	url = {https://ui.adsabs.harvard.edu/abs/1997ApJ...478..462C/abstract},
	doi = {10.1086/303805},
	language = {en},
	number = {2},
	urldate = {2025-11-17},
	journal = {The Astrophysical Journal},
	author = {Carlberg, R. G. and Yee, H. K. C. and Ellingson, E.},
	month = mar,
	year = {1997},
	pages = {462--475},
}

@article{munari_relation_2013,
	title = {The relation between velocity dispersion and mass in simulated clusters of galaxies: dependence on the tracer and the baryonic physics},
	volume = {430},
	issn = {0035-8711},
	shorttitle = {The relation between velocity dispersion and mass in simulated clusters of galaxies},
	url = {https://ui.adsabs.harvard.edu/abs/2013MNRAS.430.2638M/abstract},
	doi = {10.1093/mnras/stt049},
	language = {en},
	number = {4},
	urldate = {2025-11-17},
	journal = {Monthly Notices of the Royal Astronomical Society},
	author = {Munari, E. and Biviano, A. and Borgani, S. and Murante, G. and Fabjan, D.},
	month = apr,
	year = {2013},
	pages = {2638--2649},
}

@article{danese_velocity_1980,
	title = {On velocity dispersions of galaxies in rich clusters.},
	volume = {82},
	issn = {0004-6361},
	url = {https://ui.adsabs.harvard.edu/abs/1980A&A....82..322D/abstract},
	language = {en},
	urldate = {2025-11-17},
	journal = {Astronomy and Astrophysics},
	author = {Danese, L. and de Zotti, G. and di Tullio, G.},
	month = feb,
	year = {1980},
	pages = {322--327},
}

@article{henriques_galaxy_2015,
	title = {Galaxy formation in the {Planck} cosmology - {I}. {Matching} the observed evolution of star formation rates, colours and stellar masses},
	volume = {451},
	issn = {0035-8711},
	url = {https://ui.adsabs.harvard.edu/abs/2015MNRAS.451.2663H/abstract},
	doi = {10.1093/mnras/stv705},
	language = {en},
	number = {3},
	urldate = {2025-11-17},
	journal = {Monthly Notices of the Royal Astronomical Society},
	author = {Henriques, Bruno M. B. and White, Simon D. M. and Thomas, Peter A. and Angulo, Raul and Guo, Qi and Lemson, Gerard and Springel, Volker and Overzier, Roderik},
	month = aug,
	year = {2015},
	pages = {2663--2680},
}

@article{evrard_virial_2008,
	title = {Virial {Scaling} of {Massive} {Dark} {Matter} {Halos}: {Why} {Clusters} {Prefer} a {High} {Normalization} {Cosmology}},
	volume = {672},
	issn = {0004-637X},
	shorttitle = {Virial {Scaling} of {Massive} {Dark} {Matter} {Halos}},
	url = {https://ui.adsabs.harvard.edu/abs/2008ApJ...672..122E/abstract},
	doi = {10.1086/521616},
	language = {en},
	number = {1},
	urldate = {2025-11-17},
	journal = {The Astrophysical Journal},
	author = {Evrard, A. E. and Bialek, J. and Busha, M. and White, M. and Habib, S. and Heitmann, K. and Warren, M. and Rasia, E. and Tormen, G. and Moscardini, L. and Power, C. and Jenkins, A. R. and Gao, L. and Frenk, C. S. and Springel, V. and White, S. D. M. and Diemand, J.},
	month = jan,
	year = {2008},
	pages = {122--137},
}

@article{gobat_wfc3_2013,
	title = {{WFC3} {GRISM} {Confirmation} of the {Distant} {Cluster} {Cl} {J1449}+0856 at langzrang = 2.00: {Quiescent} and {Star}-forming {Galaxy} {Populations}},
	volume = {776},
	issn = {0004-637X},
	shorttitle = {{WFC3} {GRISM} {Confirmation} of the {Distant} {Cluster} {Cl} {J1449}+0856 at langzrang = 2.00},
	url = {https://ui.adsabs.harvard.edu/abs/2013ApJ...776....9G/abstract},
	doi = {10.1088/0004-637X/776/1/9},
	language = {en},
	number = {1},
	urldate = {2025-11-17},
	journal = {The Astrophysical Journal},
	author = {Gobat, R. and Strazzullo, V. and Daddi, E. and Onodera, M. and Carollo, M. and Renzini, A. and Finoguenov, A. and Cimatti, A. and Scarlata, C. and Arimoto, N.},
	month = oct,
	year = {2013},
	pages = {9},
}

@article{papovich_spitzer-selected_2010,
	title = {A {Spitzer}-selected {Galaxy} {Cluster} at z = 1.62},
	volume = {716},
	issn = {0004-637X},
	url = {https://ui.adsabs.harvard.edu/abs/2010ApJ...716.1503P/abstract},
	doi = {10.1088/0004-637X/716/2/1503},
	language = {en},
	number = {2},
	urldate = {2025-11-17},
	journal = {The Astrophysical Journal},
	author = {Papovich, C. and Momcheva, I. and Willmer, C. N. A. and Finkelstein, K. D. and Finkelstein, S. L. and Tran, K.-V. and Brodwin, M. and Dunlop, J. S. and Farrah, D. and Khan, S. A. and Lotz, J. and McCarthy, P. and McLure, R. J. and Rieke, M. and Rudnick, G. and Sivanandam, S. and Pacaud, F. and Pierre, M.},
	month = jun,
	year = {2010},
	pages = {1503--1513},
}

@article{yahil_velocity_1977,
	title = {The {Velocity} {Distribution} of {Galaxies} in {Clusters}},
	volume = {214},
	issn = {0004-637X},
	url = {https://ui.adsabs.harvard.edu/abs/1977ApJ...214..347Y/abstract},
	doi = {10.1086/155257},
	language = {en},
	urldate = {2025-11-17},
	journal = {The Astrophysical Journal},
	author = {Yahil, A. and Vidal, N. V.},
	month = jun,
	year = {1977},
	pages = {347--350},
}
\bibliographystyle{aasjournal}

\end{document}